\documentclass[article]{elsarticle}
\journal{Computer Methods in Applied Mechanics and Engineering}
\usepackage{lineno,hyperref}
\modulolinenumbers[5]
\usepackage{lscape}
\usepackage{graphicx}
\usepackage{amssymb}
\usepackage{amsfonts}
\usepackage[fleqn]{amsmath}

\usepackage{amsthm}
\usepackage{mathtools}
\usepackage{multirow}
\usepackage{float}
\usepackage{xcolor,soul}
\usepackage{bm}
\usepackage{enumerate}
\usepackage{yhmath}
\usepackage{indentfirst}
\usepackage{amsmath}
\usepackage{algorithm}
\usepackage{algpseudocode}
\usepackage{relsize}
\usepackage{soul}
\makeatletter
\def\cl@chapter{\@elt {theorem}}
\makeatother
\usepackage[sort&compress]{cleveref}
\let\cref\labelcref

\DeclareMathOperator*{\argmin}{arg\;min}

\renewcommand*{\arraystretch}{0.9}
\renewcommand{\tabcolsep}{0.12cm}
\usepackage{adjustbox}
\usepackage{etoolbox}
\usepackage{booktabs}
\theoremstyle{definition}
\newtheorem{remark}{Remark}

\usepackage{geometry}
\let\originalleft\left
\let\originalright\right
\renewcommand{\left}{\mathopen{}\mathclose\bgroup\originalleft}
\renewcommand{\right}{\aftergroup\egroup\originalright}
\usepackage{setspace}
\spacing{1.15}
\usepackage{scalerel}
\usepackage{empheq}
\usepackage{siunitx}
\newcommand{\specialcell}[2]{\begin{tabular}[]{@{}#1@{}}#2\end{tabular}}
\newcommand{\resizeboxlarger}[1]{
\resizebox{\ifdim\width>\textwidth\textwidth\else\width\fi}{!}{#1}}
\newcommand{\bs}[1]{\boldsymbol{#1}}
\renewcommand{\bm}[1]{\textbf{\textit{#1}}}
\newcommand{\diff}{\mathop{}\!\mathrm{d}}

\BeforeBeginEnvironment{tabular}{\renewcommand*{\arraystretch}{1.0}\scriptsize}
\makeatletter
\def\blfootnote{\gdef\@thefnmark{}\@footnotetext}
\makeatother
\usepackage{ulem}

\begin{document}
\begin{frontmatter}
\title{Very high-order accurate finite volume scheme for the streamfunction-vorticity formulation of incompressible fluid flows with polygonal meshes on arbitrary curved boundaries}
\author[ad1,ad2]{Ricardo Costa \corref{mycorrespondingauthor}}
\cortext[mycorrespondingauthor]{Corresponding author}
\ead{rcosta@dep.uminho.pt}
\author[ad3]{St\'ephane Clain}
\author[ad4,ad5]{Gaspar J. Machado}
\author[ad1,ad2]{Jo\~ao M. N\'obrega}
\address[ad1]{Institute for Polymers and Composites, University of Minho, Azur\'em \textit{Campus}, 4804-058 Guimar\~aes, Portugal}
\address[ad2]{Department of Polymer Engineering, University of Minho, Azur\'em \textit{Campus}, 4804-058 Guimar\~aes, Portugal}
\address[ad3]{Centre of Mathematics, University of Coimbra, 3000-143 Coimbra, Portugal}
\address[ad4]{Centre of Mathematics, University of Minho, Azur\'em \textit{Campus}, 4800-058 Guimar\~aes, Portugal}
\address[ad5]{Centre of Physics of Minho and Porto Universities, University of Minho, Azur\'em \textit{Campus}, 4800-058 Guimar\~aes, Portugal}
\begin{abstract}
Conventional mathematical models for simulating incompressible fluid flow problems are based on the Navier-Stokes equations expressed in terms of pressure and velocity.
In this context, pressure-velocity coupling is a key issue, and countless numerical techniques and methods have been developed over the decades to solve these equations efficiently and accurately.
In two dimensions, an alternative approach is to rewrite the Navier-Stokes equations regarding two scalar quantities: the streamfunction and the vorticity.
Compared to the primitive variables approach, this formulation does not require pressure to be computed, thereby avoiding the inherent difficulties associated with the pressure-velocity coupling.
However, deriving boundary conditions for the streamfunction and vorticity is challenging.
This work proposes an efficient, high-order accurate finite-volume discretisation of the two-dimensional incompressible Navier-Stokes equations in the streamfunction-vorticity formulation.
A detailed discussion is devoted to deriving the appropriate boundary conditions and their numerical treatment, including on arbitrary curved boundaries.
The reconstruction for off-site data method is employed to avoid the difficulties associated with generating curved meshes to preserve high-orders of convergence in arbitrary curved domains, such as sophisticated meshing algorithms, cumbersome quadrature rules, and intricate non-linear transformations.
This method approximates arbitrary curved boundaries with a conventional linear piecewise approximation, while constrained polynomial reconstructions near the boundary fulfil the prescribed conditions at the physical boundary.
Several incompressible fluid flow test cases in non-trivial 2D curved domains are presented and discussed to demonstrate the accuracy and effectiveness of the proposed methodology in achieving very high orders of convergence.
\end{abstract}
\end{frontmatter}
\section{Introduction}
\label{sec:introduction}


The incompressible Navier-Stokes system in primitive variables (pressure and velocity) is the classical formulation for two- and three-dimensional flows of incompressible viscous fluids.
However, this system poses significant numerical challenges due to its non-linear nature, pressure-velocity coupling enforced by the continuity condition to ensure a solenoidal velocity, and the absence of straightforward boundary conditions for the pressure.
For many decades, extensive literature on numerical techniques and algorithms has been developed to address these issues with regard to accuracy, stability and efficiency.
In addition to advanced numerical methods, alternative equivalent formulations of the incompressible Navier-Stokes equations (equivalent in the continuum, but generally not equivalent when temporally/spatially discretised) have also received substantial attention.
The most prominent approaches and associated issues are reviewed in the following subsections.

\subsection{Streamfunction-vorticity formulation}
\label{subsec:introduction_streamfunction_vorticity_formulation}

Conventional non-primitive formulations of the incompressible Navier-Stokes equations are based on the vorticity transport equation, derived from the curl of the momentum balance equation in primitive variables and is expressed $\left.\partial\omega\middle/\partial t+\nabla\cdot\left(u\omega\right)-\nu\nabla^2\omega=f\right.$.
Vorticity, $\omega$, defined as the curl of the velocity vector, is a pseudo-vector field that measures the local spinning motion of a continuum medium and is useful for understanding fluid dynamics, particularly about turbulence, circulation, and the formation and behaviour of vortices.
In two-dimensional flows, this formulation is particularly convenient, as the vorticity vector has a unique non-zero component, perpendicular to the plane of motion, hence reducing the vorticity transport equation to a scalar equation.

Several formulations arise depending on the second physical quantity and associated transport equation chosen to ensure the continuity condition and close the system of partial differential equations.
Among them, the streamfunction-vorticity formulation in two dimensions is the most classical form of the incompressible Navier-Stokes equations in non-primitive variables, whose development and application became more prominent following foundational papers such as those by Thom, 1933~\cite{1933_thom}, and Woods, 1954~\cite{1954_woods}.
The streamfunction $\psi$ is a potential function whose derivatives correspond to the velocity components; thus, a standard Poisson's equation, in the form $\nabla^{2}\psi=-\omega$, ensures the continuity condition and closes the system of partial differential equations.
The main advantages of the streamfunction-vorticity formulation compared with the primitive-variables counterpart are:
\begin{itemize}
\item The pressure gradient, regarded as a source of numerical complexity in the primitive variables formulation, is implicitly handled and does not need to be computed separately, avoiding the drawbacks associated with the pressure-velocity coupling, such as staggered meshes, stabilisation procedures for collocated variables arrangements, and pressure boundary conditions~\cite{2017_costa1,2017_costa2}.
\item For two-dimensional problems, the number of variables and equations in the discretised problem is reduced since the streamfunction-vorticity formulation consists of two scalar unknowns instead of three as in the primitive-variables formulation (two velocity components and pressure), thus reducing the computational burden.
\item The velocity components are obtained from the streamfunction derivatives, guaranteeing that the continuity condition is always satisfied, both locally and globally, hence eliminating the need to enforce a solenoidal velocity field separately and ensuring that the mass balance remains conserved.
\item Unlike the primitive variables formulation, the streamfunction-vorticity formulation does not require a coupled solution procedure or a projection method for a segregated solution procedure, since the governing equations can be stably solved using a simple iterative algorithm that updates the coupling terms at each iteration.
\end{itemize}

\subsection{Streamfunction-vorticity boundary conditions}
\label{subsec:introduction_streamfunction_vorticity_boundary_conditions}

Besides the advantages of the streamfunction-vorticity formulation, its numerical treatment poses significant challenges regarding boundary conditions.
For the streamfunction, two boundary conditions (Dirichlet and Neumann) can be derived from the specified normal and tangential velocity components on the boundary, resulting in an over-constrained Poisson's equation for the streamfunction.
For vorticity, the conventional practice is to prescribe homogeneous Dirichlet boundary conditions at inlets (assuming that the fluid has no tendency to rotate before entering the domain) and homogeneous Neumann boundary conditions at outlets (assuming that the domain is long enough for the flow to develop).
The most important and challenging task is to prescribe a physically meaningful vorticity boundary condition on solid walls, as no reasonable physical principle seems to exist, which makes the vorticity transport equation under-constrained.
This particular configuration of boundary conditions does not imply that the continuum problem is overdetermined for the streamfunction, but only that the vorticity must be compatible, in some sense, with the prescribed boundary conditions for the streamfunction.
Nevertheless, the lack of boundary conditions for the vorticity transport equation on solid walls, combined with the over-specification of boundary conditions for the streamfunction Poisson's equation, raises significant concerns about the numerical solution of the streamfunction–vorticity formulation.
Three classical approaches are commonly employed to address this difficulty: the pure-streamfunction formulation, local boundary conditions, and global boundary conditions, as follows.

\subsubsection{Pure-streamfunction formulation}
\label{subsubsec:introduction_pure-streamfunction_formulation}

The pure-streamfunction formulation, also known as the streamfunction-velocity formulation, involves eliminating vorticity and introducing a single scalar equation for the streamfunction.
Vorticity is then substituted in the vorticity transport equation using the streamfunction Poisson's equation, resulting in a fourth-order partial differential equation for the streamfunction~\cite{1978_tuann,1983_schreiber,1989_goodrich,1990_goodrich_1,1990_goodrich_2}.
In this case, the simultaneous specification of Dirichlet and Neumann boundary conditions is generally well-posed, as both are required to supplement a fourth-order elliptic operator.
For steady-state Stokes flows without source terms and with negligible inertial effects, in particular, the governing equation becomes the linear homogeneous biharmonic equation, $\nabla^{4}\psi=0$, and can be conveniently solved without iterative procedures.
Although the pure-streamfunction formulation appears effective, it presents several numerical challenges, such as approximations to higher derivatives compared to the streamfunction-vorticity formulation, making it more difficult to preserve convergence orders and solve the associated linear system due to poorer conditioning of the coefficient matrices.

\subsubsection{Local vorticity boundary conditions}
\label{subsubsec:introduction_local_vorticity_boundary_conditions}

Another classical technique to circumvent the lack of vorticity boundary conditions on solid walls involves exploiting the relationship between vorticity and streamfunction at the boundary to derive approximate formulas for wall vorticity.
Artificial boundary conditions for vorticity are then imposed based on the implicit streamfunction (and eventually vorticity) inner values in the vicinity, as well as the given boundary conditions for the streamfunction.
This procedure was widely investigated in the foundational works on the numerical solution of the streamfunction-vorticity formulation using the finite difference method~\cite{1933_thom,1954_woods,1963_fromm_1,1963_fromm_2,1965_pearson,1974_orszag,1979_gupta}.
Several different formulae were proposed and studies were carried out, usually based on a truncated Taylor series representing the discrete approximation of the streamfunction Poisson's equation, incorporating varying numbers of terms and considering only inner values of the streamfunction or also the vorticity, each providing different levels of accuracy and stability to the numerical procedure.
These artificial vorticity boundary conditions are local in nature, requiring only knowledge of nearby inner values and having no direct connection to other nearby boundary points through the prescribed formula (although an indirect relation exists through the common inner values).
Similar techniques based on local vorticity boundary conditions were later proposed for the finite element method~\cite{1972_cheng,1973_baker,1973_taylor,1988_tezduyar,1994_comini} and the control volume-based finite element method \cite{1989_kettleborough,1991_elkaim}, where various approximate formulae for the wall vorticity were employed, based either on finite differences or the element function basis.

Most works on the numerical solution of the streamfunction-vorticity formulation adopted segregated approaches with relaxation solution procedures, as the computational capabilities available at that time could not support a coupled solution procedure due to the memory and calculation requirements.
Consequently, the vorticity transport equation and the streamfunction Poisson's equation were usually solved sequentially, resorting to iterative algorithms wherein wall vorticity was computed between iterations, using the formulae mentioned above.
Consequently, either the Dirichlet or Neumann boundary condition is prescribed for solving the streamfunction Poisson's equations, while the approximated wall vorticity is used to impose the necessary and suitable Dirichlet boundary conditions for the vorticity transport equation.
Some authors have proposed computing wall vorticity alongside the streamfunction Poisson's equation to stabilise the calculation procedure.
Indeed, solving the streamfunction and vorticity equations separately typically requires heavy under-relaxation, especially at high Reynolds numbers, which slows down convergence.
This issue stems not only from the coupling between the streamfunction and vorticity but also, in part, from the artificial boundary conditions used for the vorticity.
In contrast, a directly coupled approach, which requires an implicit treatment of wall vorticity for boundary conditions on solid walls, can alleviate this issue, yielding a much more stable and robust procedure with faster convergence, although this was not the conventional practice at the time.


The traditional coupled approach simultaneously solves the discretised transport equations resulting from the standard discretisation of the vorticity transport and streamfunction Poisson's equations, along with the artificial vorticity boundary conditions.
In contrast, several authors have proposed coupled solution procedures that completely avoid the need for artificial vorticity boundary conditions, as in the finite difference method~\cite{1972_davis}, through an implicit method for solving the boundary-layer equations, and in the finite element method~\cite{1978_campion-renson,1981_dhatt,1984_gunzburger,1987_peeters}, which employs a mixed variational formulation for the Navier-Stokes equations in streamfunction-vorticity formulation.
Indeed, although the absence of artificial vorticity boundary conditions seems more natural in the finite element method, finite difference methods can also be implemented in a way that avoids the need for such conditions at solid walls.
According to some authors~\cite{1984_gunzburger}, the key to avoiding this specification lies not in the use of finite element methods, but rather in solving the discrete system as a coupled set of equations, rather than iterating between the vorticity transport and streamfunction Poisson's equations.
Thus, in a finite difference method, the streamfunction Poisson's equation can be discretised by imposing both Dirichlet and Neumann boundary conditions, yielding more equations than unknowns needed to determine the streamfunction.
Conversely, the vorticity transport equation can be discretised without imposing any boundary conditions, resulting in fewer equations than unknowns to determine the vorticity.
This results in an over-constrained system of discrete equations for the streamfunction that cannot be solved, whereas the system of discrete equations for the vorticity does not have a unique solution.
However, by coupling all the discrete equations, the single coupled system of equations can be solved for both the streamfunction and vorticity simultaneously and uniquely, without prescribing artificial vorticity boundary conditions.

\subsubsection{Global vorticity boundary conditions}
\label{subsubsec:introduction_global_vorticity_boundary_conditions}

A thoroughly different approach to avoid artificial vorticity boundary conditions consists of supplementing the vorticity transport equation with physical boundary conditions that are mathematically equivalent to those originally specified for the velocity.
Quartapelle et al.~\cite{1981_quartapelle_1,1981_quartapelle_2,1984_quatapelle} proved the existence of such equivalent boundary conditions, which take the form of boundary integral constraints for the vorticity.
These are derived from a simple geometric interpretation of fixing the orthogonal projection of the vorticity field onto the linear space of harmonic functions in the physical domain.
Unlike artificial vorticity boundary conditions, these physical vorticity boundary conditions are global in nature, as the entire physical boundary (or boundary points in the discrete problem) are integrated alongside the associated normal and tangential velocities.
Furthermore, the integral constraints are independent of the streamfunction, meaning the vorticity transport equation can be solved in the linear Stokes problem without knowledge of the streamfunction.
Thus, apart from the coupling due to the convective term in the non-linear problem (since velocities must be computed from the streamfunction), global vorticity boundary conditions completely eliminate the dependency of vorticity calculation and evolution on the streamfunction.
A different technique worth mentioning that follows the same philosophy of prescribing non-local physical constraints for the vorticity is due to Anderson, 1989~\cite{1989_anderson}.

Global vorticity boundary conditions restrict the coupling between vorticity and streamfunction to the non-linear convective term in the vorticity transport equation.
They are generally more accurate for complex geometries and flows, but are also more challenging to implement and are more computationally demanding than local boundary conditions.
Indeed, the boundary integral constraints proposed by Quartapelle et al., 1981~\cite{1981_quartapelle_1,1981_quartapelle_2} and 1984~\cite{1984_quatapelle}, require knowledge of all discrete harmonic functions, one per boundary point, each being a linear space with a dimension equal to the number of grid points.
In fact, the number of linearly independent harmonic functions is equal to that of the boundary points, ensuring that the integral constraints provide the correct number of independent relations needed to supplement the vorticity transport equation and create a fully determined problem.

However, while storing that much information may be feasible for two-dimensional problems, it becomes impractical for three-dimensional problems.
Conversely, numerical solutions computed using artificial vorticity boundary conditions are expected to satisfy these constraints at convergence, as the integral constraints' global effect is achieved indirectly through the iteration process~\cite{1993_quartapelle}.
Furthermore, although derived independently, Weinan et al., 1996~\cite{1996_e_1,1996_e_2}, demonstrated that boundary integral constraints are mathematically equivalent to certain local boundary vorticity formulae under specific conditions.
From a theoretical perspective, global boundary conditions are also interesting as they provide an alternative interpretation of the vorticity transport equation in the linear Stokes problem.
Thus, the study of vorticity dynamics addresses the influence of solid walls on the vorticity field without requiring knowledge of the streamfunction.

In conclusion, the most appropriate boundary conditions for vorticity on solid walls remain a controversial and long-debated issue when solving the non-primitive formulation of the Navier-Stokes equations for two-dimensional flows.
A comprehensive literature review and discussion comparing global and local vorticity boundary conditions on solid walls and associated numerical techniques are provided in Gresho, 1991~\cite{1991_gresho}, and in Napolitano et al., 1999~\cite{1999_napolitano}.

\subsection{Streamfunction-vorticity discretisation schemes}
\label{subsec:introduction_streamfunction_vorticity_discretisation_schemes}
  
Apart from classical methods, the discretisation of the Navier-Stokes equations in the streamfunction-vorticity formulation has also seen recent developments.
Hybrid finite difference/finite volume schemes on Cartesian grids have been applied in the context of immersed boundaries for the two-dimensional streamfunction-vorticity formulation~\cite{2001_calhoun,2003_li}.
Unlike most previous work, the authors directly appeal to the notion of vorticity generation to identify singular sources of vorticity at solid boundaries instead of attempting to impose boundary conditions on the vorticity.

Another topic that has received considerable attention regarding the numerical solution of the Navier-Stokes equation in the streamfunction-vorticity formulation is the development of compact finite difference schemes~\cite{1989_dennis,1991_gupta,1995_li,1995_spotz,1996_e_3,2001_li}.
Compact difference schemes restrict approximations of the underlying solution derivatives to the neighbouring cells surrounding any given grid point, usually based on point-wise values and derivatives.
Compact schemes are particularly useful for achieving high-order of convergence without increasing the stencil size, which is typically required when implementing high-order accurate finite difference formulas.
On the other hand, the application of compact finite difference schemes to the pure-streamfunction formulation is fairly recent and primarily motivated to avoid wide stencils, which are needed to approximate fourth derivatives, even for the second-order accurate case~\cite{2001_kupferman,2005_ben-artzi,2005_gupta,2006_ben-artzi,2010_ben-artzi,2017_fishelov}.
Although these schemes can achieve high-order of convergence, their practical application is essentially limited to Cartesian grids on simple geometries; otherwise, more sophisticated techniques for immersed boundaries or general curvilinear coordinates must be implemented.

\subsection{Novelties of the article}
\label{subsec:introduction_novelties_of_the_article}

The vast majority of the literature on the numerical solution of the two-dimensional incompressible Navier-Stokes equations in the streamfunction-vorticity formulation is based on Cartesian grids.
Although some relatively recent papers address the discretisation of these equations on unstructured meshes with finite cell methods and on Cartesian grids with immersed boundary methods, the treatment of complex real-world problems using non-primitive variable formulations remains largely undeveloped.
Moreover, apart from compact finite difference schemes, which are essentially limited to Cartesian grids, the development of high-order accurate discretisation schemes for the streamfunction-vorticity formulation on unstructured meshes has not yet been undertaken, which is the primary objective of the present work.
This objective entails two main challenges, described hereafter.

Firstly, the approximation of wall vorticity for prescribing suitable boundary conditions is extremely delicate, as an incorrect evaluation can lead to a significant loss of convergence order.
In particular, the adopted strategy cannot rely on classical techniques based on Taylor expansion due to the use of unstructured meshes and must consider the boundary curvature for arbitrary high orders of convergence.
Indeed, the inclusion of the boundary curvature in the wall vorticity calculation is not present in the literature.

Secondly, the development of high-order accurate methods for curved domains requires a specific treatment of boundary conditions to avoid the second-order of convergence limitation caused by the geometrical mismatch between the physical and computational boundaries when using linear piecewise meshes.
The classical approach to treating boundary conditions on arbitrary curved boundaries involves using curved mesh elements to eliminate such a geometrical mismatch and recover optimal convergence orders.

\begin{figure}[!htb]
\centering
\begin{tabular}{c@{\hskip 1.5cm}c}
\includegraphics[width=7.0cm]{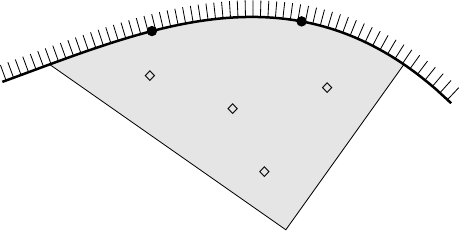}
& \includegraphics[width=7.0cm]{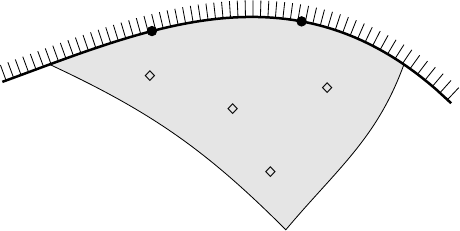}\\
(a) Cell with curved boundary edge.
& (b) Cell with curved boundary and inner edges.\\[1.0cm]
\multicolumn{2}{c}{\includegraphics[width=7.0cm]{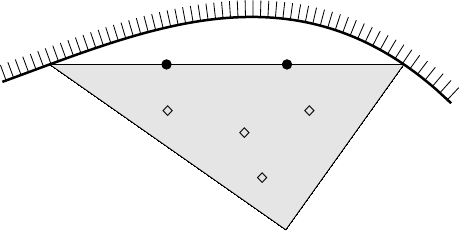}}\\
\multicolumn{2}{c}{(c) Cell with only straight edges.}
\end{tabular}
\caption{Curved and polygonal mesh elements with illustrative quadrature points on the edges (solid circles) and in the cells (empty diamonds).}
\label{fig:introduction_curved_boundary}
\end{figure}

Curved mesh approaches, such as the standard isoparametric elements method, involve curved elements that fit the physical curved domain (at least, up to a certain degree) to prevent accuracy deterioration.
To this end, sophisticated and computationally intensive algorithms are employed, and besides the boundary edges, curved inner edges may also be necessary to better accommodate the mesh elements (see Figures~\ref{fig:introduction_curved_boundary}(a) and~\ref{fig:introduction_curved_boundary}(b)).
Depending on the context, quadrature points on both curved edges and curved cells must be determined, which is a challenging task (especially in three-dimensional problems), along with complex non-linear transformations for the mapping of the elements.
On the other hand, polygonal mesh approaches avoid such difficulties by using standard mesh generation algorithms and, therefore, standard quadrature rules (see Figure~\ref{fig:introduction_curved_boundary}(c)).
However, accuracy deterioration is avoided only if specific numerical techniques are employed to recover the optimal convergence order.
The interested reader is referred to Costa et al., 2023~\cite{2018_costa}, for a comprehensive discussion and literature review.

The reconstruction for off-site data (ROD) method is a novel technique initially proposed for convection-diffusion equations~\cite{2018_costa,2019_costa1,2019_costa2,2021_costa,2022_costa1,2024_santos}, and recently extended to the incompressible Stokes and Navier–Stokes equations~\cite{2022_costa2,2023_costa}.
This method handles arbitrary curved boundaries with a linear piecewise approximation of the physical boundary and employs polynomial reconstructions with specific linear constraints to convert the boundary conditions prescribed on the physical boundary into equivalent conditions on the computational boundary.
Based on the ROD technique, this work proposes a very high-order accurate finite volume scheme on unstructured meshes for the streamfunction-vorticity formulation of incompressible fluid flows using linear piecewise meshes on arbitrary curved boundaries.

\subsection{Organisation of the article}
\label{subsec:introduction_organisation_of_the_article}

The remaining sections of the article are organised as follows.
The problem statement for the two-dimensional incompressible Navier-Stokes equations and the mathematical derivation of the streamfunction-vorticity formulation are presented in Section~2, whereas the derivation of suitable boundary conditions for the non-primitive variables is discussed in Section~3.
The numerical techniques related to the polynomial reconstruction method and the reconstruction for off-site data method are recalled in Section~4, and employed in Section~5 for the finite volume discretisation of the streamfunction-vorticity formulation.
Section~6 is dedicated to the verification of the proposed method in terms of accuracy, convergence order, and computational efficiency, featuring several benchmark test cases with analytical or reference solutions.
The article ends with the main work conclusions provided in Section~7.


\section{Mathematical formulation}
\label{sec:mathematical_formulation}

\begin{figure}[!htb]
\centering
\begin{tabular}{c@{\hskip 1.5cm}c}
\includegraphics[width=0.4\textwidth]{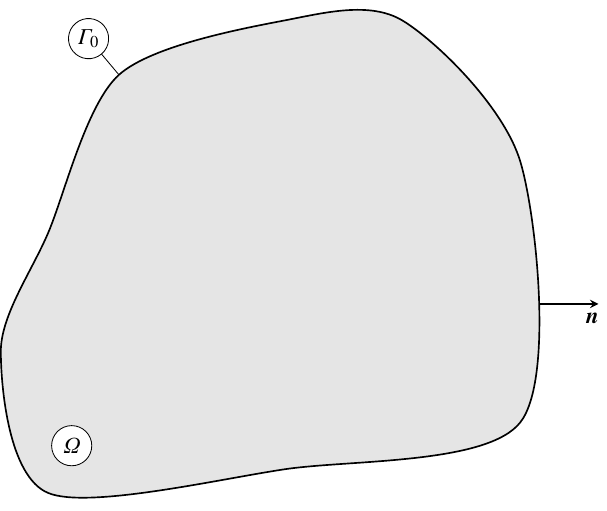}
&\includegraphics[width=0.4\textwidth]{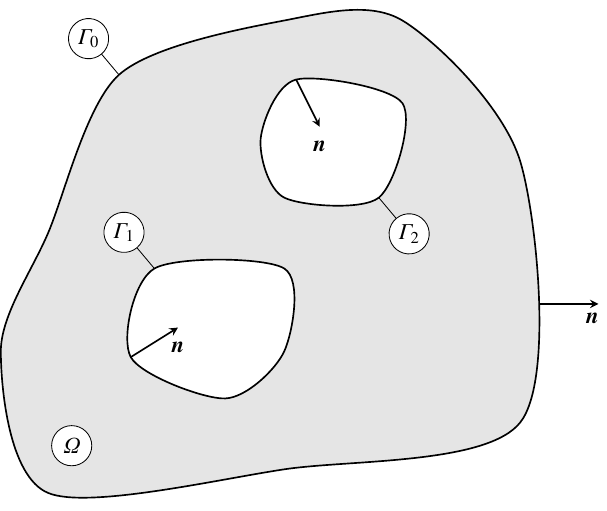}\\[0.5cm]
(a) Simply connected. & (b) Multiply connected.
\end{tabular}
\caption{Curved physical domain.}
\label{fig:mathematical_formulation_domain}
\end{figure}

The steady-state flow of an incompressible Newtonian fluid is studied in a two-dimensional simply or multiply connected physical domain $\Omega$ with arbitrary smooth curved physical boundary $\Gamma$.
Using the Cartesian coordinate system $\bm{x}\coloneqq\left(x,y\right)$, the associated outward unit normal vector is $\bm{n}\coloneqq\bm{n}\left(\bm{x}\right)\coloneqq\left(n_{x}\left(\bm{x}\right),n_{y}\left(\bm{x}\right)\right)$, and the unit tangential vector is $\bm{t}\coloneqq\bm{t}\left(\bm{x}\right)\coloneqq\left(t_{x}\left(\bm{x}\right),t_{y}\left(\bm{x}\right)\right)$ with $\bm{t},\bm{n}$ positively oriented (see Figure~\ref{fig:mathematical_formulation_domain}).
For multiply connected domains (basically, any domain with holes), each closed path representing a physical boundary is denoted as $\Gamma^{k}$, such that $\displaystyle \Gamma=\cup_{k=0}^{K}\Gamma^{k}$, where $K$ is the number of holes.
For convenience, $\Gamma^{0}$ corresponds to the external physical boundary, such that for $K=0$ (without holes), $\Gamma=\Gamma^{0}$ corresponds to a simply connected domain.

\subsection{Primitive formulation}
\label{subsec:mathematical_formulation_primitive_formulation}

The primitive formulation of the steady-state two-dimensional incompressible Navier-Stokes equations for a Newtonian fluid with constant dynamic viscosity $\mu>0$ and constant density $\rho>0$ consists of the momentum and mass balance equations, given as
\begin{alignat}{4}
\left(\bm{u}\cdot\nabla\right)\bm{u}-\nu\nabla^{2}\bm{u}+\frac{1}{\rho}\nabla p\;=\;&\bm{f},&&\qquad\text{in }\Omega,\label{eq:mathematical_formulation_momentum_equation_1}\\
\nabla\cdot\bm{u}\;=\;&0,&&\qquad\text{in }\Omega,\label{eq:mathematical_formulation_continuity_equation_1}
\end{alignat}
respectively, where $\nu=\mu/\rho$ is the constant kinematic viscosity, $\bm{u}\coloneqq\bm{u}\left(\bm{x}\right)\coloneqq\left(u_{x},u_{y}\right)$ with $u_{x}\coloneqq u_{x}\left(\bm{x}\right)$ and $u_{y}\coloneqq u_{y}\left(\bm{x}\right)$ is the unknown velocity function, $p\coloneqq p\left(\bm{x}\right)$ is the unknown pressure function, while $\bm{f}\coloneqq\bm{f} \left(\bm{x}\right) \coloneqq\left(f_{x},f_{y}\right)$ with $f_{x}\coloneqq f_{x}\left(\bm{x}\right)$ and $f_{y}\coloneqq f_{y}\left(\bm{x}\right)$ is a given body force function per unit area acting on the continuum, for instance, the gravitational force.
The system of equations~\cref{eq:mathematical_formulation_momentum_equation_1,eq:mathematical_formulation_continuity_equation_1} is often referred to as the primitive steady-state formulation of the incompressible Navier-Stokes equations, which relates the primitive pressure and velocity unknowns.

\begin{remark}
In the present work, the differential operator $\nabla\cdot\bm{A}$ for a matrix $\bm{A}$ follows the usual convention of a column-wise divergence operator, whereas $\nabla\bm{a}$ and $\nabla^{2}\bm{a}$ for a vector $\bm{a}$ follow the standard conventions for component-wise gradient and Laplace operators, respectively.
\end{remark}

\subsection{Vorticity transport equation}
\label{subsec:mathematical_formulation_vorticity_transport_equation}

In three-dimensional fluid flows, vorticity is a physical pseudovector field that describes the local rotational motion of the fluid and is defined as the curl of the velocity function.
For planar fluid flows in the $xy$-plane, it follows that the vorticity vector always points perpendicularly to the plane of motion, such as it can be represented with a scalar function $\omega\coloneqq\omega\left(\bm{x}\right)$ in two-dimensional problems, given as
\begin{equation}
\label{eq:mathematical_formulation_vorticity_definition_1}
\omega=\frac{\partial u_{y}}{\partial x}-\frac{\partial u_{x}}{\partial y}.
\end{equation}

The vorticity transport equation can be derived from taking the curl of the momentum balance equation in three-dimensions (see~\ref{appendix:derivation_of_the_vorticity_transport_equation}).
For planar fluid flows in the $xy$-plane, the components parallel to the plane of motion vanish and, therefore, in two-dimensional fluid flows the vorticity transport equation reduces to a scalar equation (the perpendicular component) for the scalar vorticity as, $\omega$, given as
\begin{equation}
\label{eq:mathematical_formulation_vorticity_equation_1}
\nabla\cdot\left(\bm{u}\omega\right)-\nu\nabla^{2}\omega\;=\;f,\qquad\text{in }\Omega,
\end{equation}
where $f\coloneqq f\left(\bm{x}\right)$ is the perpendicular component of the curl of the body force function $\bm{f}=(f_x,f_y)$ per unit area, given as
\begin{equation}
f=\frac{\partial f_{y}}{\partial x}-\frac{\partial f_{x}}{\partial y}.
\end{equation}

\begin{remark}
In two-dimensional fluid flows, the vorticity transport equation can also be obtained by subtracting the derivatives of the first and second components of Equation~\cref{eq:mathematical_formulation_momentum_equation_1} with respect to $y$ and $x$, respectively, in that order.
\end{remark}

\subsection{Incompressibility condition}
\label{subsec:mathematical_formulation_incompressibility_condition}

In two-dimensional fluid flows, the incompressibility condition can be satisfied exactly if the velocity field is expressed in terms of a scalar streamfunction $\psi\coloneqq\psi\left(\bm{x}\right)$ whose derivatives correspond to the velocity components, as
\begin{equation}
\label{eq:mathematical_formulation_streamfunction_definition}
u_{x}=\frac{\partial\psi}{\partial y}
\qquad
\text{and}
\qquad
u_{y}=-\frac{\partial\psi}{\partial x},
\qquad\text{in }\Omega,
\end{equation}
or, in compact form, given as $\bm{u}=\nabla^{\perp}\psi$.
From the above definition, it follows that the streamfunction is well-defined only up to an additive constant and always satisfies the continuity equation intrinsically since
\begin{equation}
\nabla\cdot\bm{u}=\frac{\partial u_{x}}{\partial x}+\frac{\partial u_{y}}{\partial y}=\frac{\partial}{\partial x}\left(\frac{\partial\psi}{\partial y}\right)+\frac{\partial}{\partial y}\left(-\frac{\partial\psi}{\partial x}\right)=0.
\end{equation}

The streamfunction-vorticity formulation currently has four variables ($\omega$, $\psi$, $u_{x}$, and $u_{y}$) but only three equations (the vorticity transport equation and the relationship between streamfunction and velocity components).
To close the system for the incompressible Navier-Stokes equations, a fourth equation is obtained after substituting the definition of streamfunction, in Equation~\cref{eq:mathematical_formulation_streamfunction_definition}, into the definition of vorticity, in Equation~\cref{eq:mathematical_formulation_vorticity_definition_1}, yielding
\begin{equation}
\label{eq:mathematical_formulation_streamfunction_equation_1}
\nabla^{2}\psi\;=-\omega,\qquad\text{in }\Omega.
\end{equation}

\subsection{Streamfunction-vorticity formulation}
\label{subsec:mathematical_formulation_streamfunction_vorticity_formulation}

The complete system for the streamfunction-vorticity formulation, apart from boundary conditions, is given as
\begin{alignat}{4}
\nabla\cdot\left(\bm{u}\omega\right)-\nu\nabla^{2}\omega\;=\;&f,&&\qquad\text{in }\Omega,\\
\bm{u}\;=\;&\nabla^{\perp}\psi,&&\qquad\text{in }\Omega,\\
\nabla^{2}\psi\;=\;&-\omega,&&\qquad\text{in }\Omega,
\end{alignat}
which is non-linear due to the convective term in the vorticity transport equation.
In that regard, the so-called Picard method (also known as fixed-point iteration) is employed to linearise the vorticity transport equation by considering an additional velocity function $\bm{v}\coloneqq\bm{v}\left(\bm{x}\right)\coloneqq\left(v_{x}\left(\bm{x}\right),v_{y}\left(\bm{x}\right)\right)$ and removing the second equation from the system, yielding the linearised steady-state incompressible Navier-Stokes equations in non-primitive variables, given as
\begin{alignat}{4}
\nabla\cdot\left(\bm{v}\omega\right)-\nu\nabla^{2}\omega\;=\;&f,&&\qquad\text{in }\Omega,
\label{eq:mathematical_formulation_vorticity_equation_2}\\
\nabla^{2}\psi+\omega\;=\;&0,&&\qquad\text{in }\Omega,
\label{eq:mathematical_formulation_streamfunction_equation_2}
\end{alignat}
with the unknown terms arranged on the left-hand side.

An iterative procedure is applied such that, at each fixed-point iteration, function $\bm{v}\left(\bm{x}\right)$ corresponds to the previously computed velocity function, $\bm{u}\left(\bm{x}\right)$, obtained from the streamfunction.
Therefore, the solution of the above system of equations converges to the solution of the steady-state streamfunction-vorticity formulation, provided that function $\bm{v}\left(\bm{x}\right)$ converges to $\bm{u}\left(\bm{x}\right)$.
Notice that both functions $\bm{u}\left(\bm{x}\right)$ and $\bm{v}\left(\bm{x}\right)$ stand for a velocity field, where the later is known and, therefore, becomes a physical coefficient in the linearised problem.
Introducing function $\phi\coloneqq\phi\left(\bm{x}\right)$ corresponding to the streamfunction at the previous fixed-point iteration, then $\bm{v}\left(\bm{x}\right)=\nabla^{\perp}\phi\left(\bm{x}\right)$.

\section{Boundary conditions}
\label{sec:boundary_conditions}

The most challenging task in formulating the incompressible Navier-Stokes equations in non-primitive variables is deriving the appropriate boundary conditions from those prescribed in primitive variables.
In this paper, only Dirichlet boundary conditions prescribed on the physical boundary for the velocity are considered, given as
\begin{equation}
\label{eq:boundary_conditions_velocity_dirichlet_condition_1}
\bm{u}\;=\;\bm{u}^{\textrm{B}},\qquad\text{on }\Gamma,
\end{equation}
where $\bm{u}^{\textrm{B}}\coloneqq\bm{u}^{\textrm{B}}\left(\bm{x}\right)\coloneqq\left(u^{\textrm{B}}_{x}\left(\bm{x}\right),u^{\textrm{B}}_{y}\left(\bm{x}\right)\right)$ is a given velocity function on the physical boundary.

\subsection{Local coordinate system}
\label{subsec:boundary_conditions_local_coordinate_system}

It is convenient to define the problem variables in terms of a local coordinate system to deal with arbitrary smooth curved physical boundaries.
For each point $\bm{p}$ on the boundary, a  negatively oriented vector basis $\bm{E}^{\textrm{L}}\left(\bm{p}\right)=\lbrace\bm{t}_\bm{p},\bm{n}_\bm{p}\rbrace$ with $\bm{t}_\bm{p}=\bm{t}\left(\bm{p}\right)$ and $\bm{n}_\bm{p}=\bm{n}\left(\bm{p}\right)$ is introduced for a local Cartesian coordinate system defined as
$$
\bm{x}\to \bs{\eta}_\bm{p}\coloneqq\bs{\eta}_\bm{p}\left(\bm{x}\right)\coloneqq\left(\eta\left(\bm{x}\right),\xi\left(\bm{x}\right)\right)=\big ((\bm{x}-\bm{p})\cdot \bm{t}_\bm{p},(\bm{x}-\bm{p})\cdot \bm{n}_\bm{p}\big ).
$$
Note that this coordinate system does not involve curvilinear coordinates that follow the boundary.
Moreover, the transformation from the global coordinate system to the local coordinate system (and vice-versa) is a linear orthogonal transformation.
As such, for general curved boundaries, the local coordinate system associated with a reference point $\bm{p}$ (local coordinate system origin), is aligned only with the boundary normal and tangential vectors at the reference point location.
That is, for a point $\bm{q}$ next to the reference point, the associated normal and tangential vectors, $\bm{n}\left(\bm{q}\right)$ and $\bm{t}\left(\bm{q}\right)$, are not generally aligned with the local coordinate system associated to $\bm{p}$ as it does not follow the boundary curvature.

From the definition of the local coordinate system, it follows that, for a fixed boundary point $\bm{p}$ and the associated local coordinate system, the problem variables and boundary conditions are defined as follows:
\begin{itemize}
\item The velocity field in the local coordinate system, $\bm{u}^{\textrm{L}}\coloneqq\bm{u}^{\textrm{L}}\left(\bs{\eta}\right)\coloneqq\left(u_{\eta},u_{\xi}\right )$, with $u_{\eta}\coloneqq u_{\eta}\left(\bs{\eta}\right)$ and $u_{\xi}\coloneqq u_{\xi}\left(\bs{\eta}\right)$, is defined as
\begin{equation}
\label{eq:boundary_conditions_velocity_local_definition}
u_{\eta}=\bm{u}\cdot\bm{n}_\bm{p}
\qquad
\text{and}
\qquad
u_{\xi}=\bm{u}\cdot\bm{t}_\bm{p},
\end{equation}
\item The vorticity field in the local coordinate system, $\omega^{\textrm{L}}\coloneqq\omega^{\textrm{L}}\left(\bs{\eta}\right)$, is defined as
\begin{equation}
\label{eq:boundary_conditions_vorticity_local_definition}
\omega^{\textrm{L}}=\frac{\partial u_{\xi}}{\partial\eta}-\frac{\partial u_{\eta}}{\partial\xi}.
\end{equation}

\item The streamfunction field in the local coordinate system, $\psi^{\textrm{L}}\coloneqq\psi^{\textrm{L}}\left(\bs{\eta}\right)$, is defined as
\begin{equation}
\label{eq:boundary_conditions_streamfunction_local_definition}
u_{\eta}=\frac{\partial\psi^{\textrm{L}}}{\partial\xi}
\qquad
\text{and}
\qquad
u_{\xi}=-\frac{\partial\psi^{\textrm{L}}}{\partial\eta}.
\end{equation}

\item The boundary velocity in the local coordinate system, $\bm{u}^{\textrm{L},\textrm{B}}\coloneqq\bm{u}^{\textrm{L},\textrm{B}}\left(\bs{\eta}\right)\coloneqq\left(u^{\textrm{B}}_{\eta},u^{\textrm{B}}_{\xi}\right)$, with $u^{\textrm{B}}_{\eta}\coloneqq u^{\textrm{B}}_{\eta}\left(\bs{\eta}\right)$ and $u^{\textrm{B}}_{\xi}\coloneqq u^{\textrm{B}}_{\xi}\left(\bs{\eta}\right)$, is defined as
\begin{equation}
\label{eq:boundary_conditions_velocity_dirichlet_condition_2}
u^{\textrm{B}}_{\eta}=\bm{u}^{\textrm{B}}\cdot\bm{n}_\bm{p}
\qquad
\text{and}
\qquad
u^{\textrm{B}}_{\xi}=\bm{u}^{\textrm{B}}\cdot\bm{t}_\bm{p}.
\end{equation}
\end{itemize}

For general curved boundaries, note that for any point $\bs{\eta}\in\Gamma$ in the local coordinate system associated with a reference point $\bm{p}$, the components $u^{\textrm{B}}_{\eta}\left(\bs{\eta}\right)$ and $u^{\textrm{B}}_{\xi}\left(\bs{\eta}\right)$ correspond to the normal and tangential velocity only at the local coordinate system origin, $\bm{p}$.
Finally, the vorticity definition in the local coordinate system should yield the same signal as Equation~\cref{eq:mathematical_formulation_vorticity_definition_1}, hence the orientation of the tangent vector on the boundary is not arbitrary and $\bm{t}\left(\bm{p}\right)=\left(-n_{y}\left(\bm{p}\right),n_{x}\left(\bm{p}\right)\right)$ must be chosen.

\subsection{Boundary streamfunction}
\label{subsec:boundary_conditions_boundary_streamfunction}

From the streamfunction definition, in Equation~\cref{eq:boundary_conditions_streamfunction_local_definition}, it follows that the normal velocity on the physical boundary corresponds to its tangential derivative.
Since the streamfunction is only defined up to an additive constant, assume an arbitrary point $\bm{x}_{k}$ on physical boundary subset $\Gamma^{k}$, $k=0,\ldots,K$, which corresponds to an arbitrary streamfunction value $C_{k}$.
Therefore, it follows that the boundary integration of $\bm{u}^{\textrm{B}}\left(\bm{x}\right)\cdot\bm{n}\left(\bm{x}\right)$ between $\bm{x}_{k}$ and another point $\bm{x}$ on the same physical boundary subset, gives the boundary streamfunction, $\psi^{\textrm{B}}\coloneqq\psi^{\textrm{B}}\left(\bm{x}\right)$, that is
\begin{equation}
\label{eq:boundary_conditions_boundary_streamfunction}
\psi^{\textrm{B}}(\bm{x})=\int_{\wideparen{\bm{x}_{k}\bm{x}}}\bm{u}^{\textrm{B}}\cdot\bm{n}\diff{s}+C_{k},\qquad\text{on }\Gamma^{k},
\end{equation}
where $\wideparen{\bm{x}_{k}\bm{x}}$ is the boundary path between $\bm{x}_{k}$ and $\bm{x}$ and each physical boundary subset is assumed to admit a parametrisation $\bm{x}\coloneqq\left(x\left(s\right),y\left(s\right)\right)$.
Note that, as a necessary condition for the streamfunction to exist, the total mass flux on each closed physical boundary must vanish~\cite{1979_girault}, that is
\begin{equation}
\oint_{\Gamma^{k}}\bm{u}^{\textrm{B}}\cdot\bm{n}\diff{s}=0,\qquad k=0,\ldots,K.
\end{equation}

For multiply connected domains, the presence of $K+1$ arbitrary constants $C_{k}$ indicates that the problem has multiple solutions, as different choices of these constants generally lead to different velocity fields, all of which are valid solutions.
However, since the streamfunction is only defined up to an additive constant, only one constant can be fixed arbitrarily, leaving the remaining $K$ constants undetermined.
For convenience, assume that $C_{0}$ is fixed arbitrarily and that $C_{k}$, $k=1,\ldots,K$, are to be determined.
An effective practice is to determine these constants as part of the overall solution, meaning that the problem has $K$ additional unknown scalar constants.
Then, to close the system, an integral condition on the vorticity flux along the associated physical boundary subsets is prescribed (see~\ref{appendix:derivation_of_the_compatibility_condition}), given as
\begin{equation}
\label{eq:boundary_conditions_compatibility_condition}
\oint_{\Gamma^{k}}\left(\bm{u}\omega-\nu\nabla\omega\right)\cdot\bm{n}\diff{s}=\oint_{\Gamma^{k}}\bm{f}\cdot\bm{n}\diff{s},\qquad k=1,\ldots,K.
\end{equation}
This approach provides $K$ scalar constraints (compatibility conditions) on the vorticity flux, which fixes the $K$ unknown constants for the boundary streamfunction.

Still from the streamfunction definition, in Equation~\cref{eq:boundary_conditions_streamfunction_local_definition}, it follows that the tangential velocity on the physical boundary corresponds to the boundary streamfunction normal derivative, that is
\begin{equation}
\label{eq:boundary_conditions_boundary_streamfunction_derivative}
\nabla\psi^{\textrm{B}}\cdot\bm{n}=-\bm{u}^{\textrm{B}}\cdot\bm{t},\qquad\text{on }\Gamma.
\end{equation}

\begin{remark}
In the special case of internal flows with the no-slip condition, $\bm{u}^{\textrm{B}}\left(\bm{x}\right)\cdot\bm{t}\left(\bm{x}\right)=0$, and impermeable walls, $\bm{u}^{\textrm{B}}\left(\bm{x}\right)\cdot\bm{n}\left(\bm{x}\right)=0$, on the physical boundary, the streamfunction boundary value is simply given as $\psi^{\textrm{B}}\left(\bm{x}\right)=C_{k}$ and its normal derivative as $\nabla\psi^{\textrm{B}}\cdot\bm{n}=0$.
\end{remark}


\subsection{Boundary vorticity}
\label{subsec:boundary_conditions_boundary_vorticity}

Obtaining the boundary vorticity function is more challenging since, in general, the given boundary velocity function is not sufficient to completely derive the associated boundary vorticity from the definition~\cref{eq:boundary_conditions_vorticity_local_definition}.
In this regard, the conventional practice involves substituting the known derivative term in the vorticity definition from the given boundary velocity function and providing an implicit approximation for the unknown derivative term.
Despite this hurdle, boundary vorticity is treated similarly for both simply and multiply connected physical domains.

\subsubsection{Polygonal domains}
\label{subsubsec:boundary_conditions_polygonal_domains}

Consider a rectangular physical domain with physical boundaries aligned with the global coordinate system axes.
For instance, on the top horizontal physical boundary, $\partial u_{y}\left(\bm{x}\right)/\partial x$ can be derived from the prescribed vertical boundary velocity function in the $x$-direction, given by
\begin{equation}
\frac{\partial u_{y}}{\partial x}=\frac{\partial u^{\textrm{B}}_{y}}{\partial x},\qquad\text{on }\Gamma.
\end{equation}
Conversely, although the horizontal velocity on the same boundary is known since the function $u^{\textrm{B}}_{x}\left(\bm{x}\right)$ is given, its derivative in the $y$-direction cannot typically be determined and remains unknown.
In this case, from the definition of vorticity in Equation~\cref{eq:mathematical_formulation_vorticity_definition_1}, the boundary vorticity, $\omega^{\textrm{B}}\coloneqq\omega^{\textrm{B}}\left(\bm{x}\right)$, can be written as
\begin{equation}
\omega^{\textrm{B}}=\frac{\partial u^{\textrm{B}}_{y}}{\partial x}-\frac{\partial u_{x}}{\partial y},\qquad\text{on }\Gamma,
\end{equation}
where the unknown derivative term can be determined implicitly using an appropriate numerical approximation.
The majority of the published studies employ Cartesian grids only and, therefore, a simple finite difference formula is used to approximate $\partial u_{x}\left(\bm{x}\right)/\partial y=\partial^{2}\psi\left(\bm{x}\right)/\partial y^{2}$ at the boundary by using the unknown streamfunction values inside the domain together with the boundary streamfunction value.
For unstructured meshes, more elaborated interpolation techniques are required.

For general polygonal domains, where the physical boundary sides are not necessarily aligned with the coordinate system axes, a similar approach is adopted by considering a local coordinate system for convenience.
In this case, for a reference point $\bm{p}$ on the physical boundary and associated local coordinate system, $\partial u_{\eta}\left(\bs{\eta}\right)/\partial\xi$ is determined from the prescribed normal velocity function in the $\xi$-direction, given by
\begin{equation}
\frac{\partial u_{\eta}}{\partial\xi}=\frac{\partial u^{\textrm{B}}_{\eta}}{\partial\xi},\qquad\text{at }\bm{p}\in\Gamma,
\end{equation}
while $\partial u_{\xi}\left(\bs{\eta}\right)/\partial\eta$ remains undetermined at that stage.
Then, from the definition of vorticity in Equation~\cref{eq:boundary_conditions_vorticity_local_definition}, the boundary vorticity can be written as
\begin{equation}
\omega^{\textrm{B}}=\frac{\partial u_{\xi}}{\partial\eta}-\frac{\partial u^{\textrm{B}}_{\eta}}{\partial\xi},\qquad\text{at }\bm{p}\in\Gamma,
\end{equation}
where the unknown derivative term can be determined implicitly using the same numerical approximations as for rectangular domains.

\subsubsection{Curved domains}
\label{subsubsec:boundary_conditions_curved_domains}

Consider a domain with arbitrary curved physical boundaries.
In this scenario, the approach introduced for general polygonal domains becomes inappropriate since $\partial u_{\eta}\left(\bs{\eta}\right)/\partial\xi$ cannot be determined by simply deriving the prescribed normal velocity function in the $\xi$-direction.
Indeed, the local coordinate system axes are not a parametrization of the boundary, and, apart from its origin, $\bm{p}$, they do not generally match with the boundary.
As such, $\partial/\partial\eta$ and $\partial/\partial\xi$ do not correspond to the normal and tangential derivatives on the boundary, not even at the reference point $\bm{p}$ (as the local coordinate system axes do not follow the boundary curvature anywhere).


Going back to the global coordinates with $\bm{u}\left(\bm{x}\right)$, $\bm{n}\left(\bm{x}\right)$, and $\bm{t}\left(\bm{x}\right)$ as functions of $\bm{x}$, the following identity can be derived
\begin{equation}
\frac{\partial(\bm{u}\cdot\bm{n})}{\partial\bm{t}}=\frac{\partial\bm{u}}{\partial\bm{t}}\cdot\bm{n}+\bm{u}\cdot\frac{\partial\bm{n}}{\partial\bm{t}},\qquad\text{on }\Gamma.
\end{equation}
In particular, $\partial\bm{n}\left(\bm{x}\right)/\partial\bm{t}\left(\bm{x}\right)=\kappa\left(\bm{x}\right)\bm{t}\left(\bm{x}\right)$, where $\kappa\coloneqq\kappa\left(\bm{x}\right)$ is the normal curvature, positive if the centre of curvature lies on the same side as the physical domain.
Moreover, $\left(\partial\bm{u}\left(\bm{x}\right)/\partial\bm{t}\left(\bm{x}\right)\right)\cdot\bm{n}\left(\bm{x}\right)=\left[\left(\nabla\bm{u}\left(\bm{x}\right)\right)\bm{t}\left(\bm{x}\right)\right]\cdot\bm{n}\left(\bm{x}\right)$ and, therefore, the above identity can be expressed as
\begin{equation}
\frac{\partial(\bm{u}\cdot\bm{n})}{\partial\bm{t}}=\left[\left(\nabla\bm{u}\right)\bm{t}\right]\cdot\bm{n}+\kappa\bm{u}\cdot\bm{t},\qquad\text{on }\Gamma.
\end{equation}
Since the transformation from the global coordinate system to the local coordinate system is a linear orthogonal transformation, the above identity can be rewritten in the local coordinate system associated with a reference point $\bm{p}$ on the physical boundary as
\begin{equation}
\frac{\partial(\bm{u}^{\textrm{L}}\cdot\bm{n}^{\textrm{L}})}{\partial\bm{t}^{\textrm{L}}}=\left[\left(\nabla^{\textrm{L}}\bm{u}^{\textrm{L}}\right)\bm{t}^{\textrm{L}}\right]\cdot\bm{n}^{\textrm{L}}+\kappa\bm{u}^{\textrm{L}}\cdot\bm{t}^{\textrm{L}}
\end{equation}
where $\nabla^{\textrm{L}}$ is the gradient operator and $\bm{n}^{\textrm{L}}\coloneqq\bm{n}^{\textrm{L}}\left(\bs{\eta}\right)$ and $\bm{t}^{\textrm{L}}\coloneqq\bm{t}^{\textrm{L}}\left(\bs{\eta}\right)$ are the boundary normal and tangential vectors, respectively, with respect to the local coordinates.
In particular, $\bm{n}^{\textrm{L}}\left(\bm{0}\right)=\left(1,0\right)^{\textrm{T}}$ and $\bm{t}^{\textrm{L}}\left(\bm{0}\right)=\left(0,1\right)^{\textrm{T}}$ at the local coordinate system origin, $\bm{p}$.
Thus, by further simplifying the above identity after expanding the normal and tangential vectors in the local coordinate system at the reference point $\bm{p}$, yields
\begin{equation}
\frac{\partial(\bm{u}^{\textrm{L}}\cdot\bm{n}^{\textrm{L}})}{\partial\bm{t}^{\textrm{L}}}=\frac{\partial u_{\eta}}{\partial\xi}+\kappa u_{\xi}
\quad\Rightarrow\quad
\frac{\partial u_{\eta}}{\partial\xi}=-\kappa u_{\xi}+\frac{\partial(\bm{u}^{\textrm{L}}\cdot\bm{n}^{\textrm{L}})}{\partial\bm{t}^{\textrm{L}}}
\qquad\text{at }\bm{p}\in\Gamma.
\end{equation}
Using the above identity, the boundary vorticity in the local coordinate system, $\omega^{\textrm{L},\textrm{B}}\coloneqq\omega^{\textrm{L},\textrm{B}}\left(\bs{\eta}\right)$, can be obtained as
\begin{equation}
\omega^{\textrm{L},\textrm{B}}=\frac{\partial u_{\xi}}{\partial\eta}+\kappa u^{\textrm{B}}_{\xi}-\frac{\partial(\bm{u}^{\textrm{L},\textrm{B}}\cdot\bm{n}^{\textrm{L}})}{\partial\bm{t}^{\textrm{L}}},\qquad\text{at }\bm{p}\in\Gamma,
\end{equation}
substituting $u_{\xi}\left(\bs{\eta}\right)$ by $u_{\xi}^{\textrm{B}}\left(\bs{\eta}\right)$ and $\bm{u}^{\textrm{L}}\left(\bs{\eta}\right)$ by $\bm{u}^{\textrm{L},\textrm{B}}\left(\bs{\eta}\right)$ since the the boundary velocity function is given and its derivative with respect to $\bm{t}^{\textrm{L}}\left(\bs{\eta}\right)$ can be computed analytically.
As for rectangular domains, the obtained boundary vorticity formula comprises one implicit derivative in the $\eta$-direction to be determined.
Moreover, since $u_{\xi}\left(\bs{\eta}\right)=-\partial\psi^{\textrm{L}}\left(\bs{\eta}\right)/\partial\eta$, then the boundary vorticity can be rewritten as
\begin{equation}
\label{eq:boundary_conditions_boundary_vorticity}
\omega^{\textrm{L},\textrm{B}}=-\frac{\partial^{2}\psi^{\textrm{L}}}{\partial\eta^{2}}+\kappa u^{\textrm{L},\textrm{B}}_{\xi}-\frac{\partial(\bm{u}^{\textrm{L},\textrm{B}}\cdot\bm{n}^{\textrm{L}})}{\partial\bm{t}^{\textrm{L}}},\qquad\text{at }\bm{p}\in\Gamma,
\end{equation}

\begin{remark}
In the special case of internal flows with impermeable walls, $\bm{u}^{\textrm{B}}\left(\bm{x}\right)\cdot\bm{n}\left(\bm{x}\right)=0$ on the physical boundary and, the boundary vorticity simply writes $\omega^{\textrm{L},\textrm{B}}\left(\bm{x}\right)=-\partial^{2}\psi^{\textrm{L}}\left(\bs{\eta}\right)/\partial\eta^{2}+\kappa\left(\bs{\eta}\right)u^{\textrm{B}}_{\xi}\left(\bs{\eta}\right)$.
\end{remark}

\subsection{Boundary conditions for the streamfunction-vorticity formulation}
\label{subsec:boundary_conditions_boundary_conditions_for_the_streamfunction_vorticity_formulation}

Finally, after deriving the necessary formulas for the boundary streamfunction and vorticity from the prescribed boundary velocity, the system of linear equations~\cref{eq:mathematical_formulation_vorticity_equation_2,eq:mathematical_formulation_streamfunction_equation_2} is supplemented with appropriate boundary conditions as follows:

\begin{itemize}
\item For the streamfunction, a Cauchy boundary condition (Dirichlet and Neumann boundary conditions imposed simultaneously) is prescribed on the physical boundary, given as
\begin{equation}
\label{eq:boundary_conditions_streamfunction_cauchy_condition_1}
\begin{aligned}
\psi\;=\;&\psi^{\textrm{B}}\\
\nabla\psi\cdot\bm{n}\;=\;&\nabla\psi^{\textrm{B}}\cdot\bm{n}
\end{aligned}
\qquad\text{on }\Gamma,
\end{equation}
where function $\psi^{\textrm{B}}\coloneqq\psi^{\textrm{B}}\left(\bm{x}\right)$ and its normal derivate are given in Equations~\cref{eq:boundary_conditions_boundary_streamfunction,eq:boundary_conditions_boundary_streamfunction_derivative}.


\item For the vorticity, the Dirichlet boundary condition is prescribed on the physical boundary, given as
\begin{equation}
\label{eq:boundary_conditions_boundary_vorticity_condition_1}
\omega\;=\;\omega^{\textrm{B}}\qquad\text{on }\Gamma,
\end{equation}
where function $\omega^{\textrm{B}}\coloneqq\omega^{\textrm{B}}\left(\bm{x}\right)$ in global coordinates corresponds to the boundary vorticity in local coordinates, as given in Equation~\cref{eq:boundary_conditions_boundary_vorticity}.
\end{itemize}

For multiply connected domains, the system of linear equations is also supplemented with the compatibility conditions~\cref{eq:boundary_conditions_compatibility_condition} to fix the unknown constants for the boundary streamfunction on inner physical boundary subsets.

\section{Polynomial reconstruction}
\label{sec:polynomial_reconstruction}

Before addressing the problem of discretisation, fundamental concepts and techniques to achieve very high accuracy within the finite volume framework are briefly recalled.
The polynomial reconstruction method is a powerful technique for computing local high-accurate approximations of the underlying solution and its derivatives on the mesh cell interfaces.
In this regard, the reconstruction for off-site data (ROD) method~\cite{2018_costa,2019_costa1,2019_costa2,2021_costa,2022_costa1,2022_costa2,2023_costa,2024_santos} is a further development to the polynomial reconstruction method to enable treating general boundary conditions prescribed on arbitrary curved boundaries, while the problem is solved on polygonal meshes.

\subsection{Polygonal meshes}
\label{subsec:polynomial_reconstruction_polygonal_meshes}

The physical domain $\Omega$ is represented by a polygonal domain $\Omega_{\Delta}=\bigcup_{i\in\mathcal{I}}c_{i}$ consisting of $N_{\textrm{C}}$ non-overlapping convex polygonal cells, with $N_{\textrm{E}}$ edges, and $N_{\textrm{B}}$ boundary edges.
Cells are denoted as $c_{i}$, with $i\in\mathcal{I}=\lbrace 1,\ldots,N_{\textrm{C}}\rbrace$, and an inner edge $e_{ij}=c_{i}\cap c_{j}$ is the common interface between two adjacent cells $c_{i}$ and $c_{j}$, $j\neq i$.
On the other hand, boundary edges $e_{i\textrm{B}}$, $i\in\mathcal{I}^{\textrm{B}}$, belong to the computational (surrogate) boundary $\Gamma_{\Delta}=\bigcup_{k=0}^{K}\Gamma^{k}_{\Delta}=\bigcup_{i\in\mathcal{I}}e_{i\textrm{B}}$ with the $\mathcal{I}^{\textrm{B}}\subset\mathcal{I} $ being the index set of the boundary cells and $\Gamma^{k}_{\Delta}$ the linear piecewise approximation of physical boundary subset $\Gamma^{k}$.
For simplicity, each cell has at most one boundary edge.

\begin{figure}[!htb]
\centering
\includegraphics[width=7.0cm]{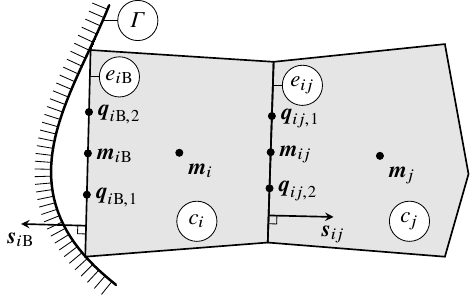}
\caption{Notation and geometric properties for the cells and edges.}
\label{fig:polynomial_reconstruction_mesh_notation}
\end{figure}

The geometric properties for the cells are provided in Table~\ref{tab:polynomial_reconstruction_mesh_notation}, whereas Figure~\ref{fig:polynomial_reconstruction_mesh_notation} provides a schematic representation of the relevant data.
Note that the inner edge $e_{ij}$ is also denoted as $e_{ji}$ and, therefore, reference and quadrature points are the same, that is, $\bm{m}_{ij}=\bm{m}_{ji}$ and $\bm{q}_{ij,r}=\bm{q}_{ji,r}$, whereas outward unit normal vectors satisfy $\bm{s}_{ij}=-\bm{s}_{ji}$.

\begin{table}[!htb]
\centering
\caption{Notation and geometric properties for the cells and edges.}
\label{tab:polynomial_reconstruction_mesh_notation}
\resizeboxlarger{
\begin{tabular}{@{}lllll@{}}
\toprule
Mesh elements & Notation & Properties & Definition & Choice\\
\midrule
\multirow{5}{*}{Cells} & \multirow{5}{*}{\specialcell{l}{$c_{i}$\\$i\in\mathcal{I}$}} & $\partial c_{i}$ & Boundary & \\
& & $\vert c_{i}\vert$ & Area & \\
& & $\bm{m}_{i}\coloneqq\left(m_{i,x},m_{i,y},m_{i,z}\right)$ & Reference point (can be any point in $c_{i}$) & Mass centre \\
& & $\bm{q}_{i,q}\coloneqq\left(q_{i,q,x},q_{i,q,y},q_{i,q,z}\right)$ & Quadrature points, $q=1,\ldots,Q$ & Gauss-Legendre \\
& & $\mathcal{N}_{i}$ & Indices of the adjacent cells and boundary subset & \\
\midrule
\multirow{5}{*}{Inner edges} & \multirow{5}{*}{\specialcell{l}{$e_{ij}$\\$i,j\in\mathcal{I}$}} & $\vert e_{ij}\vert$ & Length & \\
& & $\bm{m}_{ij}\coloneqq\left(m_{ij,x},m_{ij,y}\right)$ & Reference point (can be any point on $e_{ij}$) & Midpoint \\
& & $\bm{q}_{ij,r}\coloneqq\left(q_{ij,r,x},q_{ij,r,y}\right)$ & Quadrature points, $r=1,\ldots,R$ & Gauss-Legendre \\
& & $\bm{s}_{ij}\coloneqq\left(s_{ij,x},s_{ij,y}\right)$ & Unit normal vector from cell $c_{i}$ to cell $c_{j}$ & \\
& & $\bm{r}_{ij}\coloneqq\left(r_{ij,x},r_{ij,y}\right)$ & Unit tangential vector & \\
\midrule
\multirow{5}{*}{Boundary edges} & \multirow{5}{*}{\specialcell{l}{$e_{i\textrm{B}}$\\$i\in\mathcal{I}$}} & $\vert e_{i\textrm{B}}\vert$ & Length & \\
& & $\bm{m}_{i\textrm{B}}\coloneqq\left(m_{i\textrm{B},x},m_{i\textrm{B},y}\right)$ & Reference point (can be any point on $e_{i\textrm{B}}$) & Midpoint \\
& & $\bm{q}_{i\textrm{B},r}\coloneqq\left(q_{i\textrm{B},r,x},q_{i\textrm{B},r,y}\right)$ & Quadrature points, $r=1,\ldots,R$ & Gauss-Legendre \\
& & $\bm{s}_{i\textrm{B}}\coloneqq\left(s_{i\textrm{B},x},s_{i\textrm{B},y}\right)$ & Unit normal vector from cell $c_{i}$ & \\
& & $\bm{r}_{i\textrm{B}}\coloneqq\left(r_{i\textrm{B},x},r_{i\textrm{B},y}\right)$ & Unit tangential vector & \\
\bottomrule
\end{tabular}
}
\end{table}

\begin{figure}[!htb]
\centering
\includegraphics[width=12.0cm]{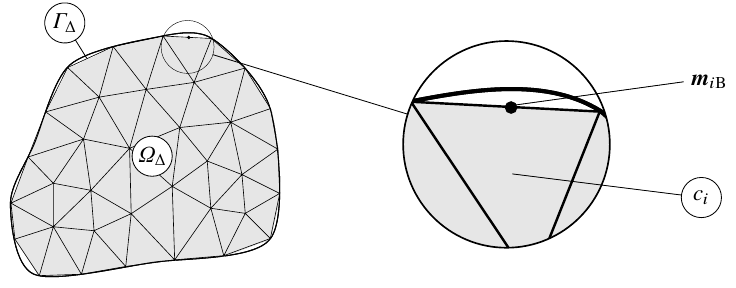}
\caption{Inset on the curved physical boundary and approximation with a polygonal computational boundary}
\label{fig:polynomial_reconstruction_curved_domain_tria_mesh_magnification}
\end{figure}

Following the motivation presented in Section~\ref{sec:introduction}, the mesh consists solely of polygonal elements, therefore avoiding the difficulties of generating and employing curved meshes for curved domains.
In this case, the computational domain, where the problem is numerically solved, does not completely match the physical domain, where the governing equations are defined.
More specifically, a geometrical mismatch exists between the computational boundary and the physical boundary, where the boundary conditions are prescribed (see Figure~\ref{fig:polynomial_reconstruction_curved_domain_tria_mesh_magnification}).
For uniform regular meshes of characteristic size $h$, the geometrical mismatch typically has an order of magnitude of $\mathcal{O}\left(h^{2}\right)$ between the physical and computational boundaries, potentially leading to a deterioration in accuracy without appropriate treatment.

\subsection{Polynomial reconstruction method}
\label{subsec:polynomial_reconstruction_polynomial_reconstruction_method}

Given a bounded scalar function $\phi\coloneqq\phi\left(\bm{x}\right)$, local polynomial approximations $\varphi$ of degree $d$ are sought, written in compact form as
\begin{equation}
\label{eq:polynomial_reconstruction_local_approximation_compact}
\varphi\left(\bs{x};\bm{a}\right)=\bm{a}\cdot\bs{p}_{d}\left(\bm{x}-\bm{m}\right)=\sum_{\alpha=0}^{d}\sum_{\beta=0}^{d-\alpha}\left[a_{\alpha\beta}\left(x-m_{x}\right)^{\alpha}\left(y-m_{y}\right)^{\beta}\right],
\end{equation}
where $\bs{p}_{d}\left(\bs{x}\right)$ is a basis vector including all two-dimensional monomials up to degree $d$, $\bs{m}\coloneqq\left(m_{x},m_{y}\right)$ is a reference point, and vector $\bm{a}\in\mathbb{R}^{n}$ gathers the polynomial coefficients, with $n=\left(d+1\right)\left(d+2\right)/2$.

For each cell, a local piecewise mean-value approximation to function $\phi\left(\bm{x}\right)$ is denoted as $\phi_{i}$, $i\in\mathcal{I}$, and is defined as
\begin{equation}
\label{eq:polynomial_reconstruction_mean_value}
\phi_{i}\approx\dfrac{1}{\left\vert c_{i}\right\vert}\int_{c_{i}}\phi\diff\bs{x}
\end{equation}
$\mathcal{S}$ denotes a stencil of the neighbouring cells indices in the vicinity of a given reference edge or cell.
Each stencil must comprise a minimum of $s=n$ cells, but in practice, $s$ is chosen higher than $n$ to increase robustness and stability.
The indices in stencil $\mathcal{S}$ are selected from geometrically close cells, for which a neighbours-of-neighbours algorithm is implemented to ensure the contiguity of the stencil, that is, without gaps.

The linear least-squares method is then employed to minimise the squared errors between the approximate cell mean-value, $\phi_{i}$, and the polynomial function $\varphi\left(\bs{x},\bm{a}\right)$ by introducing the weighted cost functional 
\begin{equation}
\label{eq:polynomial_reconstruction_cost_functional}
F\left(\bm{a};\bs{m},\mathcal{S},\bs{\omega}\right)=\sum_{\alpha\in\mathcal{S}}\omega_{\alpha}\Biggl[\dfrac{1}{\left\vert c_{\alpha}\right\vert}\displaystyle\int_{c_{\alpha}}\varphi(\bm{x};\bm{a})\diff\bs{x}-\phi_{\alpha}\Biggr]^{2},
\end{equation}
where the integrals are evaluated numerically using quadrature rules of the same order as the polynomial degree plus one.
In the cost functional, $\omega_{\alpha}=\omega\left(d_{\alpha}\right)$, $\alpha\in\mathcal{S}$, are positive weights determined from function $\omega\left(d_{\alpha}\right)$ given by
\begin{equation}
\omega\left(d_{\alpha}\right)=\dfrac{1}{\left(\sigma d_{k}\right)^{\delta}+1},
\end{equation}
where $d_{\alpha}=\left\Vert\bs{m}_{\alpha}-\bs{m}\right\Vert$ is the distance from the reference point to the cell centre, and vector $\bs{\omega}$ gathers all the weights.
Moreover, $\delta,\sigma\in\mathbb{R}$ are given parameters that control the sensitivity of the weights to the data distance, with optimal values obtained through numerical assessment.


\subsubsection{Unconstrained polynomial reconstruction}
\label{subsubsec:polynomial_reconstruction_unconstrained_polynomial_reconstruction}

If the polynomial reconstruction is computed solely by minimising functional $F\left(\bm{a};\mathcal{S}\right)$, then the procedure is referred to as unconstrained reconstruction.
The normal equations method~\cite{1971_wells} is employed in the present work to solve the least-squares problem and, for the sake of compactness, the reader is referred to~\cite{2019_costa1} for a comprehensive and detailed description of the procedure.
Unconstrained polynomial reconstructions are computed for the inner edges as the polynomial references.
In this case, $\bs{m}\coloneqq\bs{m}_{ij}$, $\mathcal{S}\coloneqq\mathcal{S}_{ij}$, and $\bs{\omega}\coloneqq\bs{\omega}_{ij}$ are assigned based on each inner edge, $e_{ij}$, and the resulting optimal polynomial reconstruction is then written as $\widetilde{\varphi}_{ij}\left(\bs{x}\right)=\varphi(\bm{x};\widetilde{\bm{a}}_{ij})=\widetilde{\bm{a}}_{ij}\cdot\bs{p}_{d}\left(\bs{x}-\bs{m}_{ij}\right)$, with $\widetilde{\bm{a}}_{ij}=\argmin_{\bm{a}}F\left(\bm{a};\bs{m}_{ij},\mathcal{S}_{ij},\bs{\omega}_{ij}\right)$, which is referred to as unconstrained polynomial reconstruction.

\subsubsection{Mean-value preservation polynomial reconstruction}
\label{subsubsec:polynomial_reconstruction_mean_value_preservation_polynomial_reconstruction}

When the polynomial reference is a cell, conservation is a critical property to satisfy, especially in convection-dominated problems, as it improves the robustness and stability of the numerical scheme while promoting the approximate solution to be physically meaningful.
To achieve this, in addition to the cost functional $F\left(\bm{a};\bs{m},\mathcal{S},\bs{\omega}\right)$, the least-squares procedure is subject to a linear constraint that must be exactly fulfilled, written in residual form as $G\left(\bm{a}\right)=0$.
Thus, the minimisation procedure involves seeking $\widehat{\bm{a}}=\argmin_{\bm{a}}F\left(\bm{a}\right)$ subject to $G\left(\bm{a}\right)=0$.
The so-called linearly constrained Lagrange multipliers method~\cite{1982_bertsekas} is employed in the present work to solve the constrained least-squares problem, and the reader is referred to~\cite{2019_costa1} for a comprehensive and detailed description of the procedure.
In this case, $\bs{m}\coloneqq\bs{m}_{i}$, $\mathcal{S}\coloneqq\mathcal{S}_{i}$, and $\bs{\omega}\coloneqq\bs{\omega}_{i}$ are assigned based on each cell, $c_{i}$.
For the conservation of the cell mean-value, the linear constraint functional from $\mathbb{R}^{n}$ to $\mathbb{R}$, denoted as $G_{i}\left(\bm{a}\right)$, is given as
\begin{equation}
G_{i}\left(\bm{a}\right)=\dfrac{1}{\left\vert c_{i}\right\vert}\int_{c_{i}}\varphi(\bm{x};\bm{a})\diff\bs{x}-\phi_{i}.
\end{equation}
The optimal polynomial reconstruction is $\widehat{\varphi}_{i}\left(\bs{x}\right)=\varphi(\bm{x};\widehat{\bm{a}}_i)=\widehat{\bm{a}}_{i}\cdot\bs{p}_{d}\left(\bs{x}-\bs{m}_{i}\right)$, with $\widehat{\bm{a}}_{i}=\argmin_{\bm{a}}F\left(\bm{a};\bs{m}_{i},\mathcal{S}_{i},\bs{\omega}_{i}\right)$ subject to $G_{i}\left(\bm{a}\right)=0$, referred to as constrained polynomial reconstruction.

\subsection{Reconstruction for off-site data method}
\label{subsec:polynomial_reconstruction_reconstruction_for_off-site_data_method}

Since arbitrary curved physical boundaries are discretised with linear piecewise edges, the associated boundary data is disconnected from the mesh structure.
In this context, the ROD method has emerged as a sophisticated evolution of the polynomial reconstruction method, enabling the imposition of boundary data in the reconstruction procedure, while preserving high-order accuracy for arbitrary curved physical boundaries.

\begin{figure}[!htb]
\centering
\begin{tabular}{c@{\hskip 1.5cm}c}
\includegraphics[width=7.0cm]{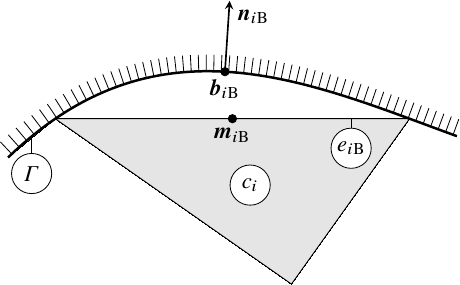}
& \includegraphics[width=7.0cm]{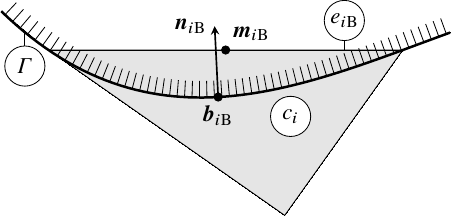}\\
(a) Concave physical boundary. & (b) Convex physical boundary.
\end{tabular}
\caption{Polygonal mesh with collocation point and associated outward unit normal vector on the curved physical boundary.}
\label{fig:polynomial_reconstruction_curved_boundary_rod}
\end{figure}

Firstly, for each edge $e_{i\textrm{B}}$ on the surrogate boundary, consider a collocation point denoted as $\bs{b}_{i\textrm{B}}\coloneqq\left(b_{i\textrm{B},x},b_{i\textrm{B},y}\right)$,  which is selected to coincide with the physical boundary in the vicinity of the boundary edge, that is, $\bs{b}_{i\textrm{B}}\in\Gamma$ (see Figure~\ref{fig:polynomial_reconstruction_curved_boundary_rod}).
Since the physical boundary is described analytically, a simple projection of the edge midpoint onto the physical boundary can be employed to determine a suitable collocation point, and the associated outward unit normal vector, denoted as $\bs{n}_{i\textrm{B}}\coloneqq\left(n_{i\textrm{B},x},n_{i\textrm{B},y}\right)$, is also determined and can be obtained from the normal vector function as $\bs{n}_{i\textrm{B}}=\bs{n}\left(\bs{b}_{i\textrm{B}}\right)$.

Secondly, for each boundary edge $e_{i\textrm{B}}$, a linear constraint functional from $\mathbb{R}^{n}$ to $\mathbb{R}$, denoted as $G_{i\textrm{B}}\left(\bm{a}\right)$, is defined at the collocation point as 
\begin{equation}
\label{eq:polynomial_reconstruction_constraint_functional_1}
G_{i\textrm{B}}\left(\bm{a}\right)=\alpha\varphi\left(\bs{b}_{i\textrm{B}};\bm{a}\right)+\beta\nabla\varphi\left(\bs{b}_{i\textrm{B}};\bm{a}\right)\cdot\bm{n}_{i\textrm{B}}-g(\bs{b}_{i\textrm{B}}),
\end{equation}
where $\alpha$ and $\beta$ are coefficients, and $g\coloneqq g\left(\bm{x}\right)$ is a scalar function that stands for the boundary condition value at the collocation point.
This linear constraint represents a Robin boundary condition that imposes any prescribed scalar boundary condition by choosing the appropriate coefficients, $\alpha$ and $\beta$.
For the Dirichlet boundary condition $\phi\left(\bm{x}\right)=g^{\textrm{D}}\left(\bm{x}\right)$, coefficients $\alpha=1$ and $\beta=0$ are chosen and the boundary condition value is $g\left(\bm{x}\right)=g^{\textrm{D}}\left(\bs{b}_{i\textrm{B}}\right)$.
For the Neumann boundary condition $\nabla\phi\left(\bm{x}\right)\cdot\bm{n}\left(\bm{x}\right)=g^{\textrm{N}}\left(\bm{x}\right)$, coefficients $\alpha=0$ and $\beta=1$ are chosen and the boundary condition value is $g\left(\bm{x}\right)=g^{\textrm{N}}\left(\bs{b}_{i\textrm{B}}\right)$.
For the Robin boundary condition $\alpha^{\textrm{R}}\left(\bm{x}\right)\phi\left(\bm{x}\right)+\beta^{\textrm{R}}\left(\bm{x}\right)\nabla\phi\left(\bm{x}\right)\cdot\bm{n}\left(\bm{x}\right)=g^{\textrm{R}}\left(\bm{x}\right)$, coefficients $\alpha=\alpha^{\textrm{R}}\left(\bm{b}_{i\textrm{B}}\right)$ and $\beta=\beta^{\textrm{R}}\left(\bm{b}_{i\textrm{B}}\right)$ are chosen and the boundary condition value is $g\left(\bm{x}\right)=\phi^{\textrm{R}}\left(\bs{b}_{i\textrm{B}}\right)$.

In case of the Cauchy boundary condition, two distinct conditions are imposed at each boundary point, and the scalar linear constraints $G_{i\textrm{B}}\left(\bm{a}\right)$ is substituted with a vectorial linear constraint, denoted as $\bm{G}_{i\textrm{B}}\left(\bm{a}\right)$, given as
\begin{equation}
\label{eq:polynomial_reconstruction_constraint_functional_2}
\bm{G}_{i\textrm{B}}\left(\bm{a}\right)
=
\begin{bmatrix}
\alpha\varphi\left(\bs{b}_{i\textrm{B}};\bm{a}\right)\\
\beta\nabla\varphi\left(\bs{b}_{i\textrm{B}};\bm{a}\right)\cdot\bm{n}_{i\textrm{B}}
\end{bmatrix}
-\bm{g}(\bs{b}_{i\textrm{B}}),
\end{equation}
where $\bm{g}\coloneqq\bm{g}\left(\bm{x}\right)$ is a vector function that stands for the Dirichlet and Neumann boundary condition values at the collocation point.

Finally, for each boundary edge $e_{i\textrm{B}}$, a constrained polynomial reconstruction is computed with $\bs{m}\coloneqq\bs{m}_{i\textrm{B}}$, $\mathcal{S}\coloneqq\mathcal{S}_{i\textrm{B}}$, and $\bs{\omega}\coloneqq\bs{\omega}_{i\textrm{B}}$ assigned to the reference boundary edge.
The minimisation procedure in this case involves seeking a unique vector $\widehat{\bm{a}}_{i\textrm{B}}\in\mathbb{R}^{n}$ that minimises the functional $F\left(\bm{a}\right)$ under the constraint equation $G_{i\textrm{B}}\left(\bm{a}\right)=0$ or $\bm{G}_{i\textrm{B}}\left(\bm{a}\right)=\bm{0}$, that is, $\widehat{\bm{a}}_{i\textrm{B}}=\argmin_{\bm{a}}F\left(\bm{a};\bs{m}_{i\textrm{B}},\mathcal{S}_{i\textrm{B}},\bs{\omega}_{i\textrm{B}}\right)$ subject to $G_{i\textrm{B}}\left(\bm{a}\right)=0$ or $\bm{G}_{i\textrm{B}}\left(\bm{a}\right)=\bm{0}$.
The optimal polynomial reconstruction is then written as $\widehat{\varphi}_{i\textrm{B}}\left(\bs{x}\right)=\varphi(\bm{x};\widehat{\bm{a}}_{i\textrm{B}})=\widehat{\bm{a}}_{i\textrm{B}}\cdot\bs{p}_{d}\left(\bs{x}-\bs{m}_{i\textrm{B}}\right)$, which is referred to as constrained polynomial reconstruction.
Similar to the conservation of the cell mean-value, the so-called linearly constrained Lagrange multipliers method~\cite{1982_bertsekas} is employed to solve the constrained least-squares problem.

\section{Finite volume discretisation}
\label{sec:finite_volume_discretisation}

A very-high order finite volume scheme to solve the system of partial differential equations~\eqref{eq:mathematical_formulation_vorticity_equation_2}--\eqref{eq:mathematical_formulation_streamfunction_equation_2} in the computational domain $\Omega_{\Delta}$ is derived.
The two main ingredients are the polynomial reconstructions and the evaluation of fluxes together with the ROD method to take into account the prescribed boundary condition on the physical boundary with the desired accuracy.

\subsection{Generic finite volume scheme}
\label{subsec:finite_volume_discretisation_generic_finite_volume_scheme}

Integrating the vorticity transport equation~\eqref{eq:mathematical_formulation_vorticity_equation_2} over each cell $c_{i}$, $i\in\mathcal{I}$, and applying the divergence theorem yields
\begin{equation}
\int_{c_{i}}\left(\nabla\cdot\left(\bm{v}\omega\right)-\nu\nabla^{2}\omega\right)\diff\bm{x}=\int_{c_{i}}\bm{f}\diff\bm{x}
\qquad
\Rightarrow
\qquad
\int_{\partial c_{i}}\left(\bm{v}\omega-\nu\nabla\omega\right)\cdot\bm{s}_{i}\diff\bm{x}=\int_{c_{i}}\bm{f}\diff\bm{x},
\label{eq:finite_volume_discretisation_integral_1}
\end{equation}
where $\partial c_{i}$ stands for the cell boundary with $\bm{s}_{i}\coloneqq\bm{s}_{i}\left(\bm{x}\right)\coloneqq\left(s_{x,i}\left(\bm{x}\right),s_{y,i}\left(\bm{x}\right)\right)$ the outward unit normal vector.
An $R$-point Gauss-Legendre quadrature rule with weights $\zeta_{r}$, $r=1,\ldots,R$, is used to approximate the integrals on the straight edges while a $Q$-point Gauss-Legendre quadrature rule with weights $\xi_{q}$, $q=1,\ldots,Q$, is used to approximate the integrals in the polygonal cells.
Then, Equation~\eqref{eq:finite_volume_discretisation_integral_1} is rewritten in the discrete residual form using the quadrature rules as
\begin{equation}
\sum_{j\in\mathcal{N}_{i}}\left\vert e_{ij}\right\vert\left[\sum_{r=1}^{R}\zeta_{r}\left(C_{ij,r}+D_{ij,r}\right)\right]-\left\vert c_{i}\right\vert f_{i}=\mathcal{O}\left(h_{i}^{\alpha}\right),
\label{eq:finite_volume_discretisation_discrete_1}
\end{equation}
where $h_{i}=\left\vert c_{i}\right\vert^{1/2}$ is the characteristic cell size and $\alpha$ is the convergence order of the quadrature rules.
Moreover, $C_{ij,r}$ and $D_{ij,r}$ stand for the physical convective and diffusive fluxes of function $\omega\left(\bm{x}\right)$, respectively, at the quadrature points of edge $e_{ij}$, and $f_{i}$ stands for the approximate mean-value of source term function $f\left(\bm{x}\right)$ in cell $c_{i}$, given as
\begin{equation}
C_{ij,r}=\left(\bm{v}\left(\bm{q}_{ij,r}\right)\cdot\bm{s}_{ij}\right)\omega\left(\bm{q}_{ij,r}\right),
\qquad
D_{ij,r}=-\nu\nabla\omega\left(\bm{q}_{ij,r}\right)\cdot\bm{s}_{ij},
\qquad
\text{and}
\qquad
f_{i}=\sum_{q=1}^{Q}\xi_{q}f\left(\bm{q}_{i,q}\right).
\label{eq:finite_volume_discretisation_exact_fluxes_1}
\end{equation}

Similarly, integrating equation~\eqref{eq:mathematical_formulation_streamfunction_equation_2} over each cell $c_{i}$, $i\in\mathcal{I}$, and applying the divergence theorem yields
\begin{equation}
\int_{c_{i}}\left(\nabla^{2}\psi+\omega\right)\diff\bm{x}=0
\qquad
\Rightarrow
\qquad
\int_{\partial c_{i}}\nabla\psi\cdot\bm{s}_{i}\diff\bm{x}+\int_{c_{i}}\omega\diff\bm{x}=0,
\label{eq:finite_volume_discretisation_integral_2}
\end{equation}
which is rewritten in the discrete residual form using the quadrature rules as
\begin{equation}
\sum_{j\in\mathcal{N}_{i}}\left\vert e_{ij}\right\vert\left[\sum_{r=1}^{R}\zeta_{r}F_{ij,r}\right]-R_{i}=\mathcal{O}\left(h_{i}^{\alpha}\right),
\label{eq:finite_volume_discretisation_discrete_2}
\end{equation}
where $F_{ij,r}$ stands for the physical diffusive flux of function $\psi\left(\bm{x}\right)$ at the quadrature points of edge $e_{ij}$, and $R_{i}$ stands for the mean-value of function $\omega\left(\bm{x}\right)$ in cell $c_{i}$, given as
\begin{equation}
F_{ij,r}=\nabla\psi\left(\bm{q}_{ij,r}\right)\cdot\bm{s}_{ij}
\qquad
\text{and}
\qquad
R_{i}=\int_{c_{i}}\omega\diff\bm{x}.
\label{eq:finite_volume_discretisation_exact_fluxes_2}
\end{equation}

\subsection{Discrete problem unknowns}
\label{subsec:data_reconstruction_discrete_problem_unknows}

The discrete problem unknown variables are the cell mean-values of the streamfunction and vorticity, denoted as $\psi_{i}$ and $\omega_{i}$, $i\in\mathcal{I}$, respectively, given as
\begin{equation}
\psi_{i}\approx\dfrac{1}{\left\vert c_{i}\right\vert}\int_{c_{i}}\psi\diff\bm{x}
\qquad
\textrm{and}
\qquad
\omega_{i}\approx\dfrac{1}{\left\vert c_{i}\right\vert}\int_{c_{i}}\omega\diff\bm{x},
\end{equation}
leading to piecewise approximations of functions $\psi\left(\bm{x}\right)$ and $\omega\left(\bm{x}\right)$, respectively.
In addition to the cell-mean values, auxiliary variables to be computed are introduced for the discrete problem formulation, namely:
\begin{itemize}
\item Velocity point-values $v_{\beta,ij,r}$, $\beta\in\lbrace x,y\rbrace$, $i,j\in\mathcal{I}$, $j\in\mathcal{N}_{i}$, $r=1,\ldots,R$, with $\bm{v}_{ij,r}\coloneqq\left(v_{x,ij,r},v_{y,ij,r}\right)$, standing for an approximation of function $v_{\beta}\left(\bm{x}\right)$ at the edge quadrature points $\bm{q}_{ij,r}$, given as
\begin{equation}
v_{\beta,ij,r}\approx v_{\beta}\left(\bm{q}_{ii,r}\right).
\end{equation}

\item Boundary vorticity point-values $\omega_{i\textrm{B}}$, $i\in\mathcal{I}^{\textrm{B}}$, standing for an approximation of function $\omega^{\textrm{B}}\left(\bm{x}\right)$ at collocation points $\bm{b}_{i\textrm{B}}$, given as
\begin{equation}
\omega_{i\textrm{B}}\approx\omega^{\textrm{B}}\left(\bm{b}_{i\textrm{B}}\right).
\end{equation}
\end{itemize}

\subsection{Polynomial reconstructions}
\label{subsec:data_reconstruction_polynomial_reconstructions}

To obtain high-accurate approximations of the physical fluxes of the finite volume scheme, polynomial reconstructions are computed based on the discrete problem cell mean-values and boundary point-values unknowns.
Each polynomial reconstruction is specifically designed depending on its purpose (see the complete list in Table~\ref{tab:finite_volume_discretisation_polynomial_reconstructions}), as follows:
\begin{itemize}
\item For each inner edge $e_{ij}$, unconstrained polynomial reconstructions $\widetilde{\psi}_{ij}\left(\bm{x}\right)$ for the streamfunction and $\widetilde{\omega}_{ij}\left(\bm{x}\right)$ for the vorticity are computed based on cell mean-values $\psi_{i}$ and $\omega_{i}$, respectively.

\item For each cell $c_{i}$, constrained polynomial reconstructions $\widehat{\omega}_{i}\left(\bm{x}\right)$ for the vorticity are computed based on cell mean-values $\omega_{i}$ to preserve the conservation of the cell mean-value in the same cell.

\item For each boundary edge $e_{i\textrm{B}}$, constrained polynomial reconstructions $\widehat{\psi}_{i\textrm{B}}\left(\bm{x}\right)$ for the streamfunction are computed based on cell mean-values $\psi_{i}$ and associated boundary values and derivatives with functions $\psi^{\textrm{B}}\left(\bm{x}\right)$ and $\nabla\psi^{\textrm{B}}\left(\bm{x}\right)\cdot\bm{n}\left(\bm{x}\right)$ to impose the prescribed Cauchy boundary condition~\cref{eq:boundary_conditions_streamfunction_cauchy_condition_1}.
Following the ROD method, a collocation point $\bm{b}_{i\textrm{B}}$ is chosen to belong to the curved physical boundary with $\bm{n}_{i\textrm{B}}$ the associated outward unit normal vector.
Then, the linear constraint~\cref{eq:polynomial_reconstruction_constraint_functional_2} with $\alpha=1$ and $\beta=1$ together with the boundary data $\bm{g}\left(\bs{b}_{i\textrm{B}}\right)=\left(\psi^{\textrm{B}}\left(\bs{b}_{i\textrm{B}}\right),\nabla\psi^{\textrm{B}}\left(\bs{b}_{i\textrm{B}}\right)\cdot\bm{n}_{i\textrm{B}}\right)$, reads
\begin{equation}
\bm{G}_{i\textrm{B}}\left(\bs{\eta}\right)
=
\begin{bmatrix}
\varphi\left(\bs{b}_{i\textrm{B}}\right)\\
\nabla\varphi\left(\bs{b}_{i\textrm{B}}\right)\cdot\bm{n}_{i\textrm{B}}
\end{bmatrix}
-
\begin{bmatrix}
\psi^{\textrm{B}}\left(\bs{b}_{i\textrm{B}}\right)\\
\nabla\psi^{\textrm{B}}\left(\bs{b}_{i\textrm{B}}\right)\cdot\bm{n}_{i\textrm{B}}
\end{bmatrix}.
\end{equation}

\item For each boundary edge $e_{i\textrm{B}}$, polynomial reconstructions $\widehat{\omega}_{i\textrm{B}}\left(\bm{x}\right)$ for the vorticity are computed based on cell mean-values $\omega_{i}$ and associated boundary values with point-values $\omega_{i\textrm{B}}$ to impose the prescribed Dirichlet boundary condition~\cref{eq:boundary_conditions_boundary_vorticity_condition_1}.
Following the ROD method, a collocation point $\bm{b}_{i\textrm{B}}$ is chosen to belong to the curved physical boundary.
Then, the linear constraint~\cref{eq:polynomial_reconstruction_constraint_functional_1} with $\alpha=1$ and $\beta=0$ together with the boundary data $g\left(\bs{b}_{i\textrm{B}}\right)=\omega_{i\textrm{B}}$, reads
\begin{equation}
G_{i\textrm{B}}\left(\bs{\eta}\right)=\varphi\left(\bs{b}_{i\textrm{B}}\right)-\omega_{i\textrm{B}}.
\end{equation}
\end{itemize}

\begin{remark}
Notice that, the boundary data for the streamfunction polynomial reconstructions associated with the boundary edges, namely $\bm{g}\left(\bs{b}_{i\textrm{B}}\right)=\left(\psi^{\textrm{B}}\left(\bs{b}_{i\textrm{B}}\right),\nabla\psi^{\textrm{B}}\left(\bs{b}_{i\textrm{B}}\right)\cdot\bm{n}_{i\textrm{B}}\right)$, can be evaluated directly from the prescribed boundary velocity using Equations~\cref{eq:boundary_conditions_boundary_streamfunction,eq:boundary_conditions_boundary_streamfunction_derivative}, apart from the constant in multiply connected domains.
On the other side, the boundary data for the vorticity polynomial reconstructions associated with the boundary edges, namely $g\left(\bs{b}_{i\textrm{B}}\right)=\omega_{i\textrm{B}}$, still requires an approximation to be computed based on Equation~\cref{eq:boundary_conditions_boundary_vorticity}.
\end{remark}


\begin{table}[!htb]
\centering
\renewcommand*{\arraystretch}{1.25}
\renewcommand{\tabcolsep}{0.2cm}
\caption{Polynomial reconstructions based on the streamfunction, vorticity, and velocity cell-mean values.}
\label{tab:finite_volume_discretisation_polynomial_reconstructions}
\resizeboxlarger{
\begin{tabular}{@{}lllll@{}}
\toprule
Underlying & Polynomial  & Reference mesh & \multirow{2}{*}{Condition type} & \multirow{2}{*}{Linear constraint}\\
function & reconstruction & element & &\\
\midrule
\multirow{4}{*}{\specialcell{@{}l@{}}{Streamfunction,\\$\phi\left(\bm{x}\right)$}} &
$\widetilde{\phi}_{ij}\left(\bm{x}\right)$ & \specialcell{@{}l@{}}{Inner edge\\[0.1cm]$e_{ij}$, $i,j\in\mathcal{I}$, $j\in\mathcal{N}_{i}$} & None & ---\\
\cmidrule{2-5}
& $\widetilde{\phi}_{i\textrm{B}}\left(\bm{x}\right)$ & \specialcell{@{}l@{}}{Boundary edge\\[0.1cm]$e_{i\textrm{B}}$, $i\in\mathcal{I}$} & None & ---\\
\midrule
\multirow{4}{*}{\specialcell{@{}l@{}}{Streamfunction,\\$\psi\left(\bm{x}\right)$}} & $\widetilde{\psi}_{ij}\left(\bm{x}\right)$ & \specialcell{@{}l@{}}{Inner edge\\[0.1cm]$e_{ij}$, $i,j\in\mathcal{I}$, $j\in\mathcal{N}_{i}$} & None & ---\\
\cmidrule{2-5}
& $\widehat{\psi}_{i\textrm{B}}\left(\bm{x}\right)$ & \specialcell{@{}l@{}}{Boundary edge\\[0.1cm]$e_{i\textrm{B}}$, $i\in\mathcal{I}$} & Cauchy & $\bm{G}_{i\textrm{B}}\left(\bs{\eta}\right)=\begin{bmatrix}\varphi\left(\bs{b}_{i\textrm{B}}\right)\\\nabla\varphi\left(\bs{b}_{i\textrm{B}}\right)\cdot\bm{n}_{i\textrm{B}}\end{bmatrix}-\begin{bmatrix}\psi^{\textrm{B}}\left(\bs{b}_{i\textrm{B}}\right)\\\nabla\psi^{\textrm{B}}\left(\bs{b}_{i\textrm{B}}\right)\cdot\bm{n}_{i\textrm{B}}\end{bmatrix}$\\
\midrule
\multirow{6}{*}{\specialcell{@{}l@{}}{Vorticity,\\$\omega\left(\bm{x}\right)$}} & $\widetilde{\omega}_{ij}\left(\bm{x}\right)$ & \specialcell{@{}l@{}}{Inner edge\\[0.1cm]$e_{ij}$, $i,j\in\mathcal{I}$, $j\in\mathcal{N}_{i}$} & None & ---\\
\cmidrule{2-5}
& $\widehat{\omega}_{i\textrm{B}}\left(\bm{x}\right)$ & \specialcell{@{}l@{}}{Boundary edge\\[0.1cm]$e_{i\textrm{B}}$, $i\in\mathcal{I}$} & Dirichlet & $G_{i\textrm{B}}\left(\bs{\eta}\right)=\varphi\left(\bs{b}_{i\textrm{B}}\right)-\omega_{i\textrm{B}}$\\
\cmidrule{2-5}
& $\widehat{\omega}_{i}\left(\bm{x}\right)$ & \specialcell{@{}l@{}}{Cell\\[0.1cm]$c_{i}$, $i\in\mathcal{I}$} & Mean-value & $G_{i}\left(\bs{\eta}\right)=\dfrac{1}{\left\vert c_{i}\right\vert}\displaystyle\int_{c_{i}}\varphi\diff\bm{x}-\omega_{i}$\\
\bottomrule
\end{tabular}
}
\end{table}

\subsection{Physical fluxes approximation}
\label{subsec:finite_volume_discretisation_physical_fluxes_approximation}

A very high-order accurate finite volume method is achieved by employing high-degree polynomial reconstructions computed from the constant piecewise approximations to evaluate the numerical fluxes.
Based on the previous polynomial reconstructions, for the vorticity transport equation, approximations to the physical fluxes in Equation~\eqref{eq:finite_volume_discretisation_exact_fluxes_1} are determined as follows:
\begin{itemize}
\item For each inner edge $e_{ij}$, physical convective and diffusive fluxes, $C_{ij,r}$ and $D_{ij,r}$, respectively, are approximated with numerical convective and diffusive fluxes, $\mathcal{C}_{ij,r}$ and $\mathcal{D}_{ij,r}$, respectively, at quadrature points $\bm{q}_{ij,r}$, $r=1,\ldots,R$, are given as
\begin{align}
&\mathcal{C}_{ij,r}=\left[\bm{v}_{ij,r}\cdot\bm{s}_{ij}\right]^{+}\widehat{\omega}_{i}\left(\bm{q}_{ij,r}\right)+\left[\bm{v}_{ij,r}\cdot\bm{s}_{ij}\right]^{-}\widehat{\omega}_{j}\left(\bm{q}_{ij,r}\right),
\label{eq:finite_volume_discretisation_vorticity_convective_flux_inner_edge1}\\
&\mathcal{D}_{ij,r}=-\nu\nabla\widetilde{\omega}_{ij}\left(\bm{q}_{ij,r}\right)\cdot\bm{s}_{ij},
\label{eq:finite_volume_discretisation_vorticity_diffusive_flux_inner_edge1}
\end{align}
with the notation $\left[a\right]^{+}=\max(a,0)$, $\left[a\right]^{-}=\min(a,0)$.

\item For each boundary edge $e_{i\textrm{B}}$, physical convective and diffusive fluxes, $C_{i\textrm{B},r}$ and $D_{i\textrm{B},r}$, respectively, are approximated with numerical convective and diffusive fluxes, $\mathcal{C}_{i\textrm{B},r}$ and $\mathcal{D}_{i\textrm{B},r}$, respectively, at quadrature points $\bm{q}_{i\textrm{B},r}$, $r=1,\ldots,R$, are given as
\begin{align}
&\mathcal{C}_{i\textrm{B},r}=\left[\bm{v}_{i\textrm{B},r}\cdot\bm{s}_{i\textrm{B}}\right]^{+}\widehat{\omega}_{i}\left(\bm{q}_{i\textrm{B},r}\right)+\left[\bm{v}_{i\textrm{B},r}\cdot\bm{s}_{i\textrm{B}}\right]^{-}\widehat{\omega}_{i\textrm{B}}\left(\bm{q}_{i\textrm{B},r}\right),
\label{eq:finite_volume_discretisation_vorticity_convective_flux_boundary_edge1}\\
&\mathcal{D}_{i\textrm{B},r}=-\nu\nabla\widehat{\omega}_{i\textrm{B}}\left(\bm{q}_{i\textrm{B},r}\right)\cdot\bm{s}_{i\textrm{B}}.
\label{eq:finite_volume_discretisation_vorticity_diffusive_flux_boundary_edge1}
\end{align}
\end{itemize}

For the convective fluxes, an upwind strategy is employed to approximate the vorticity values based on the polynomial reconstructions preserving the vorticity cell mean-values on each side of the inner edges, $e_{ij}$, that is, $\widehat{\omega}_{i}\left(\bm{x}\right)$ and $\widehat{\omega}_{j}\left(\bm{x}\right)$.
In the case of a boundary edge, $e_{i\textrm{B}}$, the polynomial reconstruction fulfilling the prescribed boundary condition, $\widehat{\omega}_{i\textrm{B}}\left(\bm{x}\right)$, together with cell mean-value preserving polynomial reconstruction $\widehat{\omega}_{i}\left(\bm{x}\right)$, are used.
For the diffusive fluxes, either unconstrained or constrained polynomial reconstructions, $\widetilde{\omega}_{ij}\left(\bm{x}\right)$ or $\widehat{\omega}_{i\textrm{B}}\left(\bm{x}\right)$, respectively, are employed to evaluate an approximation of the vorticity gradient on the edges.

For the streamfunction equation, approximations to the physical fluxes in Equation~\eqref{eq:finite_volume_discretisation_exact_fluxes_2} are determined as follows:
\begin{itemize}
\item For each inner edge $e_{ij}$, physical diffusive fluxes $F_{ij,r}$ are approximated with numerical diffusive fluxes $\mathcal{F}_{ij,r}$ at quadrature points $\bm{q}_{ij,r}$, $r=1,\ldots,R$, are given as
\begin{align}
&\mathcal{F}_{ij,r}=\nabla\widetilde{\psi}_{ij}\left(\bm{q}_{ij,r}\right)\cdot\bm{s}_{ij}.
\label{eq:finite_volume_discretisation_streamfunction_diffusive_flux_inner_edge1}
\end{align}
\item For each boundary edge $e_{i\textrm{B}}$, physical diffusive fluxes $F_{i\textrm{B},r}$ are approximated with numerical diffusive fluxes $\mathcal{F}_{i\textrm{B},r}$ at quadrature points $\bm{q}_{i\textrm{B},r}$, $r=1,\ldots,R$, are given as
\begin{align}
&\mathcal{F}_{i\textrm{B},r}=\nabla\widehat{\psi}_{i\textrm{B}}\left(\bm{q}_{i\textrm{B},r}\right)\cdot\bm{s}_{i\textrm{B}}.
\label{eq:finite_volume_discretisation_streamfunction_diffusive_flux_boundary_edge1}
\end{align}
\end{itemize}

Similar to the diffusive fluxes in the vorticity transport equation, either unconstrained or constrained polynomial reconstructions, $\widetilde{\psi}_{ij}\left(\bm{x}\right)$ or $\widehat{\psi}_{i\textrm{B}}\left(\bm{x}\right)$, respectively, are employed to evaluate an approximation of the streamfunction gradient on the edges.

Providing an approximation to $R_{i}$ in Equation~\eqref{eq:finite_volume_discretisation_exact_fluxes_2} is straightforward since it corresponds to the vorticity cell-mean value in $c_{i}$.
Then, for each cell $c_{i}$, $R_{i}$ is approximated with $\mathcal{R}_{i}$, given as
\begin{equation}
\mathcal{R}_{i}=\omega_{i}.
\label{eq:finite_volume_discretisation_streamfunction_cell_mean_value}
\end{equation}

\begin{remark}
Notice that the prescribed boundary conditions for the vorticity and streamfunction are taken into account through constrained polynomial reconstructions associated with the boundary edges.
Therefore, there is no explicit reference to either the boundary condition or the physical boundary in the numerical scheme, which handles only two situations, inner or boundary edges.
Moreover, the numerical fluxes are determined solely at quadrature points belonging to the straight edges, with no knowledge of the curved physical boundary at this stage.
\end{remark}

\subsection{Boundary vorticity approximation}
\label{subsec:finite_volume_discretisation_boundary_vorticity_approximation}

Boundary vorticity function $\omega^{\textrm{B}}\left(\bm{x}\right)$ is unknown since, in general, the prescribed boundary velocity function alone is not sufficient to compute all the terms in Equation~\cref{eq:boundary_conditions_boundary_vorticity}.
In that regard, polynomial reconstructions are employed to provide an appropriate approximation of $\omega_{i\textrm{B}}$ from numerically evaluating the second derivative of the streamfunction, while the remaining terms can be determined analytically.
More specifically, for each boundary edge $e_{i\textrm{B}}$, polynomial reconstructions $\widehat{\psi}_{i\textrm{B}}\left(\bm{x}\right)$ for the streamfunction, as for the numerical diffusive fluxes, are employed to provide an approximation to $\omega^{\textrm{B}}\left(\bm{x}\right)$ at collocation point $\bm{b}_{i\textrm{B}}$, given as
\begin{equation}
\label{eq:finite_volume_discretisation_boundary_vorticity}
\omega_{i\textrm{B}}=-\frac{\partial^{2}\widehat{\psi}_{i\textrm{B}}}{\partial\eta^{2}}\left(\bm{b}_{i\textrm{B}}\right)+\kappa_{t}\left(\bm{b}_{i\textrm{B}}\right)u^{\textrm{B}}_{\xi}\left(\bm{b}_{i\textrm{B}}\right)-\frac{\partial\bm{u}^{\textrm{B}}\cdot\bm{n}}{\partial\bm{t}}\left(\bm{b}_{i\textrm{B}}\right).
\end{equation}


\subsection{Velocity approximation}
\label{subsec:finite_volume_discretisation_velocity_approximation}

Polynomial reconstructions are employed to provide appropriate approximations to velocity point-values $\bm{v}_{ij,r}$ and $\bm{v}_{i\textrm{B},r}$ required for the computation of the numerical convective fluxes~\cref{eq:finite_volume_discretisation_vorticity_convective_flux_inner_edge1,eq:finite_volume_discretisation_vorticity_convective_flux_boundary_edge1}.
Since velocity function $\bm{v}\left(\bm{x}\right)$ can be evaluated as $\bm{v}\left(\bm{x}\right)=\nabla^{\perp}\phi\left(\bm{x}\right)$, where function $\phi\left(\bm{x}\right)$ corresponds to the streamfunction from the previous fixed-point iteration, the procedure consists in computing $\bm{v}_{ij,r}$ and $\bm{v}_{i\textrm{B},r}$ from polynomial reconstructions computed based on the streamfunction cell mean-values $\phi_{i}$, which correspond to cell mean-values $\psi_{i}$ at the previous fixed-point iteration.
More specifically, for each inner edge $e_{ij}$ and boundary edge $e_{i\textrm{B}}$, unconstrained polynomial reconstructions $\widetilde{\phi}_{ij}\left(\bm{x}\right)$ and $\widetilde{\phi}_{i\textrm{B}}\left(\bm{x}\right)$, respectively, for the streamfunction are employed to compute the velocity point-values, as follows:
\begin{itemize}
\item For each inner edge $e_{ij}$, velocity point-values $\bm{v}_{ij,r}$ at quadrature points $\bm{q}_{ij,r}$, $r=1,\ldots,R$, are given as $\bm{v}_{ij,r}=\nabla^{\perp}\widetilde{\phi}_{ij}\left(\bm{q}_{ij,r}\right)$, that is
\begin{align}
v_{x,ij,r}=\frac{\partial\widetilde{\phi}_{ij}}{\partial y}\left(\bm{q}_{ij,r}\right)
\qquad
\text{and}
\qquad
v_{y,ij,r}=-\frac{\partial\widetilde{\phi}_{ij}}{\partial x}\left(\bm{q}_{ij,r}\right).
\label{eq:finite_volume_discretisation_velocity_inner_edge1}
\end{align}
\item For each boundary edge $e_{i\textrm{B}}$, velocity point-values $\bm{v}_{i\textrm{B},r}$ at quadrature points $\bm{q}_{i\textrm{B},r}$, $r=1,\ldots,R$, are given as $\bm{v}_{i\textrm{B},r}=\nabla^{\perp}\widetilde{\phi}_{i\textrm{B}}\left(\bm{q}_{i\textrm{B},r}\right)$, that is
\begin{align}
v_{x,i\textrm{B},r}=\frac{\partial\widetilde{\phi}_{i\textrm{B}}}{\partial y}\left(\bm{q}_{i\textrm{B},r}\right)
\qquad
\text{and}
\qquad
v_{y,i\textrm{B},r}=-\frac{\partial\widetilde{\phi}_{i\textrm{B}}}{\partial x}\left(\bm{q}_{i\textrm{B},r}\right).
\label{eq:finite_volume_discretisation_velocity_boundary_edge1}
\end{align}
\end{itemize}

\subsection{Compatibility conditions}
\label{subsec:finite_volume_discretisation_compatibility_conditions}

Unknown constants $C_{k}$ associated with physical boundary subsets $\Gamma^{k}$, $k=1,\ldots,K$, are determined as part of the overall solution, with the compatibility conditions~\cref{eq:boundary_conditions_compatibility_condition} for the boundary integral of the vorticity flux imposed to close the system, which can become a cumbersome task for arbitrary curved physical boundaries.
Fortunately, these compatibility conditions can be transferred from the physical boundary, $\Gamma^{k}$, to the computational boundary, $\Gamma^{k}_{\Delta}$, thus avoiding integrating on curved boundaries, provided that the boundary data is taken into account, yielding
\begin{equation}
\sum_{e_{i\textrm{B}}\in\Gamma^{k}_{\Delta}}\left[\int_{e_{i\textrm{B}}}\left(\bm{u}\omega-\nu\nabla\omega\right)\cdot\bm{s}_{i\textrm{B}}\diff{s}\right]=\sum_{e_{i\textrm{B}}\in\Gamma^{k}_{\Delta}}\left[\int_{e_{i\textrm{B}}}\bm{f}\cdot\bm{r}_{i\textrm{B}}\diff{s}\right].
\end{equation}
Then, the integrals on the boundary edges, $e_{i\textrm{B}}$, are approximated with constrained polynomial reconstructions $\widehat{\omega}_{i\textrm{B}}\left(\bm{x}\right)$ for the vorticity computed based on cell mean-values $\omega_{i}$ and boundary vorticity point-values $\omega_{i\textrm{B}}$, evaluated at an $R$-point Gauss-Legendre quadrature rule with weights $\zeta_{r}$, $r=1,\ldots,R$, that is
\begin{equation}
\label{eq:finite_volume_discretisation_compatibility_condition}
\sum_{e_{i\textrm{B}}\in\Gamma^{k}_{\Delta}}\left\vert e_{i\textrm{B}}\right\vert\left[\sum_{r=1}^{R}\zeta_{r}\left(\bm{v}_{i\textrm{B},r}\widehat{\omega}_{i\textrm{B}}\left(\bm{q}_{i\textrm{B},r}\right)-\nu\nabla\widehat{\omega}_{i\textrm{B}}\left(\bm{q}_{i\textrm{B},r}\right)\right)\cdot\bm{s}_{i\textrm{B}}\right]=\sum_{e_{i\textrm{B}}\in\Gamma^{k}_{\Delta}}\left\vert e_{i\textrm{B}}\right\vert\left[\sum_{r=1}^{R}\zeta_{r}\bm{f}\left(\bm{q}_{i\textrm{B},r}\right)\cdot\bm{r}_{i\textrm{B}}\right].
\end{equation}

\begin{remark}
Notice that the computation of polynomial reconstructions $\widehat{\omega}_{i\textrm{B}}\left(\bm{x}\right)$ requires boundary vorticity point-values $\omega_{i\textrm{B}}$, determined from Equation~\cref{eq:finite_volume_discretisation_boundary_vorticity}.
In turn, polynomial reconstructions $\widehat{\psi}_{i\textrm{B}}$ are computed based on the associated boundary streamfunction values and derivatives, $\psi^{\textrm{B}}\left(\bm{b}_{i\textrm{B}}\right)$ and $\nabla\psi^{\textrm{B}}\left(\bm{b}_{i\textrm{B}}\right)\cdot\bm{n}_{i\textrm{B}}$.
Since $\psi^{\textrm{B}}\left(\bm{b}_{i\textrm{B}}\right)$ depends on constant $C_{k}$ of the associated physical boundary subset, $\Gamma^{k}$, then the compatibility conditions~\cref{eq:finite_volume_discretisation_compatibility_condition} effectively close the system.
This demonstrates the rationale behind using these specific compatibility conditions to fix constants $C_{k}$, provided that $\psi^{\textrm{B}}\left(\bm{b}_{i\textrm{B}}\right)$ is somehow taken into account.
\end{remark}

\subsection{Residual operators}
\label{subsec:finite_volume_discretisation_residual_operators}

Consider that the mean-value unknown variables previously introduced for the problem discretisation are gathered in vectors
\begin{equation}
\Theta^{\psi}=\left[\psi_{i}\right]_{i\in\mathcal{I}},
\qquad
\Theta^{\omega}=\left[\omega_{i}\right]_{i\in\mathcal{I}},
\qquad
\text{and}
\qquad
\Theta^{\phi}=\left[\phi_{i}\right]_{i\in\mathcal{I}},
\end{equation}
while the auxiliary point-values and constants are gathered in vectors
\begin{equation}
\Lambda^{v}=\left[v_{\beta,ij,r}\right]_{\beta\in\lbrace x,y\rbrace,i\in\mathcal{I},j\in\mathcal{N}_{i},r\in\lbrace 1,\ldots,R\rbrace},
\qquad
\Lambda^{\omega}=\left[\omega_{i\textrm{B}}\right]_{i\in\mathcal{I}^{\textrm{B}}},
\qquad
\text{and}
\qquad
\Lambda^{C}=\left[C_{k}\right]_{k=1,\ldots,K}.
\end{equation}

Then, the discretised equations are gathered in residual operators, as follows:
\begin{itemize}
\item The discrete vorticity transport equation~\eqref{eq:finite_volume_discretisation_discrete_1} for each cell $c_{i}$ provides the affine residual operator
\begin{equation}
\mathcal{G}^{\omega}_{i}\left(\Theta^{\omega},\Lambda^{\omega};\Lambda^{v}\right)=\sum_{j\in\mathcal{N}_{i}}\left\vert e_{ij}\right\vert\left[\sum_{r=1}^{R}\zeta_{r}\left(\mathcal{C}_{ij,r}+\mathcal{D}_{ij,r}\right)\right]-\left\vert c_{i}\right\vert f_{i},
\label{eq:solution_algorithm_residual_operator_1}
\end{equation}
with all residual operators gathered in vector residual operator $\mathcal{G}^{\omega}\left(\Theta^{\omega},\Lambda^{\omega};\Lambda^{v}\right)$ from $\mathbb{R}^{N_{\textrm{C}}+N_{\textrm{B}}}$ to $\mathbb{R}^{N_{\textrm{C}}}$.

\item Similarly, the discrete streamfunction equation~\eqref{eq:finite_volume_discretisation_discrete_2} for each cell $c_{i}$ provides the affine residual operator
\begin{equation}
\mathcal{G}^{\psi}_{i}\left(\Theta^{\psi},\Theta^{\omega},\Lambda^{C}\right)=\sum_{j\in\mathcal{N}_{i}}\left\vert e_{ij}\right\vert\left[\sum_{r=1}^{R}\zeta_{r}\mathcal{F}_{ij,r}\right]-\mathcal{R}_{i},
\label{eq:solution_algorithm_residual_operator_2}
\end{equation}
with all residual operators gathered in vector residual operator $\mathcal{G}^{\psi}\left(\Theta^{\psi},\Theta^{\omega},\Lambda^{C}\right)$ from $\mathbb{R}^{2N_{\textrm{C}}+K}$ to $\mathbb{R}^{N_{\textrm{C}}}$.

\item Finally, the discrete compatibility conditions~\cref{eq:finite_volume_discretisation_compatibility_condition} for each boundary subset $\Gamma^{k}_{\Delta}$ provides the affine residual operator
\begin{multline}
\mathcal{G}^{C}_{k}\left(\Theta^{\omega},\Lambda^{\omega}\right)=\sum_{e_{i\textrm{B}}\in\Gamma^{k}_{\Delta}}\left\vert e_{i\textrm{B}}\right\vert\left[\sum_{r=1}^{R}\zeta_{r}\left(\bm{v}_{i\textrm{B},r}\widehat{\omega}_{i\textrm{B}}\left(\bm{q}_{i\textrm{B},r}\right)-\nu\nabla\widehat{\omega}_{i\textrm{B}}\left(\bm{q}_{i\textrm{B},r}\right)\right)\cdot\bm{s}_{i\textrm{B}}\right]\\
-\sum_{e_{i\textrm{B}}\in\Gamma^{k}_{\Delta}}\left\vert e_{i\textrm{B}}\right\vert\left[\sum_{r=1}^{R}\zeta_{r}\bm{f}\left(\bm{q}_{i\textrm{B},r}\right)\cdot\bm{r}_{i\textrm{B}}\right],
\label{eq:solution_algorithm_residual_operator_3}
\end{multline}
with all residual operators gathered in vector residual operator $\mathcal{G}^{C}\left(\Theta^{\omega},\Lambda^{\omega}\right)$ from $\mathbb{R}^{N_{\textrm{C}}+N_{\textrm{B}}}$ to $\mathbb{R}^{K}$.
\end{itemize}

\subsection{Static condensation}
\label{subsec:finite_volume_discretisation_static_condensation}

In the residual operators~\cref{eq:solution_algorithm_residual_operator_1} and~\cref{eq:solution_algorithm_residual_operator_3}, each vorticity boundary point-value can be substituted with the prescribed numerical approximation, in Equation~\cref{eq:finite_volume_discretisation_boundary_vorticity}, which is a function of the unknown streamfunction cell mean-values in vector $\Theta^{\psi}$ and unknown constants $C_{k}$ in vector $\Lambda^{C}$ (due to the boundary streamfunction value in multiply connected domains).
Input vector $\Lambda^{v}$ in the residual operator~\cref{eq:solution_algorithm_residual_operator_1} can also be replaced with vector vector $\Theta^{\phi}$ since each velocity point-value $v_{\beta,ij,r}$ can be substituted with the prescribed numerical approximation, either Equation~\cref{eq:finite_volume_discretisation_velocity_inner_edge1} or~\cref{eq:finite_volume_discretisation_velocity_boundary_edge1}, which is a function of the streamfunction cell mean-values computed at the previous fixed point iteration.
In this case, the vector residual operators can be rewritten as $\mathcal{G}^{\omega}\left(\Theta^{\omega},\Theta^{\psi},\Lambda^{C};\Theta^{\phi}\right)$ from $\mathbb{R}^{2N_{\textrm{C}}+K}$ to $\mathbb{R}^{N_{\textrm{C}}}$ and $\mathcal{G}^{C}\left(\Theta^{\omega},\Theta^{\psi},\Lambda^{C}\right)$ from $\mathbb{R}^{2N_{\textrm{C}}+K}$ to $\mathbb{R}^{K}$.

This static condensation procedure allows rewriting the discrete problem only in terms of cell mean-values for the streamfunction and vorticity and boundary constants for the streamfunction in multiply connected domains.
Notice that vector $\Theta^{\phi}$ is not part of the solution at the current fixed point iteration, since its values correspond to the previously computed streamfunction cell mean-values.
Putting together the above residual operators, a global affine residual operator for the discrete problem is given as 
\begin{equation}
\mathcal{H}\left(\Theta^{\omega},\Theta^{\psi},\Lambda^{C};\Theta^{\phi}\right)=
\begin{bmatrix}
\mathcal{G}^{\omega}\left(\Theta^{\omega},\Theta^{\psi},\Lambda^{C};\Theta^{\phi}\right)\\
\mathcal{G}^{\psi}\left(\Theta^{\psi},\Theta^{\omega},\Lambda^{C}\right)\\
\mathcal{G}^{C}\left(\Theta^{\omega},\Theta^{\psi},\Lambda^{C}\right)
\end{bmatrix},
\end{equation}
from $\mathbb{R}^{2N_{\textrm{C}}+K}$ to $\mathbb{R}^{2N_{\textrm{C}}+K}$.

The solution of the non-primitive formulation of the steady-state incompressible Navier-Stokes problem is obtained iteratively solving $\mathcal{H}\left(\Theta^{\omega},\Theta^{\psi},\Lambda^{C};\Theta^{\phi}\right)=\bs{0}$, with $\Theta^{\phi}\leftarrow\Theta^{\psi}$ assigned between iterations.
The coefficient matrix and associated right-hand side for the above global affine residual operator are determined to improve the calculation efficiency.
Then, the GMRES method, supplemented with a preconditioning matrix, is used to solve the system of linear equations.

\begin{remark}
Notice that, the non-linearity of the global affine residual operator $\mathcal{H}\left(\Theta^{\omega},\Theta^{\psi},\Lambda^{C};\Theta^{\phi}\right)$ comes from the convective term in the vorticity transport equation, requiring the computation of the intermediate auxiliary variables for the velocity.
\end{remark}

\section{Verification benchmark}
\label{sec:verification_benchmark}

This section outlines the verification conducted on the modelling code implemented in accordance with the very high-order method presented in the previous sections.
The verification process is based on simulations of specific benchmark test cases with known or manufactured analytical solutions, following the methodology described in the next subsection.

\subsection{Verification methodology}
\label{subsec:verification_benchmark_verification_methodology}

The evaluation of the computed approximate solution is based on terms of accuracy, convergence order, stability, and robustness.
Each benchmark test case is numerically solved with progressively finer polygonal meshes generated for the associated physical domain, allowing for the determination the method's convergence order under mesh refinement.

\subsubsection{Errors and convergence orders}
\label{subsubsec:verification_benchmark_errors_and_convergence_orders}

From the given analytic functions $\omega\left(\bm{x}\right)$ and $\psi\left(\bm{x}\right)$, the exact mean-values of the vorticity and streamfunction in the cells are evaluated as
\begin{equation}
\overline{\omega}_{i}=\dfrac{1}{\left\vert c_{i}\right\vert}\int_{c_{i}}\omega\left(\bm{x}\right)\diff\bm{x},
\qquad
i\in\mathcal{I},
\qquad
\textrm{and}
\qquad
\overline{\psi}_{i}=\dfrac{1}{\left\vert c_{i}\right\vert}\int_{c_{i}}\psi\left(\bm{x}\right)\diff\bm{x},
\qquad
i\in\mathcal{I},
\end{equation}
respectively.
Similarly, from the given analytic function $\bm{u}\left(\bm{x}\right)$, the exact mean-values of the velocity on the edges are evaluated as
\begin{equation}
\overline{\bm{u}}_{ij}=\dfrac{1}{\left\vert e_{ij}\right\vert}\int_{e_{ij}}\bm{u}\left(\bm{x}\right)\diff\bm{x},
\qquad
i,\in\mathcal{I},\qquad j\in\mathcal{N}_{i},
\end{equation}
where vector $\overline{\bm{u}}_{ij}\coloneqq\left(\overline{u}_{x,ij},\overline{u}_{y,ij}\right)$ stands for the exact mean-value of the velocity components in the $x$- and $y$-directions.
Finally, from the given analytic function $\omega^{\textrm{B}}\left(\bm{x}\right)$, the exact point-values of boundary vorticity at the collocation points are evaluated as
\begin{equation}
\overline{\omega}_{i\textrm{B}}=\omega\left(\bm{b}_{i\textrm{B}}\right),
\qquad
i\in\mathcal{I}^{\textrm{B}}.
\end{equation}
The corresponding approximate variables are denoted as $\omega^{\ast}_{i}$, $\psi^{\ast}_{i}$, $\bm{u}^{\ast}_{ij}\coloneqq\left(u^{\ast}_{x,ij},u^{\ast}_{y,ij}\right)$, and $\omega^{\textrm{B},\ast}_{i\textrm{B}}$ respectively, and are obtained by successively solving the discrete system of linear equations until convergence of the fixed point.

With the exact and approximate values in hand, the associated errors are determined in the $L^{1}$- and $L^{\infty}$-norms, generically denoted as $E_{1}$ and $E_{\infty}$, given as
\begin{equation}
E_{1}=\dfrac{\displaystyle\sum_{i\in\mathcal{I}}\left\vert\xi^{\ast}_{i}-\overline{\xi}_{i}\right\vert\left\vert c_{i}\right\vert}{\displaystyle\sum_{i\in\mathcal{I}}\left\vert c_{i}\right\vert}
\qquad
\text{and}
\qquad
E_{\infty}=\displaystyle\max_{i\in\mathcal{I}}\left\vert\xi^{\ast}_{i}-\overline{\xi}_{i}\right\vert,
\end{equation}
where $\xi$ is taken as either the vorticity or streamfunction.
For the velocity and boundary vorticity, the associated errors are computed similarly, adapting the above expressions for the edges and collocation points accordingly.

To determine the converge order, two meshes are considered, denoted as $\mathcal{M}_{1}$ and $\mathcal{M}_{2}$, for the same physical domain having different characteristic sizes, $N_{\textrm{C},1}$ and $N_{\textrm{C},2}$ for cells, respectively, $N_{\textrm{E},1}$ and $N_{\textrm{E},2}$ for edges, respectively, and $N_{\textrm{B},1}$ and $N_{\textrm{B},2}$ for boundary edges, respectively.
With the associated relative errors in the $L^{1}$-norm denoted as $E_{1,1}$ and $E_{1,2}$, and in the $L^{\infty}$-norm are denoted as $E_{\infty,1}$ and $E_{\infty,2}$, the convergence orders for the relative errors in the $L^{1}$- and $L^{\infty}$-norms, denoted as $O_{1}$ and $O_{\infty}$, respectively, are determined as
\begin{equation}
O_{1}=2\dfrac{\left\vert\ln\left(\dfrac{E_{1,1}}{E_{1,2}}\right)\right\vert}{\left\vert\ln\left(\dfrac{N_{\textrm{C},1}}{N_{\textrm{C},2}}\right)\right\vert}
\qquad
\text{and}
\qquad
O_{\infty}=2\dfrac{\left\vert\ln\left(\dfrac{E_{\infty,1}}{E_{\infty,2}}\right)\right\vert}{\left\vert\ln\left(\dfrac{N_{\textrm{C},1}}{N_{\textrm{C},2}}\right)\right\vert},
\end{equation}

For compactness, the $L^{1}$-norm errors for the velocity presented thereafter correspond to the sum of the individual components errors, while the $L^{\infty}$-norm errors correspond to the maximum of the individual components errors.
The associated convergence orders are then determined accordingly.

\subsubsection{Methods and boundary treatment}
\label{subsubsec:verification_benchmark_methods_and_boundary_treatment}

Seeking to investigate and improve the numerical accuracy and convergence order of the proposed techniques, various methods are explored regarding the boundary treatment.
More specifically, for the location of the collocation points in Equations~\cref{eq:polynomial_reconstruction_constraint_functional_1,eq:polynomial_reconstruction_constraint_functional_2} and the associated boundary condition values, the following methods are considered:
\begin{itemize}
\item Naive method -- collocation points correspond to the edge midpoints (computational boundary).
For constant boundary conditions, the boundary condition values to be fulfilled correspond to the prescribed constant; otherwise, some extrapolation of the boundary condition function from the physical boundary to the computational boundary is necessary.
For simplicity, for each edge, the projection of the edge midpoint onto the physical boundary is computed, and the associated value is obtained from the prescribed boundary condition function at the edge midpoint.

\item Exact method -- collocation points correspond to the edge midpoints (computational boundary).
In contrast to the naive method, the exact or manufactured solution is used to obtain the boundary condition values to be fulfilled; hence the computational boundary acts as a physical boundary with the appropriate values to be satisfied.
In practice, such an approach cannot be applied since the analytical solution is not known, but it allows for a comparison of the ROD method with that of an ideal situation, that is, without a geometrical gap between the physical and computational boundaries.

\item ROD method -- collocation points are derived from the analytical boundary description (physical boundary), where either one or two collocation points for each polynomial reconstruction associated with the boundary edges are used.
For one collocation point, a projection of the edge midpoint onto the physical boundary is computed, whereas for two collocation points, two points with barycentric coordinates $\left(1/3,2/3\right)$ and $\left(2/3,1/3\right)$ relative to the vertices of the edge are projected onto the physical boundary.
The boundary condition values to be fulfilled are straightforwardly obtained from the prescribed boundary condition function.
\end{itemize}

Regarding the polynomial reconstructions for the vorticity and streamfunction, the following methods are considered:
\begin{itemize}
\item $\mathbb{P}_{d}$--$\mathbb{P}_{d}$ method -- $d$-degree polynomials are employed for both the vorticity and streamfunction polynomial reconstructions.
\item $\mathbb{P}_{d}$--$\mathbb{P}_{d+1}$ method -- $d$-degree polynomials are employed for the vorticity reconstructions and $\left(d+1\right)$-degree polynomials are employed for the streamfunction polynomial reconstructions.
\item $\mathbb{P}_{d}$--$\mathbb{P}_{d+2}$ method -- $d$-degree polynomials are employed for the vorticity reconstructions and $\left(d+2\right)$-degree polynomials are employed for the streamfunction polynomial reconstructions.
\end{itemize}

\begin{remark}
Following previous works, for the ROD method to achieve optimal accuracy and convergence order, $\left(d+1\right)$-degree polynomials are employed for the polynomial reconstructions associated with the boundary edges imposing a Cauchy boundary condition, whenever $d$-degree polynomials are employed for those associated with the cells and inner edges.
\end{remark}


\subsection{Circular domain}
\label{subsec:verification_benchmark_t1}

This test case addresses an internal creeping flow (Stokes equations) in a circular domain with radius $r_{\textrm{E}}>0$ centred at the origin.
The manufactured solution for the vorticity and streamfunction in polar coordinates, $\omega\coloneqq\omega(r,\theta)$ and $\psi\coloneqq\psi(r,\theta)$, respectively, are given as
\begin{equation}
\omega=\dfrac{2u_{\textrm{E}}}{r_{\textrm{E}}}\left(r^{2}+1\right)\exp\left(r^{2}-r_{\textrm{E}}^{2}\right)
\qquad
\text{and}
\qquad
\psi=\dfrac{u_{\textrm{E}}}{2r_{\textrm{E}}\exp\left(r_{\textrm{E}}^{2}\right)}\left(\exp\left(r_{\textrm{E}}^{2}\right)-\exp\left(r^{2}\right)\right),
\end{equation}
where $u_{\textrm{E}}\in\mathbb{R}$ is the linear velocity magnitude on the physical boundary.
The velocity components in polar coordinates, $u_{r}\coloneqq u_{r}\left(r,\theta\right)$ and $u_{\theta}\coloneqq u_{\theta}\left(r,\theta\right)$, respectively, are obtained from the streamfunction and given as
\begin{equation}
u_{r}=0
\qquad
\textrm{and}
\qquad
u_{\theta}=\dfrac{u_{\textrm{E}}}{r_{\textrm{E}}}r\exp\left(r^{2}-r_{\textrm{E}}^{2}\right).
\end{equation}
The source term function in polar coordinates, $f\coloneqq f\left(r,\theta\right)$, derives from the vorticity transport equation with the above-manufactured solution and is given as
\begin{equation}
f=-\dfrac{4\nu u_{\textrm{E}}}{r_{\textrm{E}}}r\left(r^{2}+2\right)\exp\left(r^{2}-r_{\textrm{E}}^{2}\right).
\end{equation}
Note that although the exact solution is given in polar coordinates, the problem is solved in Cartesian coordinates.

\begin{figure}[!htb]
\centering
\begin{tabular}{@{}c@{}c@{}}
\includegraphics[width=0.49\textwidth,trim=0cm 0cm 0cm 0cm,clip=true]{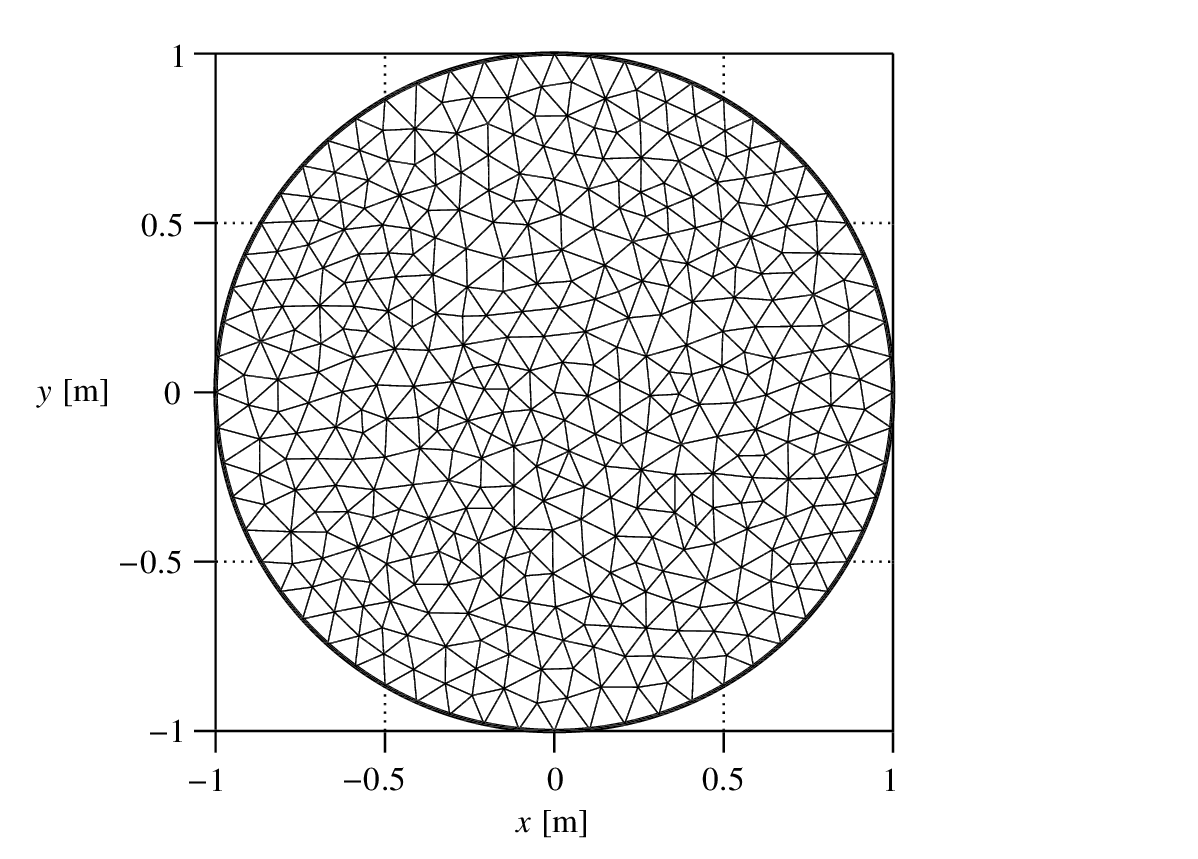} &
\includegraphics[width=0.49\textwidth,trim=0cm 0cm 0cm 0cm,clip=true]{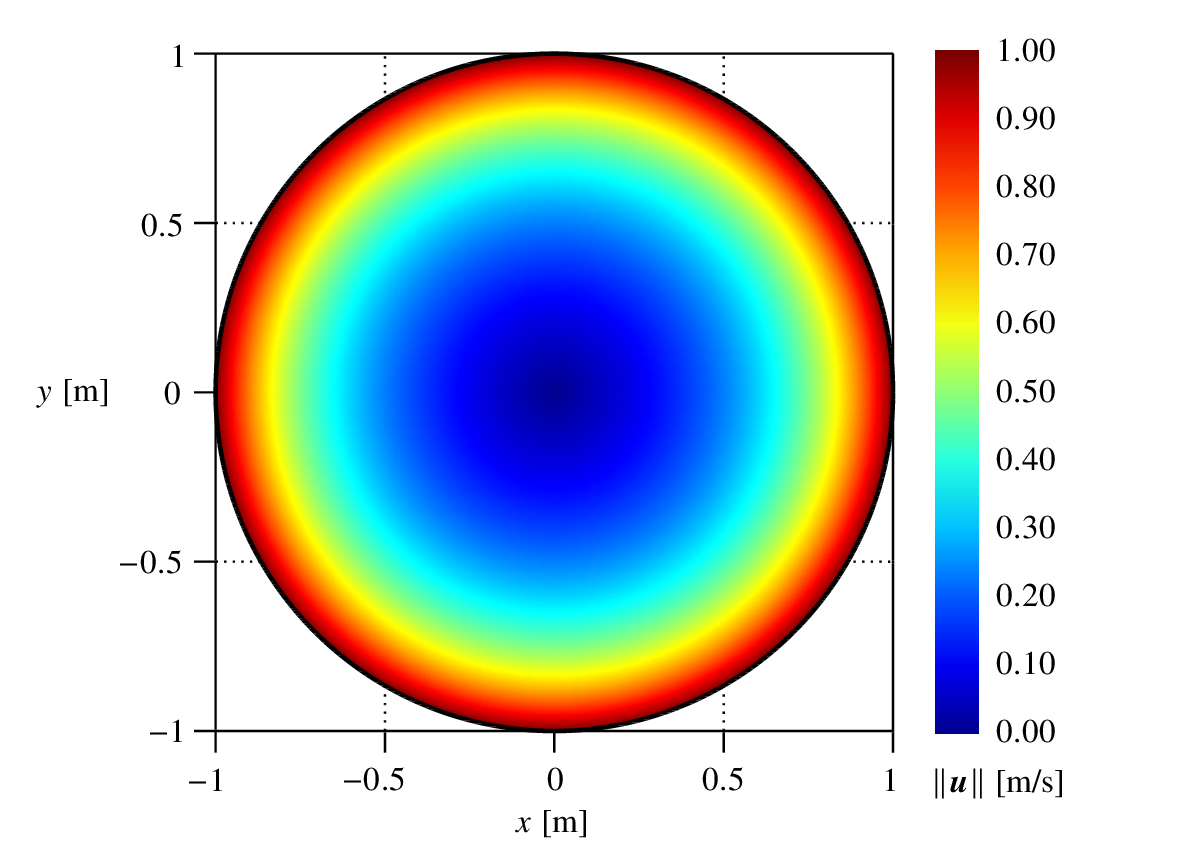}\\[0.2cm]
\small (a) Coarse mesh. & \small (b) Velocity magnitude.\\
\includegraphics[width=0.49\textwidth,trim=0cm 0cm 0cm 0cm,clip=true]{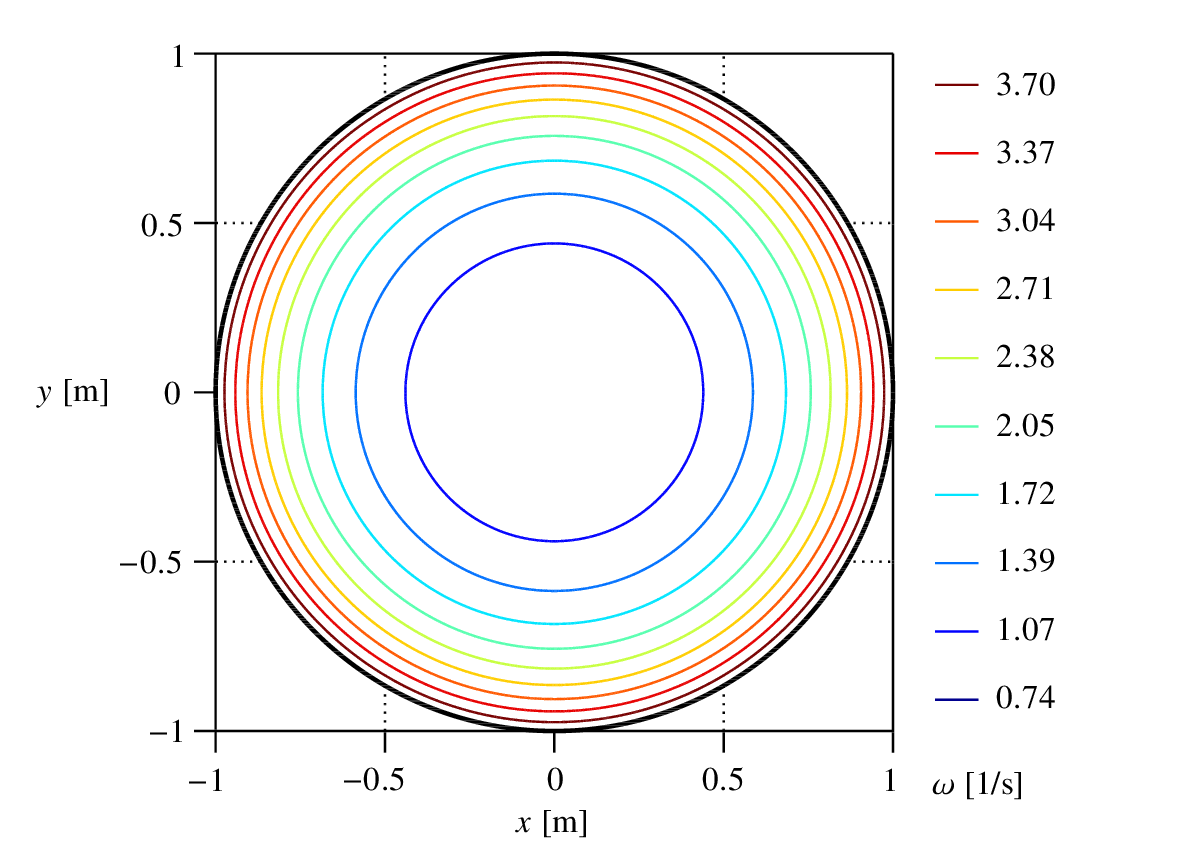} &
\includegraphics[width=0.49\textwidth,trim=0cm 0cm 0cm 0cm,clip=true]{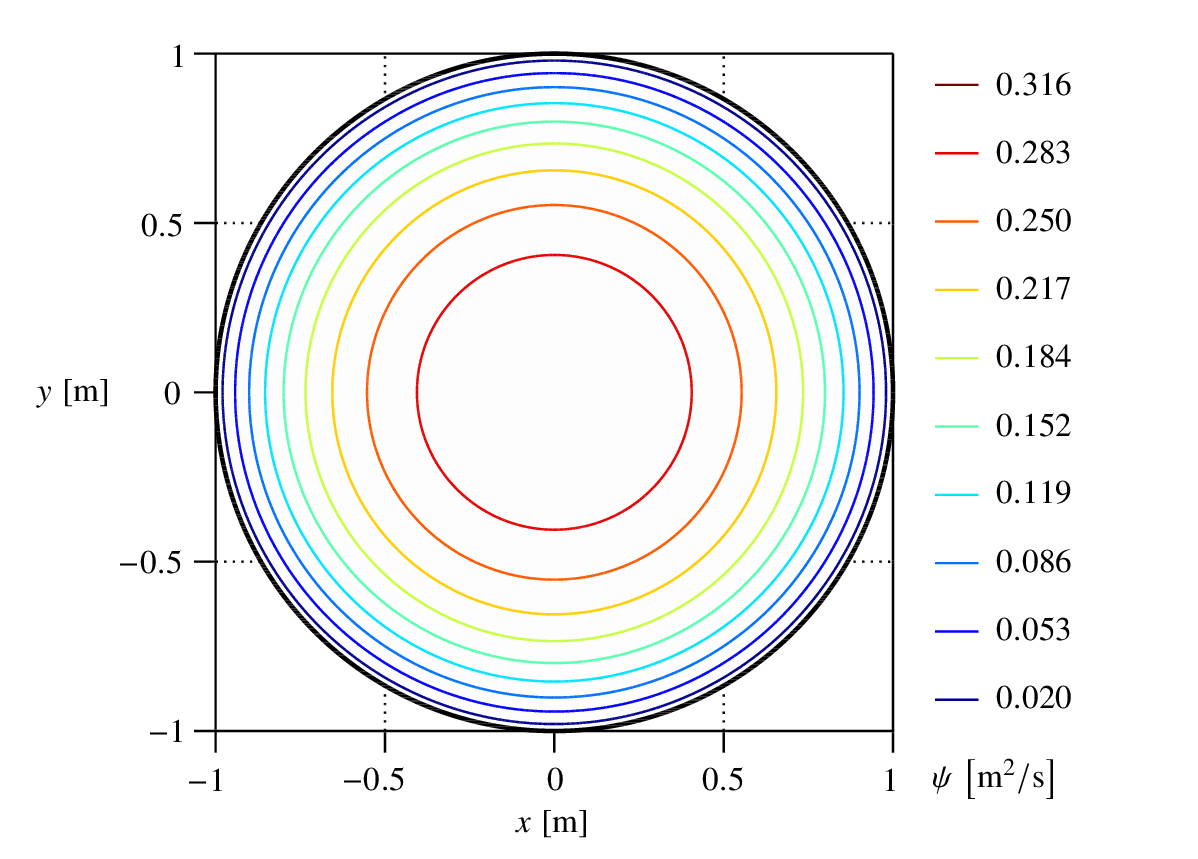}\\[0.2cm]
\small (c) Vorticity contours. & \small (d) Streamlines.
\end{tabular}
\caption{Coarse mesh and analytical solution for the flow in a circular domain test case (for proper interpretation of the colour scale, the reader is referred to the electronic version of this article).}
\label{fig:verification_benchmark_t1}
\end{figure}

For the domain, the physical boundary has a radius of $r_{\textrm{E}}=1$~\si[per-mode=symbol]{\m} (normal curvature of $\kappa_{\textrm{E}}=1$).
For the fluid, a dynamic viscosity of $\mu=1$~\si[per-mode=symbol]{\kg\per\m\per\second}, a density of $\rho=1$~\si[per-mode=symbol]{\kg\per\m^{3}}, and a velocity magnitude of $u_{\textrm{E}}=1$~\si[per-mode=symbol]{\m\per\second} are considered.
Successively finer uniform Delaunay triangular meshes are used to discretise the physical domain, and the simulations are carried out for the $\mathbb{P}_{d}$--$\mathbb{P}_{d}$ method with $d=1,3,5$, while the boundary is treated with the naive, exact, or ROD methods.
Figure~\ref{fig:verification_benchmark_t1} illustrates the physical domain with a coarse mesh and the analytical solutions for the velocity, vorticity, and streamfunction fields.


\subsubsection{Boundary treatment assessment}
\label{subsec:verification_benchmark_t11}

To assess the different boundary treatment methods without the influence of the boundary vorticity approximation, the exact values derived from the corresponding analytic solution are imposed instead at this stage.
In all the simulations, the $\mathbb{P}_{d}$--$\mathbb{P}_{d}$ method is employed, meaning that the same polynomial degree is considered for both the vorticity and streamfunction polynomial reconstructions.

The errors and convergence orders for the approximate vorticity and streamfunction are reported in Tables~\ref{tab:verification_benchmark_t11_errorstable_omega} and~\ref{tab:verification_benchmark_t11_errorstable_psi}, respectively.
In both cases, the naive method provides a second-order convergence limitation regardless of the employed polynomial degree for both error norms, since the prescribed boundary conditions are fulfilled on the computational boundary rather than on the physical boundary.
Conversely, the ROD method effectively achieves optimal convergence, obtaining second-, fourth-, and sixth-order with $d=1,3,5$, respectively, in both error norms.
Moreover, the ROD method provides comparable accuracy and convergence orders to the exact method, supporting its effectiveness relative to the ideal scenario.

\begin{table}[!htb]
\centering
\footnotesize
\caption{Vorticity errors and convergence orders obtained in the flow in a circular domain test case with exact boundary vorticity.}
\label{tab:verification_benchmark_t11_errorstable_omega}
\resizeboxlarger{
\begin{tabular}{@{}r@{}r@{}rrrr@{}r@{}rrrr@{}r@{}rrrr@{}}
\toprule
& \phantom{aaa} & \multicolumn{4}{@{}l@{}}{$d=1$} & \phantom{aaa} & \multicolumn{4}{@{}l@{}}{$d=3$} & \phantom{aaa} & \multicolumn{4}{@{}l@{}}{$d=5$} \\
\cmidrule{3-6}\cmidrule{8-11}\cmidrule{13-16}
$N_{\textrm{C}}$ & & $E_{1}$ & $O_{1}$ & $E_{\infty}$ & $O_{\infty}$ & & $E_{1}$ & $O_{1}$ & $E_{\infty}$ & $O_{\infty}$ & & $E_{1}$ & $O_{1}$ & $E_{\infty}$ & $O_{\infty}$ \\
\midrule
\multicolumn{16}{@{}l}{Naive} \\
\midrule
1,096 & & 6.25E$-$03 & --- & 2.12E$-$02 & --- & & 1.14E$-$02 & --- & 1.16E$-$02 & --- & & 1.14E$-$02 & --- & 1.14E$-$02 & --- \\
2,478 & & 3.04E$-$03 & 1.76 & 1.33E$-$02 & 1.15 & & 5.07E$-$03 & 1.99 & 5.12E$-$03 & 2.01 & & 5.07E$-$03 & 1.98 & 5.07E$-$03 & 1.98 \\
5,366 & & 1.61E$-$03 & 1.64 & 7.04E$-$03 & 1.65 & & 2.31E$-$03 & 2.03 & 2.33E$-$03 & 2.04 & & 2.31E$-$03 & 2.03 & 2.31E$-$03 & 2.03 \\
11,624 & & 6.95E$-$04 & 2.18 & 3.53E$-$03 & 1.78 & & 1.06E$-$03 & 2.01 & 1.07E$-$03 & 2.02 & & 1.06E$-$03 & 2.01 & 1.06E$-$03 & 2.01 \\
26,890 & & 3.23E$-$04 & 1.83 & 1.49E$-$03 & 2.07 & & 4.67E$-$04 & 1.96 & 4.68E$-$04 & 1.96 & & 4.67E$-$04 & 1.96 & 4.67E$-$04 & 1.96 \\
\midrule
\multicolumn{16}{@{}l}{Exact} \\
\midrule
1,096 & & 1.76E$-$02 & --- & 3.26E$-$02 & --- & & 7.87E$-$05 & --- & 3.93E$-$04 & --- & & 1.52E$-$05 & --- & 2.54E$-$05 & --- \\
2,478 & & 8.11E$-$03 & 1.91 & 1.84E$-$02 & 1.41 & & 1.54E$-$05 & 4.00 & 1.04E$-$04 & 3.25 & & 1.89E$-$06 & 5.10 & 3.15E$-$06 & 5.12 \\
5,366 & & 3.92E$-$03 & 1.88 & 9.35E$-$03 & 1.75 & & 3.20E$-$06 & 4.07 & 2.34E$-$05 & 3.88 & & 2.19E$-$07 & 5.58 & 3.78E$-$07 & 5.48 \\
11,624 & & 1.76E$-$03 & 2.08 & 4.60E$-$03 & 1.84 & & 6.47E$-$07 & 4.13 & 6.33E$-$06 & 3.38 & & 2.52E$-$08 & 5.60 & 3.89E$-$08 & 5.89 \\
26,890 & & 7.90E$-$04 & 1.91 & 1.95E$-$03 & 2.04 & & 1.18E$-$07 & 4.06 & 1.21E$-$06 & 3.93 & & 2.04E$-$09 & 5.99 & 3.43E$-$09 & 5.79 \\
\midrule
\multicolumn{16}{@{}l}{ROD} \\
\midrule
1,096 & & 1.75E$-$02 & --- & 3.25E$-$02 & --- & & 7.85E$-$05 & --- & 3.93E$-$04 & --- & & 1.51E$-$05 & --- & 2.53E$-$05 & --- \\
2,478 & & 8.08E$-$03 & 1.90 & 1.83E$-$02 & 1.40 & & 1.54E$-$05 & 3.99 & 1.04E$-$04 & 3.25 & & 1.89E$-$06 & 5.10 & 3.14E$-$06 & 5.12 \\
5,366 & & 3.91E$-$03 & 1.88 & 9.33E$-$03 & 1.75 & & 3.19E$-$06 & 4.07 & 2.33E$-$05 & 3.88 & & 2.19E$-$07 & 5.58 & 3.78E$-$07 & 5.48 \\
11,624 & & 1.75E$-$03 & 2.08 & 4.59E$-$03 & 1.83 & & 6.47E$-$07 & 4.13 & 6.32E$-$06 & 3.38 & & 2.52E$-$08 & 5.60 & 3.88E$-$08 & 5.89 \\
26,890 & & 7.89E$-$04 & 1.91 & 1.95E$-$03 & 2.04 & & 1.18E$-$07 & 4.06 & 1.21E$-$06 & 3.93 & & 2.05E$-$09 & 5.99 & 3.43E$-$09 & 5.79 \\
\bottomrule
\end{tabular}

}
\end{table}

\begin{table}[!htb]
\centering
\footnotesize
\caption{Streamfunction errors and convergence orders obtained in the flow in a circular domain test case with exact boundary vorticity.}
\label{tab:verification_benchmark_t11_errorstable_psi}
\resizeboxlarger{
\begin{tabular}{@{}r@{}r@{}rrrr@{}r@{}rrrr@{}r@{}rrrr@{}}
\toprule
& \phantom{aaa} & \multicolumn{4}{@{}l@{}}{$d=1$} & \phantom{aaa} & \multicolumn{4}{@{}l@{}}{$d=3$} & \phantom{aaa} & \multicolumn{4}{@{}l@{}}{$d=5$} \\
\cmidrule{3-6}\cmidrule{8-11}\cmidrule{13-16}
$N_{\textrm{C}}$ & & $E_{1}$ & $O_{1}$ & $E_{\infty}$ & $O_{\infty}$ & & $E_{1}$ & $O_{1}$ & $E_{\infty}$ & $O_{\infty}$ & & $E_{1}$ & $O_{1}$ & $E_{\infty}$ & $O_{\infty}$ \\
\midrule
\multicolumn{16}{@{}l}{Naive} \\
\midrule
1,096 & & 3.29E$-$04 & --- & 1.32E$-$03 & --- & & 7.87E$-$04 & --- & 1.89E$-$03 & --- & & 7.86E$-$04 & --- & 1.88E$-$03 & --- \\
2,478 & & 1.60E$-$04 & 1.77 & 8.63E$-$04 & 1.05 & & 3.51E$-$04 & 1.98 & 8.43E$-$04 & 1.98 & & 3.51E$-$04 & 1.98 & 8.43E$-$04 & 1.97 \\
5,366 & & 8.59E$-$05 & 1.61 & 4.57E$-$04 & 1.64 & & 1.60E$-$04 & 2.03 & 3.85E$-$04 & 2.03 & & 1.60E$-$04 & 2.03 & 3.85E$-$04 & 2.03 \\
11,624 & & 3.68E$-$05 & 2.19 & 2.31E$-$04 & 1.77 & & 7.38E$-$05 & 2.01 & 1.77E$-$04 & 2.01 & & 7.38E$-$05 & 2.01 & 1.77E$-$04 & 2.01 \\
26,890 & & 1.70E$-$05 & 1.84 & 9.65E$-$05 & 2.08 & & 3.24E$-$05 & 1.96 & 7.78E$-$05 & 1.96 & & 3.24E$-$05 & 1.96 & 7.78E$-$05 & 1.96 \\
\midrule
\multicolumn{16}{@{}l}{Exact} \\
\midrule
1,096 & & 1.04E$-$03 & --- & 2.24E$-$03 & --- & & 4.32E$-$06 & --- & 2.07E$-$05 & --- & & 1.45E$-$06 & --- & 3.50E$-$06 & --- \\
2,478 & & 4.86E$-$04 & 1.86 & 1.21E$-$03 & 1.52 & & 8.82E$-$07 & 3.89 & 5.32E$-$06 & 3.33 & & 1.73E$-$07 & 5.21 & 4.17E$-$07 & 5.21 \\
5,366 & & 2.37E$-$04 & 1.86 & 6.22E$-$04 & 1.71 & & 1.73E$-$07 & 4.22 & 1.22E$-$06 & 3.82 & & 1.94E$-$08 & 5.66 & 4.65E$-$08 & 5.68 \\
11,624 & & 1.05E$-$04 & 2.10 & 3.13E$-$04 & 1.77 & & 4.88E$-$08 & 3.27 & 3.22E$-$07 & 3.44 & & 2.22E$-$09 & 5.61 & 5.24E$-$09 & 5.65 \\
26,890 & & 4.74E$-$05 & 1.89 & 1.34E$-$04 & 2.03 & & 7.98E$-$09 & 4.32 & 6.16E$-$08 & 3.94 & & 2.09E$-$10 & 5.64 & 4.94E$-$10 & 5.63 \\
\midrule
\multicolumn{16}{@{}l}{ROD} \\
\midrule
1,096 & & 1.03E$-$03 & --- & 2.23E$-$03 & --- & & 4.29E$-$06 & --- & 2.07E$-$05 & --- & & 1.46E$-$06 & --- & 3.49E$-$06 & --- \\
2,478 & & 4.84E$-$04 & 1.86 & 1.20E$-$03 & 1.51 & & 8.87E$-$07 & 3.86 & 5.32E$-$06 & 3.33 & & 1.73E$-$07 & 5.22 & 4.16E$-$07 & 5.21 \\
5,366 & & 2.36E$-$04 & 1.86 & 6.21E$-$04 & 1.71 & & 1.72E$-$07 & 4.25 & 1.22E$-$06 & 3.82 & & 1.94E$-$08 & 5.67 & 4.65E$-$08 & 5.68 \\
11,624 & & 1.05E$-$04 & 2.10 & 3.13E$-$04 & 1.77 & & 4.89E$-$08 & 3.25 & 3.21E$-$07 & 3.44 & & 2.22E$-$09 & 5.61 & 5.24E$-$09 & 5.64 \\
26,890 & & 4.74E$-$05 & 1.89 & 1.34E$-$04 & 2.03 & & 8.00E$-$09 & 4.32 & 6.16E$-$08 & 3.94 & & 2.10E$-$10 & 5.63 & 4.97E$-$10 & 5.62 \\
\bottomrule
\end{tabular}

}
\end{table}

The errors and convergence orders for the approximate velocity are reported in Table~\ref{tab:verification_benchmark_t11_errorstable_u}.
Apart from the naive method, which fails to achieve the optimal convergence orders with $d=3,5$, the first-, third-, and fifth-order of convergence is obtained with $d=1,3,5$, respectively, especially in the $L^{\infty}$-norm, while slightly higher convergence orders are obtained in the $L^{1}$-norm.
In this case, there is a reduction in the convergence orders compared to those obtained for the streamfunction, which is expected since the velocity is obtained from the streamfunction first derivatives.

\begin{table}[!htb]
\centering
\footnotesize
\caption{Velocity errors and convergence orders obtained in the flow in a circular domain test case with exact boundary vorticity.}
\label{tab:verification_benchmark_t11_errorstable_u}
\resizeboxlarger{
\begin{tabular}{@{}r@{}r@{}rrrr@{}r@{}rrrr@{}r@{}rrrr@{}}
\toprule
& \phantom{aaa} & \multicolumn{4}{@{}l@{}}{$d=1$} & \phantom{aaa} & \multicolumn{4}{@{}l@{}}{$d=3$} & \phantom{aaa} & \multicolumn{4}{@{}l@{}}{$d=5$} \\
\cmidrule{3-6}\cmidrule{8-11}\cmidrule{13-16}
$N_{\textrm{E}}$ & & $E_{1}$ & $O_{1}$ & $E_{\infty}$ & $O_{\infty}$ & & $E_{1}$ & $O_{1}$ & $E_{\infty}$ & $O_{\infty}$ & & $E_{1}$ & $O_{1}$ & $E_{\infty}$ & $O_{\infty}$ \\
\midrule
\multicolumn{16}{@{}l}{Naive} \\
\midrule
1,680 & & 1.65E$-$02 & --- & 1.68E$-$01 & --- & & 3.79E$-$03 & --- & 5.65E$-$03 & --- & & 3.83E$-$03 & --- & 5.67E$-$03 & --- \\
3,771 & & 9.08E$-$03 & 1.48 & 1.27E$-$01 & 0.68 & & 1.69E$-$03 & 2.00 & 2.55E$-$03 & 1.97 & & 1.70E$-$03 & 2.01 & 2.53E$-$03 & 1.99 \\
8,129 & & 5.12E$-$03 & 1.49 & 8.56E$-$02 & 1.04 & & 7.72E$-$04 & 2.04 & 1.19E$-$03 & 1.98 & & 7.74E$-$04 & 2.05 & 1.16E$-$03 & 2.04 \\
17,554 & & 3.00E$-$03 & 1.39 & 6.12E$-$02 & 0.87 & & 3.55E$-$04 & 2.02 & 5.40E$-$04 & 2.05 & & 3.56E$-$04 & 2.02 & 5.32E$-$04 & 2.02 \\
40,513 & & 1.75E$-$03 & 1.28 & 4.26E$-$02 & 0.87 & & 1.56E$-$04 & 1.97 & 2.39E$-$04 & 1.95 & & 1.56E$-$04 & 1.97 & 2.34E$-$04 & 1.97 \\
\midrule
\multicolumn{16}{@{}l}{Exact} \\
\midrule
1,680 & & 1.83E$-$02 & --- & 1.73E$-$01 & --- & & 1.41E$-$04 & --- & 1.84E$-$03 & --- & & 1.00E$-$05 & --- & 1.19E$-$04 & --- \\
3,771 & & 9.70E$-$03 & 1.57 & 1.30E$-$01 & 0.71 & & 3.69E$-$05 & 3.31 & 6.99E$-$04 & 2.40 & & 1.10E$-$06 & 5.47 & 1.57E$-$05 & 5.01 \\
8,129 & & 5.33E$-$03 & 1.56 & 8.67E$-$02 & 1.05 & & 1.04E$-$05 & 3.29 & 3.38E$-$04 & 1.89 & & 1.28E$-$07 & 5.59 & 2.45E$-$06 & 4.85 \\
17,554 & & 3.07E$-$03 & 1.43 & 6.17E$-$02 & 0.88 & & 2.97E$-$06 & 3.26 & 8.16E$-$05 & 3.69 & & 1.60E$-$08 & 5.42 & 4.03E$-$07 & 4.68 \\
40,513 & & 1.78E$-$03 & 1.31 & 4.28E$-$02 & 0.88 & & 7.78E$-$07 & 3.21 & 2.38E$-$05 & 2.94 & & 1.74E$-$09 & 5.30 & 6.02E$-$08 & 4.55 \\
\midrule
\multicolumn{16}{@{}l}{ROD} \\
\midrule
1,680 & & 1.83E$-$02 & --- & 1.73E$-$01 & --- & & 1.41E$-$04 & --- & 1.84E$-$03 & --- & & 1.00E$-$05 & --- & 1.19E$-$04 & --- \\
3,771 & & 9.70E$-$03 & 1.57 & 1.30E$-$01 & 0.71 & & 3.69E$-$05 & 3.31 & 6.98E$-$04 & 2.40 & & 1.10E$-$06 & 5.46 & 1.57E$-$05 & 5.01 \\
8,129 & & 5.32E$-$03 & 1.56 & 8.67E$-$02 & 1.05 & & 1.04E$-$05 & 3.29 & 3.38E$-$04 & 1.89 & & 1.28E$-$07 & 5.59 & 2.44E$-$06 & 4.85 \\
17,554 & & 3.07E$-$03 & 1.43 & 6.17E$-$02 & 0.88 & & 2.97E$-$06 & 3.26 & 8.15E$-$05 & 3.69 & & 1.60E$-$08 & 5.42 & 4.03E$-$07 & 4.68 \\
40,513 & & 1.78E$-$03 & 1.31 & 4.28E$-$02 & 0.88 & & 7.78E$-$07 & 3.21 & 2.38E$-$05 & 2.94 & & 1.74E$-$09 & 5.30 & 6.02E$-$08 & 4.55 \\
\bottomrule
\end{tabular}

}
\end{table}

From the approximated streamfunction cell mean-values and associated boundary data, the approximate boundary vorticity is computed \textit{a posteriori} using Equation~\cref{eq:finite_volume_discretisation_boundary_vorticity} (while the exact boundary vorticity was used for the simulations) and the associated errors and convergence orders are reported in Table~\ref{tab:verification_benchmark_t11_errorstable_omegabound}.
As observed, the ROD method achieves second- and fourth-order of convergence with $d=3,5$, respectively, for both error norms, whereas it fails to converge with $d=1$ since the approximate boundary vorticity is obtained from the streamfunction second derivatives and, therefore, a reduction of two orders is expected compared to those obtained for the streamfunction.

\begin{table}[!htb]
\centering
\footnotesize
\caption{Boundary vorticity errors and convergence orders obtained in the flow in a circular domain test case with exact boundary vorticity.}
\label{tab:verification_benchmark_t11_errorstable_omegabound}
\resizeboxlarger{
\begin{tabular}{@{}r@{}r@{}rrrr@{}r@{}rrrr@{}r@{}rrrr@{}}
\toprule
& \phantom{aaa} & \multicolumn{4}{@{}l@{}}{$d=1$} & \phantom{aaa} & \multicolumn{4}{@{}l@{}}{$d=3$} & \phantom{aaa} & \multicolumn{4}{@{}l@{}}{$d=5$} \\
\cmidrule{3-6}\cmidrule{8-11}\cmidrule{13-16}
$N_{\textrm{B}}$ & & $E_{1}$ & $O_{1}$ & $E_{\infty}$ & $O_{\infty}$ & & $E_{1}$ & $O_{1}$ & $E_{\infty}$ & $O_{\infty}$ & & $E_{1}$ & $O_{1}$ & $E_{\infty}$ & $O_{\infty}$ \\
\midrule
\multicolumn{16}{@{}l}{Naive} \\
\midrule
72 & & 3.00E$+$00 & --- & 3.00E$+$00 & --- & & 6.55E$-$02 & --- & 7.23E$-$02 & --- & & 5.62E$-$03 & --- & 6.43E$-$03 & --- \\
108 & & 3.00E$+$00 & --- & 3.00E$+$00 & --- & & 3.21E$-$02 & 1.76 & 3.94E$-$02 & 1.50 & & 2.07E$-$03 & 2.47 & 2.30E$-$03 & 2.54 \\
160 & & 3.00E$+$00 & --- & 3.00E$+$00 & --- & & 1.52E$-$02 & 1.91 & 1.91E$-$02 & 1.85 & & 8.56E$-$04 & 2.24 & 9.23E$-$04 & 2.32 \\
236 & & 3.00E$+$00 & --- & 3.00E$+$00 & --- & & 7.51E$-$03 & 1.81 & 9.67E$-$03 & 1.75 & & 3.73E$-$04 & 2.14 & 3.90E$-$04 & 2.21 \\
356 & & 3.00E$+$00 & --- & 3.00E$+$00 & --- & & 3.20E$-$03 & 2.07 & 4.39E$-$03 & 1.92 & & 1.59E$-$04 & 2.07 & 1.64E$-$04 & 2.10 \\
\midrule
\multicolumn{16}{@{}l}{Exact} \\
\midrule
72 & & 2.99E$+$00 & --- & 2.99E$+$00 & --- & & 6.17E$-$02 & --- & 6.85E$-$02 & --- & & 1.82E$-$03 & --- & 2.63E$-$03 & --- \\
108 & & 3.00E$+$00 & --- & 3.00E$+$00 & --- & & 3.04E$-$02 & 1.74 & 3.77E$-$02 & 1.47 & & 3.74E$-$04 & 3.89 & 6.05E$-$04 & 3.63 \\
160 & & 3.00E$+$00 & --- & 3.00E$+$00 & --- & & 1.44E$-$02 & 1.90 & 1.83E$-$02 & 1.84 & & 8.54E$-$05 & 3.76 & 1.52E$-$04 & 3.52 \\
236 & & 3.00E$+$00 & --- & 3.00E$+$00 & --- & & 7.15E$-$03 & 1.80 & 9.31E$-$03 & 1.73 & & 1.83E$-$05 & 3.96 & 3.60E$-$05 & 3.71 \\
356 & & 3.00E$+$00 & --- & 3.00E$+$00 & --- & & 3.05E$-$03 & 2.07 & 4.24E$-$03 & 1.92 & & 3.41E$-$06 & 4.09 & 8.64E$-$06 & 3.47 \\
\midrule
\multicolumn{16}{@{}l}{ROD} \\
\midrule
72 & & 3.00E$+$00 & --- & 3.00E$+$00 & --- & & 6.41E$-$02 & --- & 7.09E$-$02 & --- & & 1.89E$-$03 & --- & 2.72E$-$03 & --- \\
108 & & 3.00E$+$00 & --- & 3.00E$+$00 & --- & & 3.12E$-$02 & 1.78 & 3.85E$-$02 & 1.51 & & 3.86E$-$04 & 3.92 & 6.18E$-$04 & 3.66 \\
160 & & 3.00E$+$00 & --- & 3.00E$+$00 & --- & & 1.46E$-$02 & 1.92 & 1.85E$-$02 & 1.86 & & 8.73E$-$05 & 3.78 & 1.54E$-$04 & 3.53 \\
236 & & 3.00E$+$00 & --- & 3.00E$+$00 & --- & & 7.23E$-$03 & 1.81 & 9.40E$-$03 & 1.75 & & 1.86E$-$05 & 3.98 & 3.64E$-$05 & 3.72 \\
356 & & 3.00E$+$00 & --- & 3.00E$+$00 & --- & & 3.07E$-$03 & 2.08 & 4.26E$-$03 & 1.92 & & 3.45E$-$06 & 4.10 & 8.70E$-$06 & 3.48 \\
\bottomrule
\end{tabular}

}
\end{table}


\subsubsection{Polynomial reconstructions assessment}
\label{subsec:verification_benchmark_t12}

To assess the different combinations of polynomial degrees for the vorticity and streamfunction polynomial reconstructions, the exact boundary vorticity is substituted with the prescribed approximation in Equation~\cref{eq:finite_volume_discretisation_boundary_vorticity} and the ROD method is employed for the boundary treatment in all simulations.
The errors and convergence orders for the approximate vorticity and streamfunction are reported in Tables~\ref{tab:verification_benchmark_t12_errorstable_omega} and~\ref{tab:verification_benchmark_t12_errorstable_psi}, respectively, while the results for the approximate velocity and boundary vorticity are reported in Tables~\ref{tab:verification_benchmark_t12_errorstable_u} and~\ref{tab:verification_benchmark_t12_errorstable_omegabound}, respectively.
The discussion follows:
\begin{itemize}
\item The $\mathbb{P}_{d}$--$\mathbb{P}_{d}$ method does not converge with $d=1$, either for the vorticity and streamfunction, whereas the obtained convergence orders are suboptimal with $d=3,5$.
The convergence deterioration is more pronounced for the vorticity, where it achieves only the third- and fifth-orders of convergence with $d=3,5$ in the $L^{1}$-norm, and the second- and fourth-orders of convergence in the $L^{\infty}$-norm, respectively.
Such a deterioration in convergence order is expected since the boundary vorticity is approximated with second- and fourth-orders of convergence with $d=3,5$, respectively, in both error norms, while with $d=1$ it does not converge at all.
Consequently, there is substantial accuracy deterioration, especially in the vicinity of the boundary, which is reflected in the vorticity and, ultimately, also in the streamfunction.
For the velocity, the third- and fifth-orders of convergence are achieved with $d=3,5$, respectively, in both error norms, while with $d=1$ it does not converge, following the expected behaviour considering the obtained convergence orders for the streamfunction.

\item Increasing the polynomial degree for the streamfunction polynomial reconstructions in the $\mathbb{P}_{d}$--$\mathbb{P}_{d+1}$ method improves accuracy and increases convergence order for the boundary vorticity, with first-, third-, and fifth-orders of convergence obtained with $d=1,3,5$, respectively, in both error norms.
Consequently, accuracy and convergence orders for the vorticity are substantially improved, achieving second-, fourth-, and sixth-orders of convergence in the $L^{1}$-norm and above first-, third-, and fifth-orders of convergence in the $L^{\infty}$-norm with $d=1,3,5$, respectively.
For the streamfunction, optimal convergence orders are achieved in both error norms, resulting in improved accuracy and convergence order for the velocity compared to the $\mathbb{P}_{d}$--$\mathbb{P}_{d}$ method, which achieves second-order of convergence with $d=1$ and almost fourth- and sixth-order of convergence with $d=3,5$, respectively, in both error norms.

\item Further increasing the polynomial degree for the streamfunction polynomial reconstructions in the $\mathbb{P}_{d}$--$\mathbb{P}_{d+2}$ method slightly improves boundary vorticity accuracy in comparison with the $\mathbb{P}_{d}$--$\mathbb{P}_{d+1}$ method, in general.
Thus, slightly more accurate vorticity and streamfunction cell mean-values are also computed, although such small improvements can often be limited in more practical domains due to the higher condition numbers of the resulting system of linear equations.
There is, however, a substantial accuracy improvement in the velocity accuracy.
\end{itemize}

\begin{table}[!htb]
\centering
\footnotesize
\caption{Vorticity errors and convergence orders obtained in the flow in a circular domain test case with approximate boundary vorticity.}
\label{tab:verification_benchmark_t12_errorstable_omega}
\resizeboxlarger{
\begin{tabular}{@{}r@{}r@{}rrrr@{}r@{}rrrr@{}r@{}rrrr@{}}
\toprule
& \phantom{aaa} & \multicolumn{4}{@{}l@{}}{$d=1$} & \phantom{aaa} & \multicolumn{4}{@{}l@{}}{$d=3$} & \phantom{aaa} & \multicolumn{4}{@{}l@{}}{$d=5$} \\
\cmidrule{3-6}\cmidrule{8-11}\cmidrule{13-16}
$N_{\textrm{C}}$ & & $E_{1}$ & $O_{1}$ & $E_{\infty}$ & $O_{\infty}$ & & $E_{1}$ & $O_{1}$ & $E_{\infty}$ & $O_{\infty}$ & & $E_{1}$ & $O_{1}$ & $E_{\infty}$ & $O_{\infty}$ \\
\midrule
\multicolumn{16}{@{}l}{$\mathbb{P}_{d}$--$\mathbb{P}_{d}$} \\
\midrule
1,096 & & 3.45E$-$01 & --- & 1.73E$+$01 & --- & & 1.87E$-$04 & --- & 2.96E$-$03 & --- & & 3.78E$-$06 & --- & 3.95E$-$05 & --- \\
2,478 & & 3.45E$+$01 & --- & 1.00E$+$04 & --- & & 5.21E$-$05 & 3.14 & 2.04E$-$03 & 0.91 & & 4.85E$-$07 & 5.04 & 8.33E$-$06 & 3.82 \\
5,366 & & 1.44E$-$01 & --- & 1.23E$+$01 & --- & & 1.48E$-$05 & 3.26 & 7.73E$-$04 & 2.51 & & 7.47E$-$08 & 4.84 & 1.86E$-$06 & 3.88 \\
11,624 & & 3.38E$-$01 & --- & 1.54E$+$02 & --- & & 4.84E$-$06 & 2.89 & 3.43E$-$04 & 2.11 & & 1.19E$-$08 & 4.74 & 5.81E$-$07 & 3.02 \\
26,890 & & 6.01E$-$02 & --- & 2.87E$+$01 & --- & & 1.36E$-$06 & 3.03 & 1.99E$-$04 & 1.30 & & 1.58E$-$09 & 4.83 & 9.73E$-$08 & 4.26 \\
\midrule
\multicolumn{16}{@{}l}{$\mathbb{P}_{d}$--$\mathbb{P}_{d+1}$} \\
\midrule
1,096 & & 4.46E$-$03 & --- & 2.73E$-$02 & --- & & 9.09E$-$05 & --- & 5.04E$-$04 & --- & & 3.75E$-$06 & --- & 2.57E$-$05 & --- \\
2,478 & & 2.17E$-$03 & 1.77 & 1.72E$-$02 & 1.14 & & 1.62E$-$05 & 4.23 & 1.38E$-$04 & 3.17 & & 3.75E$-$07 & 5.65 & 3.16E$-$06 & 5.13 \\
5,366 & & 9.77E$-$04 & 2.06 & 8.80E$-$03 & 1.73 & & 3.08E$-$06 & 4.29 & 3.17E$-$05 & 3.82 & & 3.49E$-$08 & 6.14 & 5.49E$-$07 & 4.53 \\
11,624 & & 4.62E$-$04 & 1.94 & 8.40E$-$03 & 0.12 & & 6.30E$-$07 & 4.11 & 1.04E$-$05 & 2.87 & & 3.66E$-$09 & 5.84 & 5.98E$-$08 & 5.73 \\
26,890 & & 1.96E$-$04 & 2.05 & 5.99E$-$03 & 0.81 & & 1.17E$-$07 & 4.01 & 2.27E$-$06 & 3.64 & & 3.08E$-$10 & 5.90 & 7.48E$-$09 & 4.96 \\
\midrule
\multicolumn{16}{@{}l}{$\mathbb{P}_{d}$--$\mathbb{P}_{d+2}$} \\
\midrule
1,096 & & 3.92E$-$03 & --- & 1.35E$-$02 & --- & & 8.75E$-$05 & --- & 3.75E$-$04 & --- & & 3.88E$-$06 & --- & 1.23E$-$05 & --- \\
2,478 & & 1.97E$-$03 & 1.69 & 9.24E$-$03 & 0.93 & & 1.56E$-$05 & 4.22 & 9.83E$-$05 & 3.28 & & 3.80E$-$07 & 5.69 & 1.72E$-$06 & 4.83 \\
5,366 & & 8.94E$-$04 & 2.04 & 4.60E$-$03 & 1.80 & & 3.05E$-$06 & 4.23 & 2.10E$-$05 & 3.99 & & 3.51E$-$08 & 6.17 & 2.21E$-$07 & 5.31 \\
11,624 & & 4.18E$-$04 & 1.97 & 2.75E$-$03 & 1.33 & & 6.05E$-$07 & 4.19 & 4.76E$-$06 & 3.84 & & 3.60E$-$09 & 5.89 & 2.86E$-$08 & 5.29 \\
26,890 & & 1.76E$-$04 & 2.06 & 1.26E$-$03 & 1.87 & & 1.14E$-$07 & 3.97 & 1.15E$-$06 & 3.38 & & 3.61E$-$10 & 5.48 & 1.65E$-$08 & 1.31 \\
\bottomrule
\end{tabular}

}
\end{table}

\begin{table}[!htb]
\centering
\footnotesize
\caption{Streamfunction errors and convergence orders obtained in the flow in a circular domain test case with approximate boundary vorticity.}
\label{tab:verification_benchmark_t12_errorstable_psi}
\resizeboxlarger{
\begin{tabular}{@{}r@{}r@{}rrrr@{}r@{}rrrr@{}r@{}rrrr@{}}
\toprule
& \phantom{aaa} & \multicolumn{4}{@{}l@{}}{$d=1$} & \phantom{aaa} & \multicolumn{4}{@{}l@{}}{$d=3$} & \phantom{aaa} & \multicolumn{4}{@{}l@{}}{$d=5$} \\
\cmidrule{3-6}\cmidrule{8-11}\cmidrule{13-16}
$N_{\textrm{C}}$ & & $E_{1}$ & $O_{1}$ & $E_{\infty}$ & $O_{\infty}$ & & $E_{1}$ & $O_{1}$ & $E_{\infty}$ & $O_{\infty}$ & & $E_{1}$ & $O_{1}$ & $E_{\infty}$ & $O_{\infty}$ \\
\midrule
\multicolumn{16}{@{}l}{$\mathbb{P}_{d}$--$\mathbb{P}_{d}$} \\
\midrule
1,096 & & 6.10E$-$03 & --- & 9.07E$-$02 & --- & & 1.40E$-$05 & --- & 2.69E$-$05 & --- & & 3.46E$-$07 & --- & 7.57E$-$07 & --- \\
2,478 & & 3.03E$-$01 & --- & 7.31E$+$00 & --- & & 3.80E$-$06 & 3.21 & 6.82E$-$06 & 3.36 & & 6.39E$-$08 & 4.14 & 1.07E$-$07 & 4.80 \\
5,366 & & 9.56E$-$04 & --- & 1.90E$-$02 & --- & & 1.05E$-$06 & 3.32 & 1.65E$-$06 & 3.67 & & 8.61E$-$09 & 5.19 & 1.43E$-$08 & 5.20 \\
11,624 & & 3.51E$-$03 & --- & 7.91E$-$02 & --- & & 2.09E$-$07 & 4.19 & 3.76E$-$07 & 3.83 & & 1.19E$-$09 & 5.12 & 1.71E$-$09 & 5.49 \\
26,890 & & 1.55E$-$04 & --- & 4.22E$-$03 & --- & & 4.40E$-$08 & 3.72 & 7.55E$-$08 & 3.83 & & 1.13E$-$10 & 5.61 & 1.71E$-$10 & 5.49 \\
\midrule
\multicolumn{16}{@{}l}{$\mathbb{P}_{d}$--$\mathbb{P}_{d+1}$} \\
\midrule
1,096 & & 2.70E$-$04 & --- & 6.80E$-$04 & --- & & 1.02E$-$05 & --- & 2.04E$-$05 & --- & & 3.33E$-$07 & --- & 7.39E$-$07 & --- \\
2,478 & & 1.41E$-$04 & 1.58 & 3.78E$-$04 & 1.44 & & 2.14E$-$06 & 3.83 & 3.80E$-$06 & 4.12 & & 3.45E$-$08 & 5.55 & 6.80E$-$08 & 5.85 \\
5,366 & & 6.07E$-$05 & 2.19 & 1.97E$-$04 & 1.69 & & 4.52E$-$07 & 4.02 & 7.14E$-$07 & 4.33 & & 3.93E$-$09 & 5.63 & 6.71E$-$09 & 5.99 \\
11,624 & & 2.88E$-$05 & 1.93 & 9.74E$-$05 & 1.82 & & 9.62E$-$08 & 4.00 & 1.42E$-$07 & 4.18 & & 3.77E$-$10 & 6.06 & 5.78E$-$10 & 6.35 \\
26,890 & & 1.15E$-$05 & 2.20 & 4.05E$-$05 & 2.10 & & 1.74E$-$08 & 4.08 & 2.45E$-$08 & 4.19 & & 2.56E$-$11 & 6.41 & 4.11E$-$11 & 6.30 \\
\midrule
\multicolumn{16}{@{}l}{$\mathbb{P}_{d}$--$\mathbb{P}_{d+2}$} \\
\midrule
1,096 & & 5.37E$-$04 & --- & 8.86E$-$04 & --- & & 4.76E$-$06 & --- & 7.72E$-$06 & --- & & 2.09E$-$07 & --- & 3.77E$-$07 & --- \\
2,478 & & 2.78E$-$04 & 1.62 & 4.52E$-$04 & 1.65 & & 5.70E$-$07 & 5.20 & 9.13E$-$07 & 5.23 & & 1.51E$-$08 & 6.44 & 2.63E$-$08 & 6.52 \\
5,366 & & 1.28E$-$04 & 2.01 & 2.10E$-$04 & 1.99 & & 4.69E$-$08 & 6.47 & 1.01E$-$07 & 5.69 & & 7.70E$-$10 & 7.71 & 1.60E$-$09 & 7.25 \\
11,624 & & 6.03E$-$05 & 1.94 & 9.89E$-$05 & 1.94 & & 1.14E$-$08 & 3.65 & 3.93E$-$08 & 2.45 & & 4.78E$-$11 & 7.19 & 1.27E$-$10 & 6.55 \\
26,890 & & 2.56E$-$05 & 2.05 & 4.30E$-$05 & 1.98 & & 2.98E$-$09 & 3.21 & 1.02E$-$08 & 3.22 & & 4.14E$-$12 & 5.83 & 1.14E$-$11 & 5.76 \\
\bottomrule
\end{tabular}

}
\end{table}

\begin{table}[!htb]
\centering
\footnotesize
\caption{Boundary vorticity errors and convergence orders obtained in the flow in a circular domain test case approximate boundary vorticity.}
\label{tab:verification_benchmark_t12_errorstable_omegabound}
\resizeboxlarger{
\begin{tabular}{@{}r@{}r@{}rrrr@{}r@{}rrrr@{}r@{}rrrr@{}}
\toprule
& \phantom{aaa} & \multicolumn{4}{@{}l@{}}{$d=1$} & \phantom{aaa} & \multicolumn{4}{@{}l@{}}{$d=3$} & \phantom{aaa} & \multicolumn{4}{@{}l@{}}{$d=5$} \\
\cmidrule{3-6}\cmidrule{8-11}\cmidrule{13-16}
$N_{\textrm{B}}$ & & $E_{1}$ & $O_{1}$ & $E_{\infty}$ & $O_{\infty}$ & & $E_{1}$ & $O_{1}$ & $E_{\infty}$ & $O_{\infty}$ & & $E_{1}$ & $O_{1}$ & $E_{\infty}$ & $O_{\infty}$ \\
\midrule
\multicolumn{16}{@{}l}{$\mathbb{P}_{d}$--$\mathbb{P}_{d}$} \\
\midrule
72 & & 2.35E$+$00 & --- & 2.38E$+$01 & --- & & 2.48E$-$03 & --- & 4.86E$-$03 & --- & & 2.07E$-$05 & --- & 6.08E$-$05 & --- \\
108 & & 5.20E$+$02 & --- & 1.70E$+$04 & --- & & 9.46E$-$04 & 2.38 & 2.99E$-$03 & 1.20 & & 4.26E$-$06 & 3.90 & 1.25E$-$05 & 3.90 \\
160 & & 2.18E$+$00 & --- & 1.72E$+$01 & --- & & 3.19E$-$04 & 2.77 & 1.07E$-$03 & 2.62 & & 1.00E$-$06 & 3.68 & 2.79E$-$06 & 3.82 \\
236 & & 5.86E$+$00 & --- & 2.01E$+$02 & --- & & 1.94E$-$04 & 1.27 & 5.63E$-$04 & 1.65 & & 2.33E$-$07 & 3.75 & 8.65E$-$07 & 3.01 \\
356 & & 2.43E$+$00 & --- & 4.08E$+$01 & --- & & 7.15E$-$05 & 2.43 & 2.82E$-$04 & 1.68 & & 4.39E$-$08 & 4.06 & 1.54E$-$07 & 4.19 \\
\midrule
\multicolumn{16}{@{}l}{$\mathbb{P}_{d}$--$\mathbb{P}_{d+1}$} \\
\midrule
72 & & 1.73E$-$02 & --- & 4.83E$-$02 & --- & & 1.15E$-$04 & --- & 3.85E$-$04 & --- & & 1.54E$-$05 & --- & 3.71E$-$05 & --- \\
108 & & 9.37E$-$03 & 1.52 & 2.71E$-$02 & 1.43 & & 3.82E$-$05 & 2.71 & 1.06E$-$04 & 3.19 & & 1.98E$-$06 & 5.06 & 4.18E$-$06 & 5.38 \\
160 & & 5.19E$-$03 & 1.50 & 1.59E$-$02 & 1.36 & & 8.51E$-$06 & 3.82 & 2.70E$-$05 & 3.47 & & 2.35E$-$07 & 5.42 & 8.00E$-$07 & 4.20 \\
236 & & 3.88E$-$03 & 0.75 & 1.31E$-$02 & 0.49 & & 4.37E$-$06 & 1.71 & 1.53E$-$05 & 1.46 & & 3.19E$-$08 & 5.14 & 8.46E$-$08 & 5.78 \\
356 & & 2.46E$-$03 & 1.11 & 8.44E$-$03 & 1.07 & & 1.07E$-$06 & 3.41 & 3.01E$-$06 & 3.95 & & 3.24E$-$09 & 5.56 & 1.15E$-$08 & 4.87 \\
\midrule
\multicolumn{16}{@{}l}{$\mathbb{P}_{d}$--$\mathbb{P}_{d+2}$} \\
\midrule
72 & & 1.61E$-$02 & --- & 2.81E$-$02 & --- & & 3.24E$-$05 & --- & 1.06E$-$04 & --- & & 1.51E$-$05 & --- & 1.88E$-$05 & --- \\
108 & & 7.58E$-$03 & 1.86 & 1.92E$-$02 & 0.94 & & 1.67E$-$05 & 1.64 & 4.52E$-$05 & 2.11 & & 1.89E$-$06 & 5.13 & 2.54E$-$06 & 4.94 \\
160 & & 3.87E$-$03 & 1.71 & 1.08E$-$02 & 1.46 & & 3.21E$-$06 & 4.19 & 1.16E$-$05 & 3.46 & & 2.19E$-$07 & 5.48 & 2.98E$-$07 & 5.45 \\
236 & & 2.12E$-$03 & 1.55 & 6.25E$-$03 & 1.42 & & 9.51E$-$07 & 3.13 & 4.19E$-$06 & 2.62 & & 2.49E$-$08 & 5.59 & 4.22E$-$08 & 5.03 \\
356 & & 1.09E$-$03 & 1.62 & 3.25E$-$03 & 1.59 & & 1.94E$-$07 & 3.87 & 5.96E$-$07 & 4.74 & & 6.00E$-$09 & 3.46 & 2.70E$-$08 & 1.08 \\
\bottomrule
\end{tabular}

}
\end{table}

\begin{table}[!htb]
\centering
\footnotesize
\caption{Velocity errors and convergence orders obtained in the flow in a circular domain test case with approximate boundary vorticity.}
\label{tab:verification_benchmark_t12_errorstable_u}
\resizeboxlarger{
\begin{tabular}{@{}r@{}r@{}rrrr@{}r@{}rrrr@{}r@{}rrrr@{}}
\toprule
& \phantom{aaa} & \multicolumn{4}{@{}l@{}}{$d=1$} & \phantom{aaa} & \multicolumn{4}{@{}l@{}}{$d=3$} & \phantom{aaa} & \multicolumn{4}{@{}l@{}}{$d=5$} \\
\cmidrule{3-6}\cmidrule{8-11}\cmidrule{13-16}
$N_{\textrm{E}}$ & & $E_{1}$ & $O_{1}$ & $E_{\infty}$ & $O_{\infty}$ & & $E_{1}$ & $O_{1}$ & $E_{\infty}$ & $O_{\infty}$ & & $E_{1}$ & $O_{1}$ & $E_{\infty}$ & $O_{\infty}$ \\
\midrule
\multicolumn{16}{@{}l}{$\mathbb{P}_{d}$--$\mathbb{P}_{d}$} \\
\midrule
1,680 & & 5.46E$-$02 & --- & 8.85E$-$01 & --- & & 1.41E$-$04 & --- & 1.99E$-$03 & --- & & 6.08E$-$06 & --- & 1.08E$-$04 & --- \\
3,771 & & 2.66E$+$00 & --- & 1.90E$+$02 & --- & & 3.68E$-$05 & 3.32 & 6.48E$-$04 & 2.78 & & 6.88E$-$07 & 5.39 & 1.41E$-$05 & 5.04 \\
8,129 & & 1.29E$-$02 & --- & 3.19E$-$01 & --- & & 1.09E$-$05 & 3.16 & 3.35E$-$04 & 1.72 & & 9.33E$-$08 & 5.20 & 2.30E$-$06 & 4.72 \\
17,554 & & 2.71E$-$02 & --- & 1.69E$+$00 & --- & & 2.97E$-$06 & 3.39 & 8.06E$-$05 & 3.70 & & 1.29E$-$08 & 5.14 & 3.84E$-$07 & 4.64 \\
40,513 & & 3.08E$-$03 & --- & 1.58E$-$01 & --- & & 7.80E$-$07 & 3.20 & 2.28E$-$05 & 3.02 & & 1.54E$-$09 & 5.09 & 5.79E$-$08 & 4.53 \\
\midrule
\multicolumn{16}{@{}l}{$\mathbb{P}_{d}$--$\mathbb{P}_{d+1}$} \\
\midrule
1,680 & & 2.83E$-$03 & --- & 1.95E$-$02 & --- & & 2.75E$-$05 & --- & 2.97E$-$04 & --- & & 1.24E$-$06 & --- & 2.04E$-$05 & --- \\
3,771 & & 1.25E$-$03 & 2.03 & 9.41E$-$03 & 1.80 & & 6.32E$-$06 & 3.64 & 6.86E$-$05 & 3.62 & & 1.27E$-$07 & 5.63 & 2.26E$-$06 & 5.44 \\
8,129 & & 5.59E$-$04 & 2.09 & 5.03E$-$03 & 1.63 & & 1.62E$-$06 & 3.55 & 1.98E$-$05 & 3.23 & & 1.47E$-$08 & 5.62 & 2.21E$-$07 & 6.05 \\
17,554 & & 2.53E$-$04 & 2.06 & 2.27E$-$03 & 2.07 & & 3.84E$-$07 & 3.74 & 4.34E$-$06 & 3.94 & & 1.71E$-$09 & 5.58 & 2.83E$-$08 & 5.34 \\
40,513 & & 1.06E$-$04 & 2.08 & 1.11E$-$03 & 1.70 & & 8.16E$-$08 & 3.70 & 7.84E$-$07 & 4.09 & & 1.83E$-$10 & 5.35 & 3.32E$-$09 & 5.12 \\
\midrule
\multicolumn{16}{@{}l}{$\mathbb{P}_{d}$--$\mathbb{P}_{d+2}$} \\
\midrule
1,680 & & 7.30E$-$04 & --- & 3.49E$-$03 & --- & & 1.33E$-$05 & --- & 1.26E$-$04 & --- & & 5.53E$-$07 & --- & 3.93E$-$06 & --- \\
3,771 & & 3.32E$-$04 & 1.95 & 1.31E$-$03 & 2.43 & & 1.44E$-$06 & 5.49 & 1.69E$-$05 & 4.96 & & 4.18E$-$08 & 6.39 & 4.91E$-$07 & 5.14 \\
8,129 & & 1.55E$-$04 & 1.99 & 8.41E$-$04 & 1.15 & & 2.52E$-$07 & 4.54 & 2.80E$-$06 & 4.68 & & 2.82E$-$09 & 7.02 & 3.14E$-$08 & 7.16 \\
17,554 & & 7.14E$-$05 & 2.01 & 4.32E$-$04 & 1.73 & & 5.90E$-$08 & 3.77 & 4.25E$-$07 & 4.90 & & 3.06E$-$10 & 5.77 & 4.12E$-$09 & 5.27 \\
40,513 & & 3.08E$-$05 & 2.01 & 1.76E$-$04 & 2.15 & & 1.28E$-$08 & 3.65 & 7.21E$-$08 & 4.24 & & 2.39E$-$11 & 6.10 & 1.86E$-$09 & 1.90 \\
\bottomrule
\end{tabular}

}
\end{table}


\subsection{Annular domain}
\label{subsec:verification_benchmark_t2}

This test case addresses an internal creeping flow (Stokes equations) confined between two infinitely long cylinders rotating at specific constant angular velocities.
The interior and exterior cylinders have radii $r_{\textrm{I}}>0$ and $r_{\textrm{E}}>0$, respectively, and rotate with linear velocity magnitudes $u_{\textrm{I}}$ and $u_{\textrm{E}}$, respectively.
The exterior cylinder is assumed to be centred at the origin, while the two cylinders are not necessarily coaxial, resulting in a non-null eccentricity (the distance between the centres of the two cylinders) defined as $e=r_{\textrm{E}}-r_{\textrm{I}}$.
Since the cylinder lengths are assumed to be infinite, the flow occurs only along the azimuthal axis, and, under these conditions, the flow has a known analytic solution for the vorticity and streamfuntion~\cite{1950_wannier}, given as
\begin{align}
&\omega=2B\dfrac{\left(-x+y+s\right)\left(x+y+s\right)}{c^{2}}+2C\dfrac{\left(x+y-s\right)\left(x-y+s\right)}{d^{2}}-4E-8Fs\dfrac{\left(x^{2}-y^{2}+s^{2}\right)}{cd},\text{ and}\\
&\psi=A\log\left(\dfrac{a}{b}\right)+By\dfrac{s+y}{a}+Cy\dfrac{s-y}{b}+Dy+E\left(x^{2}+y^{2}+s^{2}\right)+Fy\log\left(\dfrac{a}{b}\right),
\end{align}
where the parameters $a$, $b$, $c$, and $d$ are expressed as $a=x^{2}+\left(s+y\right)^{2}$, $b=x^{2}+\left(s-y\right)^{2}$, $c=x^{2}+y^{2}+s^{2}+2sy$, and $d=x^{2}+y^{2}+s^{2}-2sy$, and the coefficients $A$, $B$, $C$, $D$, $E$, $F$, $G$, and $H$ are expressed as
\begin{align}
&A=-G\dfrac{d_{\textrm{I}}d_{\textrm{E}}-s^{2}}{2},
\qquad
B=G\left(d_{\textrm{I}}+s\right)\left(d_{\textrm{E}}+s\right),
\qquad
C=G\left(d_{\textrm{I}}-s\right)\left(d_{\textrm{E}}-s\right),\\
&D=H\Biggl(d_{\textrm{I}}\log\left(\dfrac{d_{\textrm{E}}+s}{d_{\textrm{E}}-s}\right)-d_{\textrm{E}}\log\left(\dfrac{d_{\textrm{I}}+s}{d_{\textrm{I}}-s}\right)-2Hs\dfrac{r_{\textrm{E}}^{2}-r_{\textrm{I}}^{2}}{r_{\textrm{E}}^{2}+r_{\textrm{I}}^{2}}\Biggr)\left(r_{\textrm{I}}u_{\textrm{I}}+r_{\textrm{E}}u_{\textrm{E}}\right)-\dfrac{r_{\textrm{I}}^{2}r_{\textrm{E}}^{2}\left(r_{\textrm{I}}^{-1}u_{\textrm{I}}-r_{\textrm{E}}^{-1}u_{\textrm{E}}\right)}{\left(r_{\textrm{I}}^{2}+r_{\textrm{E}}^{2}\right)e},\\
&E=\dfrac{H}{2}\log\Biggl(\dfrac{\left(d_{\textrm{I}}+s\right)\left(d_{\textrm{E}}-s\right)}{\left(d_{\textrm{I}}-s\right)\left(d_{\textrm{E}}+s\right)}\Biggr)\left(r_{\textrm{I}}u_{\textrm{I}}+r_{\textrm{E}}u_{\textrm{E}}\right),
\qquad
F=He\left(r_{\textrm{I}}u_{\textrm{I}}+r_{\textrm{E}}u_{\textrm{E}}\right),\\
&G=2H\dfrac{d_{\textrm{E}}^{2}-d_{\textrm{I}}^{2}}{r_{\textrm{I}}^{2}+r_{\textrm{E}}^{2}}\left(r_{\textrm{I}}u_{\textrm{I}}+r_{\textrm{E}}u_{\textrm{E}}\right),
\qquad
\text{and}
\qquad
H=\Biggl(\left(r_{\textrm{I}}^{2}+r_{\textrm{E}}^{2}\right)\log\Biggl(\dfrac{\left(d_{\textrm{I}}+s\right)\left(d_{\textrm{E}}-s\right)}{\left(d_{\textrm{I}}-s\right)\left(d_{\textrm{E}}+s\right)}\Biggr)-4se\Biggr)^{-1},
\end{align}
and the distances $s$, $d_{\textrm{I}}$, and $d_{\textrm{E}}$ are given as
\begin{align}
&s=\Biggl(\dfrac{\left(r_{\textrm{E}}-r_{\textrm{I}}-e\right)\left(r_{\textrm{E}}-r_{\textrm{I}}+e\right)\left(r_{\textrm{E}}+r_{\textrm{I}}-e\right)\left(r_{\textrm{E}}+r_{\textrm{I}}+e\right)}{4e^{2}}\Biggr)^{1/2},\\
&d_{\textrm{I}}=\dfrac{r_{\textrm{E}}^{2}-r_{\textrm{I}}^{2}}{2e}-\dfrac{e}{2},
\qquad
\text{and}
\qquad
d_{\textrm{E}}=\dfrac{r_{\textrm{E}}^{2}-r_{\textrm{I}}^{2}}{2e}+\dfrac{e}{2}.
\end{align}
The velocity components are obtained from the streamfunction and read as
\begin{align}
\begin{split}
&u_{x}=4As\dfrac{x^{2}-y^{2}+s^{2}}{cd}+B\dfrac{sx^{2}+sy^{2}+2x^{2}y+2s^{2}y+s^{3}}{c^{2}}+C\dfrac{sx^{2}+sy^{2}-2x^{2}y-2s^{2}y+s^{3}}{d^{2}}\\
&\qquad+D+2Ey+F\dfrac{2y\left(\left(s-y\right)a+\left(s+y\right)b\right)+ab\log\left(\dfrac{a}{b}\right)}{ab},\text{ and}
\end{split}\\
&u_{y}=A\dfrac{8sxy}{cd}+B\dfrac{2xy\left(s+y\right)}{c^{2}}+C\dfrac{2xy\left(s-y\right)}{d^{2}}-2Ex+F\dfrac{8sxy^{2}}{cd}.
\end{align}

\begin{figure}[!htb]
\centering
\begin{tabular}{@{}c@{}c@{}}
\includegraphics[width=0.49\textwidth,trim=0cm 0cm 0cm 0cm,clip=true]{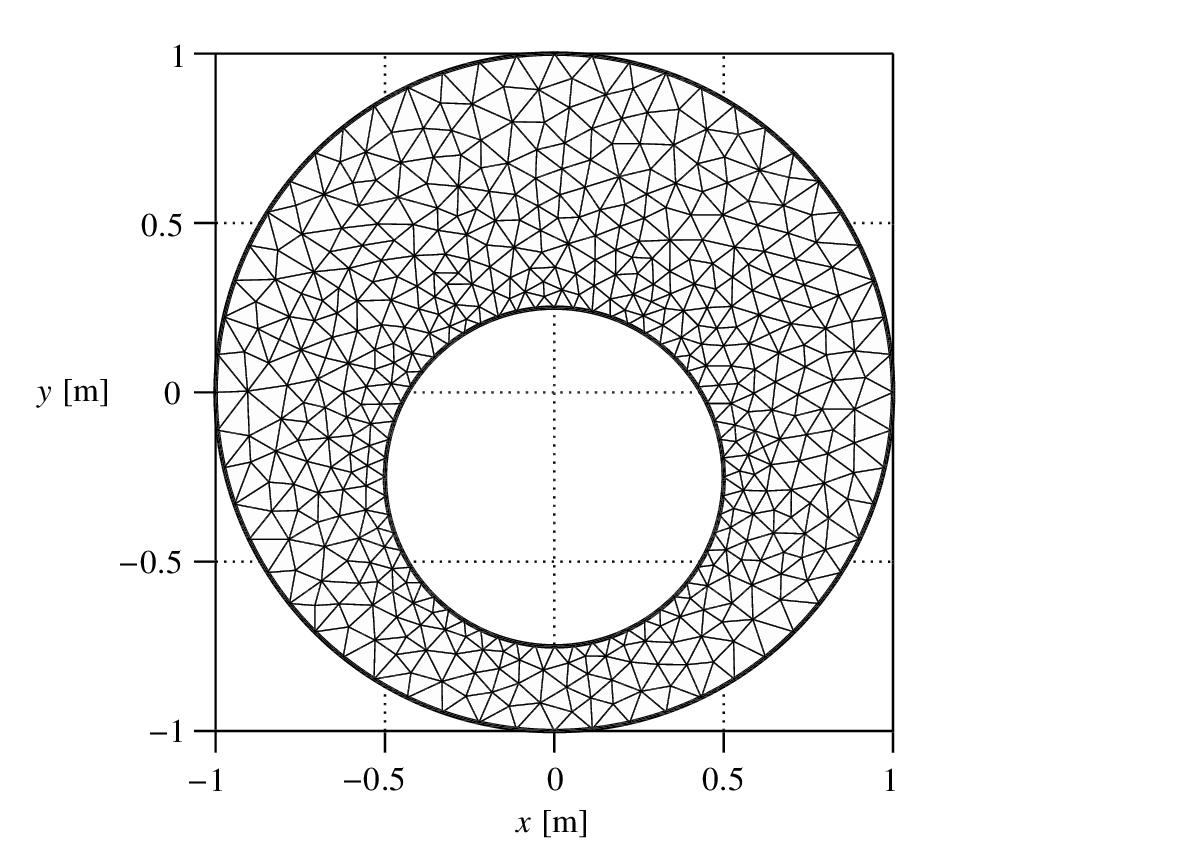} &
\includegraphics[width=0.49\textwidth,trim=0cm 0cm 0cm 0cm,clip=true]{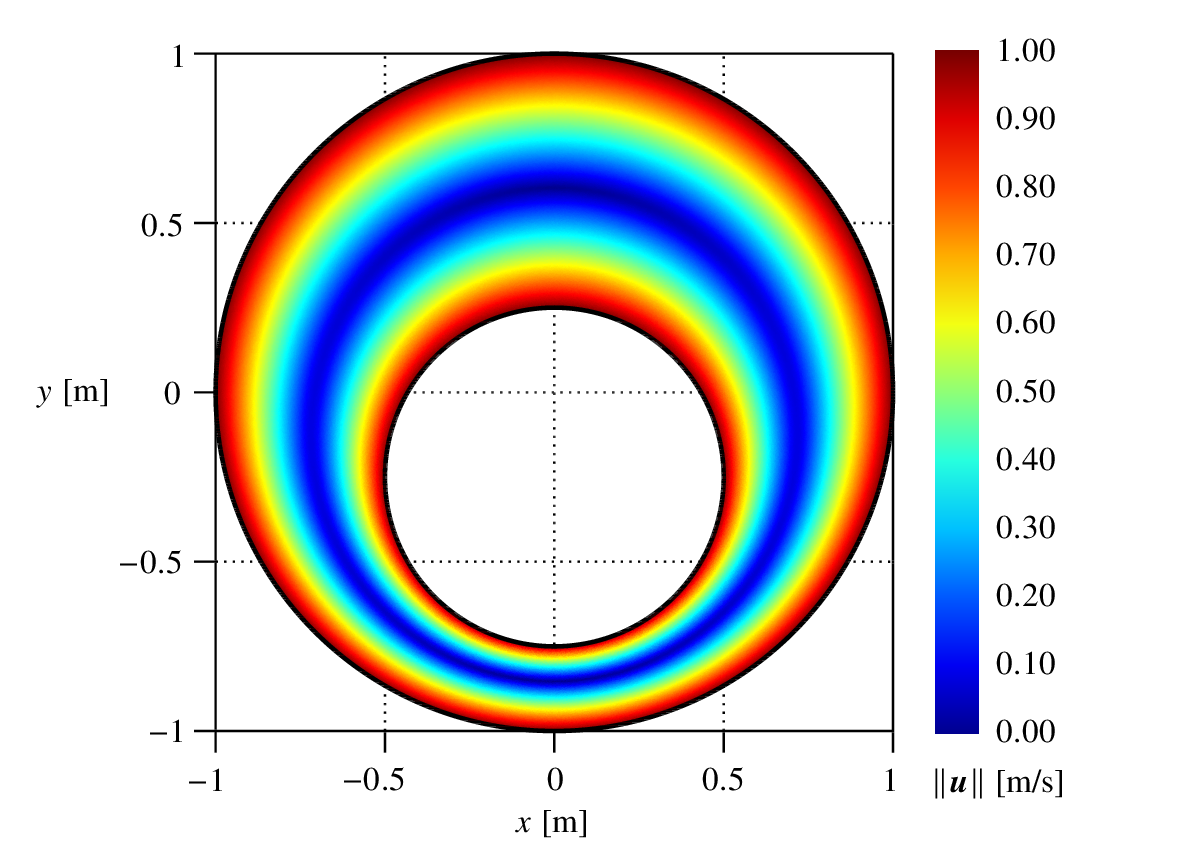}\\[0.2cm]
\small (a) Coarse mesh. & \small (b) Velocity magnitude.\\
\includegraphics[width=0.49\textwidth,trim=0cm 0cm 0cm 0cm,clip=true]{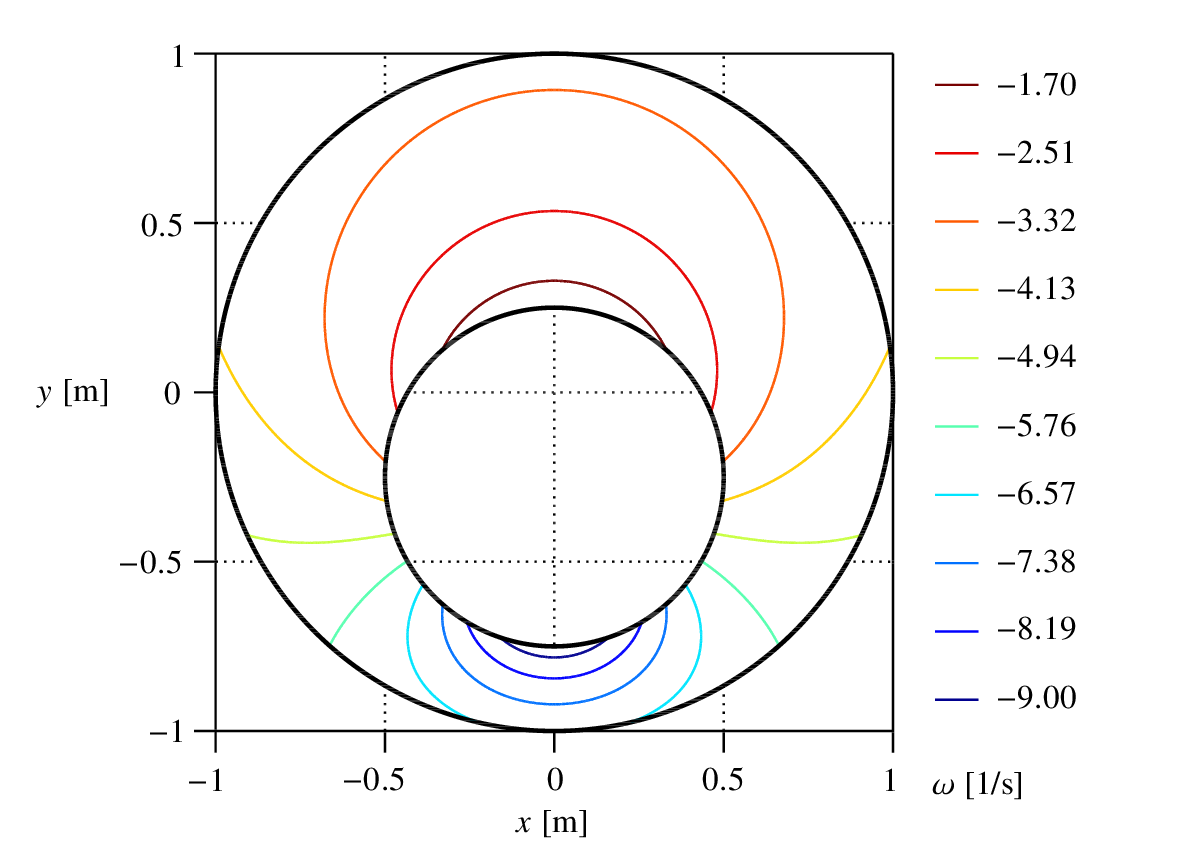} &
\includegraphics[width=0.49\textwidth,trim=0cm 0cm 0cm 0cm,clip=true]{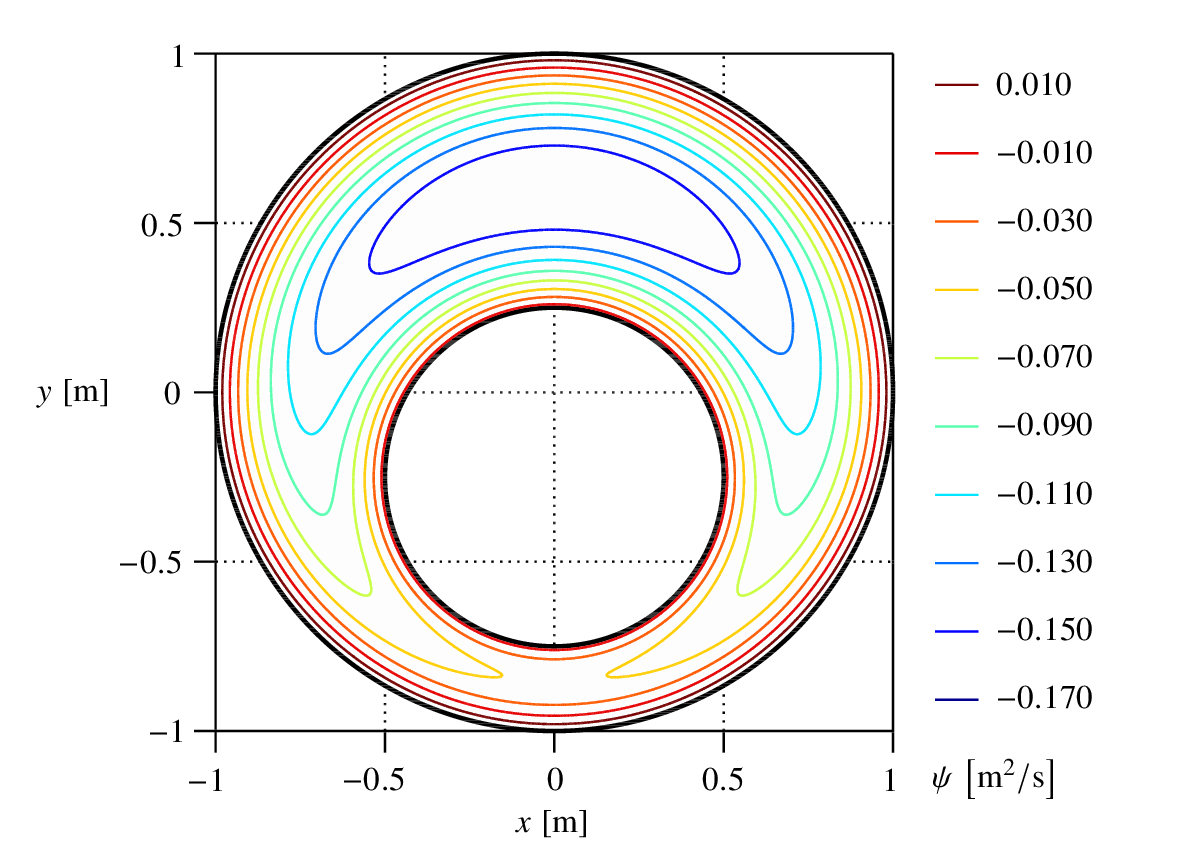}\\[0.2cm]
\small (c) Vorticity contours. & \small (d) Streamlines.
\end{tabular}
\caption{Coarse mesh and analytical solution for the flow in an annular domain test case (for proper interpretation of the colour scale, the reader is referred to the electronic version of this article).}
\label{fig:verification_benchmark_t2}
\end{figure}

For the domain, the interior physical boundary has a radius of $r_{\textrm{I}}=1/2$~\si[per-mode=symbol]{\m} (normal curvature of $\kappa_{\textrm{I}}=2$), centre at $\left(0,-1/4\right)$~\si[per-mode=symbol]{\m}, and rotates with $u_{\textrm{I}}=-1$~\si[per-mode=symbol]{\m\per\second} (counter-clockwise).
The exterior physical boundary has a radius of $r_{\textrm{E}}=1$~\si[per-mode=symbol]{\m} (normal curvature of $\kappa_{\textrm{E}}=1$), centre at the origin, and rotates with $u_{\textrm{E}}=1$~\si[per-mode=symbol]{\m\per\second} (clockwise).
For the fluid, a dynamic viscosity of $\mu=1$~\si[per-mode=symbol]{\kg\per\m\per\second} and a density of $\rho=1$~\si[per-mode=symbol]{\kg\per\m^{3}} were used.
Successively finer uniform Delaunay triangular meshes are used to discretise the physical domain, and the simulations are carried out for the $\mathbb{P}_{d}$--$\mathbb{P}_{d}$, $\mathbb{P}_{d}$--$\mathbb{P}_{d+1}$, and $\mathbb{P}_{d}$--$\mathbb{P}_{d+2}$ methods with $d=1,3,5$.
Figure~\ref{fig:verification_benchmark_t2} illustrates the physical domain with a coarse mesh and the analytic solutions for the velocity, vorticity, and streamfunction fields.


The errors and convergence orders for the approximate vorticity, streamfunction, boundary vorticity, and velocity are reported in Tables~\ref{tab:verification_benchmark_t21_errorstable_omega},~\ref{tab:verification_benchmark_t21_errorstable_psi},~\ref{tab:verification_benchmark_t21_errorstable_omegabound}, and~\ref{tab:verification_benchmark_t21_errorstable_u}, respectively.
The results exhibit the same behaviour as the previous test case, and, in general, the $\mathbb{P}_{d}$--$\mathbb{P}_{d}$ method does not achieve optimal convergence orders in both error norms for the vorticity and streamfunction.
Again, in accordance with the previous test case, the $\mathbb{P}_{d}$--$\mathbb{P}_{d+1}$ method substantially improves the boundary vorticity accuracy and convergence order, which turns out to be effective in recovering the optimal convergence orders for the vorticity and streamfunction, except for the $L^{\infty}$-norm for the streamfunction.
Moreover, the same convergence orders as for the streamfunction are achieved for the velocity, indicating that the accuracy deterioration in the streamfunction polynomial reconstruction derivatives is not significant.
In turn, the $\mathbb{P}_{d}$--$\mathbb{P}_{d+2}$ method further improves the boundary vorticity accuracy, achieving fourth- and sixth-orders of convergence with $d=3,5$, respectively, in both error norms.
On the contrary, the strategy is not effective with $d=1$, which maintains a first-order of convergence in both error norms.
Consequently, there is also no improvement in the convergence order for the vorticity, streamfunction, and velocity with $d=1$ compared to the $\mathbb{P}_{d}$--$\mathbb{P}_{d+1}$ method.
However, it is effective in recovering the optimal order for the vorticity in the $L^{\infty}$-norm with $d=3,5$, while super-optimal convergence orders are obtained in the $L^{1}$-norm and for the streamfunction and velocity in both error norms.

\begin{table}[!htb]
\centering
\footnotesize
\caption{Vorticity errors and convergence orders obtained in the flow in an annular domain test case.}
\label{tab:verification_benchmark_t21_errorstable_omega}
\resizeboxlarger{
\begin{tabular}{@{}r@{}r@{}rrrr@{}r@{}rrrr@{}r@{}rrrr@{}}
\toprule
& \phantom{aaa} & \multicolumn{4}{@{}l@{}}{$d=1$} & \phantom{aaa} & \multicolumn{4}{@{}l@{}}{$d=3$} & \phantom{aaa} & \multicolumn{4}{@{}l@{}}{$d=5$} \\
\cmidrule{3-6}\cmidrule{8-11}\cmidrule{13-16}
$N_{\textrm{C}}$ & & $E_{1}$ & $O_{1}$ & $E_{\infty}$ & $O_{\infty}$ & & $E_{1}$ & $O_{1}$ & $E_{\infty}$ & $O_{\infty}$ & & $E_{1}$ & $O_{1}$ & $E_{\infty}$ & $O_{\infty}$ \\
\midrule
\multicolumn{16}{@{}l}{$\mathbb{P}_{d}$--$\mathbb{P}_{d}$} \\
\midrule
1,178 & & 4.65E$-$01 & --- & 3.65E$+$01 & --- & & 5.15E$-$04 & --- & 9.62E$-$03 & --- & & 3.38E$-$04 & --- & 9.78E$-$03 & --- \\
2,374 & & 7.04E$-$01 & --- & 4.88E$+$01 & --- & & 1.74E$-$04 & 3.10 & 7.24E$-$03 & 0.81 & & 5.75E$-$05 & 5.05 & 1.45E$-$03 & 5.44 \\
5,236 & & 1.43E$+$00 & --- & 4.56E$+$02 & --- & & 5.22E$-$05 & 3.04 & 3.43E$-$03 & 1.89 & & 6.40E$-$06 & 5.55 & 3.12E$-$04 & 3.89 \\
10,378 & & 1.15E$+$00 & --- & 3.47E$+$02 & --- & & 1.53E$-$05 & 3.59 & 1.43E$-$03 & 2.56 & & 8.74E$-$07 & 5.82 & 9.65E$-$05 & 3.43 \\
22,072 & & 2.71E$-$01 & --- & 1.81E$+$02 & --- & & 5.39E$-$06 & 2.76 & 1.01E$-$03 & 0.92 & & 1.26E$-$07 & 5.14 & 1.81E$-$05 & 4.44 \\
\midrule
\multicolumn{16}{@{}l}{$\mathbb{P}_{d}$--$\mathbb{P}_{d+1}$} \\
\midrule
1,178 & & 2.05E$-$03 & --- & 3.62E$-$02 & --- & & 3.19E$-$04 & --- & 5.32E$-$03 & --- & & 3.75E$-$04 & --- & 1.40E$-$02 & --- \\
2,374 & & 1.19E$-$03 & 1.55 & 2.55E$-$02 & 1.00 & & 7.03E$-$05 & 4.31 & 1.71E$-$03 & 3.24 & & 3.07E$-$05 & 7.14 & 1.11E$-$03 & 7.23 \\
5,236 & & 5.60E$-$04 & 1.90 & 1.76E$-$02 & 0.93 & & 1.28E$-$05 & 4.31 & 8.25E$-$04 & 1.84 & & 2.30E$-$06 & 6.56 & 1.03E$-$04 & 6.02 \\
10,378 & & 2.80E$-$04 & 2.03 & 1.19E$-$02 & 1.16 & & 2.67E$-$06 & 4.58 & 2.88E$-$04 & 3.08 & & 2.12E$-$07 & 6.96 & 2.46E$-$05 & 4.17 \\
22,072 & & 1.41E$-$04 & 1.81 & 9.11E$-$03 & 0.70 & & 5.81E$-$07 & 4.04 & 8.75E$-$05 & 3.15 & & 2.44E$-$08 & 5.73 & 2.94E$-$06 & 5.64 \\
\midrule
\multicolumn{16}{@{}l}{$\mathbb{P}_{d}$--$\mathbb{P}_{d+2}$} \\
\midrule
1,178 & & 1.84E$-$03 & --- & 2.62E$-$02 & --- & & 4.64E$-$04 & --- & 1.25E$-$02 & --- & & 2.97E$-$04 & --- & 6.67E$-$03 & --- \\
2,374 & & 9.94E$-$04 & 1.76 & 1.74E$-$02 & 1.16 & & 4.52E$-$05 & 6.65 & 1.41E$-$03 & 6.22 & & 2.68E$-$05 & 6.86 & 1.04E$-$03 & 5.32 \\
5,236 & & 5.11E$-$04 & 1.68 & 1.49E$-$02 & 0.39 & & 5.06E$-$06 & 5.54 & 3.66E$-$04 & 3.42 & & 1.70E$-$06 & 6.98 & 9.38E$-$05 & 6.07 \\
10,378 & & 2.46E$-$04 & 2.14 & 1.38E$-$02 & 0.22 & & 7.92E$-$07 & 5.42 & 7.24E$-$05 & 4.73 & & 1.14E$-$07 & 7.88 & 7.59E$-$06 & 7.35 \\
22,072 & & 1.24E$-$04 & 1.81 & 1.12E$-$02 & 0.55 & & 1.34E$-$07 & 4.72 & 1.18E$-$05 & 4.82 & & 7.49E$-$09 & 7.22 & 1.38E$-$06 & 4.51 \\
\bottomrule
\end{tabular}

}
\end{table}

\begin{table}[!htb]
\centering
\footnotesize
\caption{Streamfunction errors and convergence orders obtained in the flow in an annular domain test case.}
\label{tab:verification_benchmark_t21_errorstable_psi}
\resizeboxlarger{
\begin{tabular}{@{}r@{}r@{}rrrr@{}r@{}rrrr@{}r@{}rrrr@{}}
\toprule
& \phantom{aaa} & \multicolumn{4}{@{}l@{}}{$d=1$} & \phantom{aaa} & \multicolumn{4}{@{}l@{}}{$d=3$} & \phantom{aaa} & \multicolumn{4}{@{}l@{}}{$d=5$} \\
\cmidrule{3-6}\cmidrule{8-11}\cmidrule{13-16}
$N_{\textrm{C}}$ & & $E_{1}$ & $O_{1}$ & $E_{\infty}$ & $O_{\infty}$ & & $E_{1}$ & $O_{1}$ & $E_{\infty}$ & $O_{\infty}$ & & $E_{1}$ & $O_{1}$ & $E_{\infty}$ & $O_{\infty}$ \\
\midrule
\multicolumn{16}{@{}l}{$\mathbb{P}_{d}$--$\mathbb{P}_{d}$} \\
\midrule
1,178 & & 4.88E$-$03 & --- & 7.60E$-$02 & --- & & 3.85E$-$05 & --- & 1.42E$-$04 & --- & & 4.52E$-$06 & --- & 2.59E$-$05 & --- \\
2,374 & & 8.28E$-$03 & --- & 8.87E$-$02 & --- & & 1.04E$-$05 & 3.74 & 3.26E$-$05 & 4.21 & & 1.91E$-$06 & 2.45 & 6.52E$-$06 & 3.93 \\
5,236 & & 1.49E$-$02 & --- & 4.02E$-$01 & --- & & 2.47E$-$06 & 3.63 & 7.29E$-$06 & 3.79 & & 2.32E$-$07 & 5.33 & 6.28E$-$07 & 5.91 \\
10,378 & & 6.85E$-$03 & --- & 2.80E$-$01 & --- & & 6.89E$-$07 & 3.73 & 1.85E$-$06 & 4.01 & & 3.14E$-$08 & 5.85 & 8.29E$-$08 & 5.92 \\
22,072 & & 6.67E$-$04 & --- & 3.58E$-$02 & --- & & 1.89E$-$07 & 3.43 & 4.88E$-$07 & 3.53 & & 3.82E$-$09 & 5.59 & 1.04E$-$08 & 5.50 \\
\midrule
\multicolumn{16}{@{}l}{$\mathbb{P}_{d}$--$\mathbb{P}_{d+1}$} \\
\midrule
1,178 & & 7.95E$-$04 & --- & 2.11E$-$03 & --- & & 1.84E$-$05 & --- & 4.19E$-$05 & --- & & 1.53E$-$05 & --- & 4.34E$-$05 & --- \\
2,374 & & 3.81E$-$04 & 2.10 & 1.19E$-$03 & 1.63 & & 2.66E$-$06 & 5.53 & 6.38E$-$06 & 5.37 & & 5.65E$-$07 & 9.42 & 2.15E$-$06 & 8.57 \\
5,236 & & 1.75E$-$04 & 1.96 & 5.28E$-$04 & 2.07 & & 4.17E$-$07 & 4.68 & 8.36E$-$07 & 5.14 & & 5.08E$-$08 & 6.09 & 1.49E$-$07 & 6.75 \\
10,378 & & 8.99E$-$05 & 1.95 & 2.92E$-$04 & 1.73 & & 8.77E$-$08 & 4.56 & 1.75E$-$07 & 4.57 & & 4.71E$-$09 & 6.95 & 1.43E$-$08 & 6.86 \\
22,072 & & 4.13E$-$05 & 2.06 & 1.41E$-$04 & 1.93 & & 1.94E$-$08 & 3.99 & 3.85E$-$08 & 4.02 & & 4.34E$-$10 & 6.32 & 1.30E$-$09 & 6.35 \\
\midrule
\multicolumn{16}{@{}l}{$\mathbb{P}_{d}$--$\mathbb{P}_{d+2}$} \\
\midrule
1,178 & & 8.46E$-$04 & --- & 2.22E$-$03 & --- & & 7.30E$-$06 & --- & 3.39E$-$05 & --- & & 5.34E$-$06 & --- & 3.06E$-$05 & --- \\
2,374 & & 4.14E$-$04 & 2.04 & 1.30E$-$03 & 1.53 & & 1.32E$-$06 & 4.89 & 4.36E$-$06 & 5.85 & & 5.11E$-$07 & 6.70 & 1.71E$-$06 & 8.23 \\
5,236 & & 1.92E$-$04 & 1.95 & 5.50E$-$04 & 2.17 & & 1.46E$-$07 & 5.55 & 4.29E$-$07 & 5.86 & & 3.34E$-$08 & 6.90 & 1.23E$-$07 & 6.65 \\
10,378 & & 9.69E$-$05 & 1.99 & 3.04E$-$04 & 1.74 & & 1.89E$-$08 & 5.99 & 5.33E$-$08 & 6.09 & & 2.53E$-$09 & 7.55 & 6.62E$-$09 & 8.54 \\
22,072 & & 4.56E$-$05 & 2.00 & 1.48E$-$04 & 1.91 & & 2.66E$-$09 & 5.19 & 7.71E$-$09 & 5.13 & & 1.31E$-$10 & 7.85 & 4.67E$-$10 & 7.03 \\
\bottomrule
\end{tabular}

}
\end{table}

\begin{table}[!htb]
\centering
\footnotesize
\caption{Boundary vorticity errors and convergence orders obtained in the flow in an annular domain test case.}
\label{tab:verification_benchmark_t21_errorstable_omegabound}
\resizeboxlarger{
\begin{tabular}{@{}r@{}r@{}rrrr@{}r@{}rrrr@{}r@{}rrrr@{}}
\toprule
& \phantom{aaa} & \multicolumn{4}{@{}l@{}}{$d=1$} & \phantom{aaa} & \multicolumn{4}{@{}l@{}}{$d=3$} & \phantom{aaa} & \multicolumn{4}{@{}l@{}}{$d=5$} \\
\cmidrule{3-6}\cmidrule{8-11}\cmidrule{13-16}
$N_{\textrm{B}}$ & & $E_{1}$ & $O_{1}$ & $E_{\infty}$ & $O_{\infty}$ & & $E_{1}$ & $O_{1}$ & $E_{\infty}$ & $O_{\infty}$ & & $E_{1}$ & $O_{1}$ & $E_{\infty}$ & $O_{\infty}$ \\
\midrule
\multicolumn{16}{@{}l}{$\mathbb{P}_{d}$--$\mathbb{P}_{d}$} \\
\midrule
128 & & 2.44E$+$00 & --- & 4.95E$+$01 & --- & & 2.83E$-$03 & --- & 1.38E$-$02 & --- & & 1.65E$-$03 & --- & 1.34E$-$02 & --- \\
184 & & 3.92E$+$00 & --- & 7.34E$+$01 & --- & & 1.38E$-$03 & 1.97 & 1.03E$-$02 & 0.81 & & 3.24E$-$04 & 4.49 & 2.03E$-$03 & 5.20 \\
272 & & 9.36E$+$00 & --- & 6.03E$+$02 & --- & & 6.98E$-$04 & 1.75 & 5.19E$-$03 & 1.75 & & 4.94E$-$05 & 4.81 & 4.76E$-$04 & 3.71 \\
384 & & 1.20E$+$01 & --- & 4.75E$+$02 & --- & & 3.23E$-$04 & 2.24 & 2.10E$-$03 & 2.62 & & 1.30E$-$05 & 3.87 & 1.51E$-$04 & 3.33 \\
560 & & 5.41E$+$00 & --- & 2.34E$+$02 & --- & & 1.56E$-$04 & 1.92 & 1.52E$-$03 & 0.86 & & 2.35E$-$06 & 4.53 & 2.66E$-$05 & 4.61 \\
\midrule
\multicolumn{16}{@{}l}{$\mathbb{P}_{d}$--$\mathbb{P}_{d+1}$} \\
\midrule
128 & & 7.26E$-$03 & --- & 5.58E$-$02 & --- & & 1.27E$-$03 & --- & 8.19E$-$03 & --- & & 1.67E$-$03 & --- & 1.96E$-$02 & --- \\
184 & & 5.72E$-$03 & 0.65 & 4.06E$-$02 & 0.88 & & 3.93E$-$04 & 3.23 & 2.48E$-$03 & 3.29 & & 1.72E$-$04 & 6.26 & 1.44E$-$03 & 7.20 \\
272 & & 3.33E$-$03 & 1.38 & 2.30E$-$02 & 1.45 & & 1.18E$-$04 & 3.09 & 1.18E$-$03 & 1.90 & & 2.00E$-$05 & 5.51 & 1.80E$-$04 & 5.31 \\
384 & & 2.89E$-$03 & 0.41 & 1.80E$-$02 & 0.71 & & 4.09E$-$05 & 3.06 & 4.23E$-$04 & 2.98 & & 2.99E$-$06 & 5.51 & 3.68E$-$05 & 4.61 \\
560 & & 1.36E$-$03 & 1.99 & 1.26E$-$02 & 0.95 & & 1.22E$-$05 & 3.20 & 1.33E$-$04 & 3.06 & & 4.70E$-$07 & 4.91 & 4.25E$-$06 & 5.72 \\
\midrule
\multicolumn{16}{@{}l}{$\mathbb{P}_{d}$--$\mathbb{P}_{d+2}$} \\
\midrule
128 & & 1.07E$-$02 & --- & 5.46E$-$02 & --- & & 2.01E$-$03 & --- & 1.99E$-$02 & --- & & 1.46E$-$03 & --- & 1.10E$-$02 & --- \\
184 & & 6.17E$-$03 & 1.53 & 5.01E$-$02 & 0.24 & & 2.98E$-$04 & 5.25 & 2.12E$-$03 & 6.17 & & 1.39E$-$04 & 6.48 & 1.55E$-$03 & 5.40 \\
272 & & 4.59E$-$03 & 0.76 & 2.75E$-$02 & 1.53 & & 4.57E$-$05 & 4.80 & 5.06E$-$04 & 3.67 & & 1.14E$-$05 & 6.40 & 1.66E$-$04 & 5.72 \\
384 & & 2.98E$-$03 & 1.25 & 2.43E$-$02 & 0.35 & & 1.09E$-$05 & 4.16 & 1.21E$-$04 & 4.15 & & 1.12E$-$06 & 6.73 & 1.19E$-$05 & 7.64 \\
560 & & 2.09E$-$03 & 0.95 & 1.93E$-$02 & 0.61 & & 1.87E$-$06 & 4.68 & 1.74E$-$05 & 5.14 & & 1.07E$-$07 & 6.21 & 2.05E$-$06 & 4.67 \\
\bottomrule
\end{tabular}

}
\end{table}

\begin{table}[!htb]
\centering
\footnotesize
\caption{Velocity errors and convergence orders obtained in the flow in an annular domain test case.}
\label{tab:verification_benchmark_t21_errorstable_u}
\resizeboxlarger{
\begin{tabular}{@{}r@{}r@{}rrrr@{}r@{}rrrr@{}r@{}rrrr@{}}
\toprule
& \phantom{aaa} & \multicolumn{4}{@{}l@{}}{$d=1$} & \phantom{aaa} & \multicolumn{4}{@{}l@{}}{$d=3$} & \phantom{aaa} & \multicolumn{4}{@{}l@{}}{$d=5$} \\
\cmidrule{3-6}\cmidrule{8-11}\cmidrule{13-16}
$N_{\textrm{E}}$ & & $E_{1}$ & $O_{1}$ & $E_{\infty}$ & $O_{\infty}$ & & $E_{1}$ & $O_{1}$ & $E_{\infty}$ & $O_{\infty}$ & & $E_{1}$ & $O_{1}$ & $E_{\infty}$ & $O_{\infty}$ \\
\midrule
\multicolumn{16}{@{}l}{$\mathbb{P}_{d}$--$\mathbb{P}_{d}$} \\
\midrule
1,831 & & 7.14E$-$02 & --- & 1.11E$+$00 & --- & & 5.00E$-$04 & --- & 1.13E$-$02 & --- & & 2.58E$-$04 & --- & 1.45E$-$02 & --- \\
3,653 & & 7.16E$-$02 & --- & 9.00E$-$01 & --- & & 1.19E$-$04 & 4.15 & 4.99E$-$03 & 2.36 & & 3.62E$-$05 & 5.69 & 2.71E$-$03 & 4.86 \\
7,990 & & 1.10E$-$01 & --- & 5.02E$+$00 & --- & & 2.88E$-$05 & 3.63 & 1.32E$-$03 & 3.40 & & 3.83E$-$06 & 5.74 & 3.86E$-$04 & 4.98 \\
15,759 & & 7.09E$-$02 & --- & 3.89E$+$00 & --- & & 8.45E$-$06 & 3.61 & 4.49E$-$04 & 3.18 & & 5.56E$-$07 & 5.68 & 6.28E$-$05 & 5.34 \\
33,388 & & 1.23E$-$02 & --- & 1.16E$+$00 & --- & & 2.46E$-$06 & 3.28 & 1.75E$-$04 & 2.51 & & 6.84E$-$08 & 5.58 & 1.24E$-$05 & 4.32 \\
\midrule
\multicolumn{16}{@{}l}{$\mathbb{P}_{d}$--$\mathbb{P}_{d+1}$} \\
\midrule
1,831 & & 2.41E$-$03 & --- & 5.57E$-$02 & --- & & 2.73E$-$04 & --- & 8.48E$-$03 & --- & & 2.47E$-$04 & --- & 1.60E$-$02 & --- \\
3,653 & & 1.09E$-$03 & 2.30 & 3.36E$-$02 & 1.46 & & 4.81E$-$05 & 5.03 & 2.60E$-$03 & 3.42 & & 1.68E$-$05 & 7.78 & 1.41E$-$03 & 7.03 \\
7,990 & & 4.55E$-$04 & 2.23 & 1.33E$-$02 & 2.36 & & 7.49E$-$06 & 4.75 & 4.80E$-$04 & 4.32 & & 1.19E$-$06 & 6.76 & 1.47E$-$04 & 5.77 \\
15,759 & & 1.86E$-$04 & 2.63 & 5.01E$-$03 & 2.88 & & 1.49E$-$06 & 4.76 & 1.43E$-$04 & 3.56 & & 1.21E$-$07 & 6.73 & 1.71E$-$05 & 6.35 \\
33,388 & & 9.98E$-$05 & 1.66 & 3.84E$-$03 & 0.71 & & 2.82E$-$07 & 4.43 & 3.87E$-$05 & 3.49 & & 1.03E$-$08 & 6.57 & 3.42E$-$06 & 4.28 \\
\midrule
\multicolumn{16}{@{}l}{$\mathbb{P}_{d}$--$\mathbb{P}_{d+2}$} \\
\midrule
1,831 & & 2.16E$-$03 & --- & 3.35E$-$02 & --- & & 3.25E$-$04 & --- & 1.71E$-$02 & --- & & 2.21E$-$04 & --- & 1.70E$-$02 & --- \\
3,653 & & 9.48E$-$04 & 2.39 & 9.45E$-$03 & 3.67 & & 3.71E$-$05 & 6.28 & 2.96E$-$03 & 5.07 & & 1.06E$-$05 & 8.79 & 1.00E$-$03 & 8.21 \\
7,990 & & 4.04E$-$04 & 2.18 & 6.72E$-$03 & 0.87 & & 3.91E$-$06 & 5.75 & 4.09E$-$04 & 5.06 & & 5.17E$-$07 & 7.72 & 8.73E$-$05 & 6.23 \\
15,759 & & 1.94E$-$04 & 2.16 & 2.69E$-$03 & 2.70 & & 5.74E$-$07 & 5.65 & 7.09E$-$05 & 5.16 & & 3.80E$-$08 & 7.69 & 8.05E$-$06 & 7.02 \\
33,388 & & 8.87E$-$05 & 2.08 & 1.43E$-$03 & 1.69 & & 7.11E$-$08 & 5.57 & 1.37E$-$05 & 4.38 & & 2.21E$-$09 & 7.58 & 6.36E$-$07 & 6.76 \\
\bottomrule
\end{tabular}

}
\end{table}


\subsection{Rose-shaped domain}
\label{subsec:verification_benchmark_t3}

This test case addresses an internal flow (Navier-Stokes equations) confined between two rose-shaped boundaries, which correspond to a diffeomorphic transformation of an annular domain with interior and exterior physical boundaries with radii $r_{\textrm{I}}$ and $r_{\textrm{E}}$, respectively.
The resulting interior and exterior rose-shaped boundaries are parametrised in polar coordinates as
\begin{equation}
\Gamma_{\textrm{I}}:\left(r,\theta\right)=\left(R_{\textrm{I}}\left(\theta\right),\theta\right)
\qquad
\text{and}
\qquad
\Gamma_{\textrm{E}}:\left(r,\theta\right)=\left(R_{\textrm{E}}\left(\theta\right),\theta\right),
\end{equation}
respectively, where functions $R_{\textrm{I}}\coloneqq R_{\textrm{I}}\left(\theta\right)$ and $R_{\textrm{E}}\coloneqq R_{\textrm{E}}\left(\theta\right)$ correspond to a periodic radius perturbation of frequency $\alpha_{\textrm{I}},\alpha_{\textrm{E}}\in\mathbb{N}^{+}$ and magnitude $\beta_{\textrm{I}},\beta_{\textrm{E}}\in\mathbb{R}$, given as
\begin{equation}
R_{\textrm{I}}=r_{\textrm{I}}-\beta_{\textrm{I}}+\beta_{\textrm{I}}\cos\left(\alpha_{\textrm{I}}\theta\right)
\qquad
\text{and}
\qquad
R_{\textrm{E}}=r_{\textrm{E}}-\beta_{\textrm{E}}+\beta_{\textrm{E}}\cos\left(\alpha_{\textrm{E}}\theta\right),
\end{equation}
respectively.
The normal curvatures of the interior and exterior physical boundaries, $\kappa_{\textrm{I}}\coloneqq\kappa_{\textrm{I}}\left(\theta\right)$ and $\kappa_{\textrm{E}}\coloneqq\kappa_{\textrm{E}}\left(\theta\right)$, respectively, can be determined analytically (see~\ref{appendix:calculation_of_line_curvature}), given as
\begin{equation}
\kappa_{\textrm{I}}=-\dfrac{R_{\textrm{I}}^{2}+2\left(R_{\textrm{I}}'\right)^{2}-R_{\textrm{I}}R_{\textrm{I}}''}{\left(R_{\textrm{I}}^{2}+\left(R_{\textrm{I}}'\right)^{2}\right)^{3/2}}
\qquad
\text{and}
\qquad
\kappa_{\textrm{E}}=\dfrac{R_{\textrm{E}}^{2}+2\left(R_{\textrm{E}}'\right)^{2}-R_{\textrm{E}}R_{\textrm{E}}''}{\left(R_{\textrm{E}}^{2}+\left(R_{\textrm{E}}'\right)^{2}\right)^{3/2}},
\end{equation}
where $R_{\textrm{I}}'\coloneqq R_{\textrm{I}}'\left(\theta\right)$ and $R_{\textrm{E}}'\coloneqq R_{\textrm{E}}'\left(\theta\right)$ are the first derivatives and $R_{\textrm{I}}''\coloneqq R_{\textrm{I}}''\left(\theta\right)$ and $R_{\textrm{E}}''\coloneqq R_{\textrm{E}}''\left(\theta\right)$ are the second derivatives of $R_{\textrm{I}}\left(\theta\right)$ and $R_{\textrm{E}}\left(\theta\right)$ with respect to $\theta$, that is
\begin{align}
&R_{\textrm{I}}'=\frac{\partial R_{\textrm{I}}}{\partial\theta}=-\alpha_{\textrm{I}}\beta_{\textrm{I}}\sin\left(\alpha_{\textrm{I}}\theta\right),
&&R_{\textrm{I}}''=\frac{\partial R^{2}_{\textrm{I}}}{\partial\theta^{2}}=-\alpha_{\textrm{I}}^{2}\beta_{\textrm{I}}\cos\left(\alpha_{\textrm{I}}\theta\right),\\
&R_{\textrm{E}}'=\frac{\partial R_{\textrm{E}}}{\partial\theta}=-\alpha_{\textrm{E}}\beta_{\textrm{E}}\sin\left(\alpha_{\textrm{E}}\theta\right),
&\text{and }
&R_{\textrm{E}}''=\frac{\partial R^{2}_{\textrm{E}}}{\partial\theta^{2}}=-\alpha_{\textrm{I}}^{2}\beta_{\textrm{E}}\cos\left(\alpha_{\textrm{E}}\theta\right).
\end{align}
Notice that such non-trivial parametrisation poses significant difficulties in curved mesh approaches to determining quadrature rules and implementing non-linear transformations.

The manufactured solution for the vorticity and streamfunction in polar coordinates, $\omega\coloneqq\omega(r,\theta)$ and $\psi\coloneqq\psi(r,\theta)$, respectively, read as
\begin{align}
&\omega=\dfrac{u}{ar^{2}}\left(R_{\textrm{I}}''\left(r-R_{\textrm{E}}\right)+R_{\textrm{E}}''\left(r-R_{\textrm{I}}\right)-2R_{\textrm{I}}'R_{\textrm{E}}'-r\left(4r-R_{\textrm{I}}-R_{\textrm{E}}\right)\right),\\
&\psi=\dfrac{u}{a}\left(r-R_{\textrm{I}}\right)\left(r-R_{\textrm{E}}\right),
\end{align}
where $u\in\mathbb{R}$ is a reference linear velocity magnitude on the physical boundary and $a\in\mathbb{R}$ is a normalization factor.
The velocity components in polar coordinates, $u_{r}\coloneqq u_{r}\left(r,\theta\right)$ and $u_{\theta}\coloneqq u_{\theta}\left(r,\theta\right)$, are obtained from the streamfunction and read as
\begin{equation}
u_{r}=\dfrac{u}{ar}\left(R_{\textrm{I}}'\left(R_{\textrm{E}}-r\right)+R_{\textrm{E}}'\left(R_{\textrm{I}}-r\right)\right)
\qquad
\textrm{and}
\qquad
u_{\theta}=\dfrac{u}{a}\left(R_{\textrm{I}}+R_{\textrm{E}}-2r\right).
\end{equation}
The source term function in polar coordinates, $f\coloneqq f\left(r,\theta\right)$, derives from the vorticity transport equation with the above-manufactured solution, the expression of which is omitted for compactness.
Notice that although the exact solution is given in polar coordinates, the problem is solved in Cartesian coordinates.

\begin{figure}[!htb]
\centering
\begin{tabular}{@{}c@{}c@{}}
\includegraphics[width=0.49\textwidth,trim=0cm 0cm 0cm 0cm,clip=true]{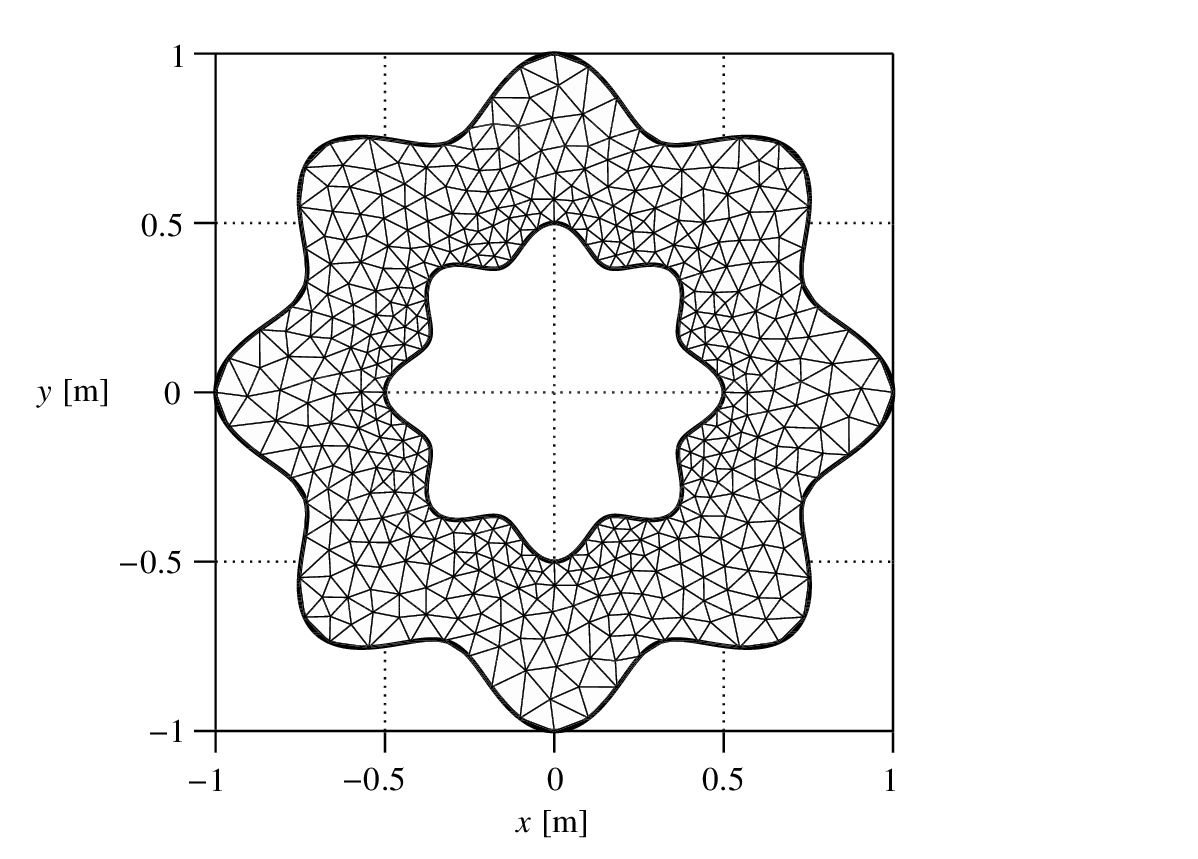} &
\includegraphics[width=0.49\textwidth,trim=0cm 0cm 0cm 0cm,clip=true]{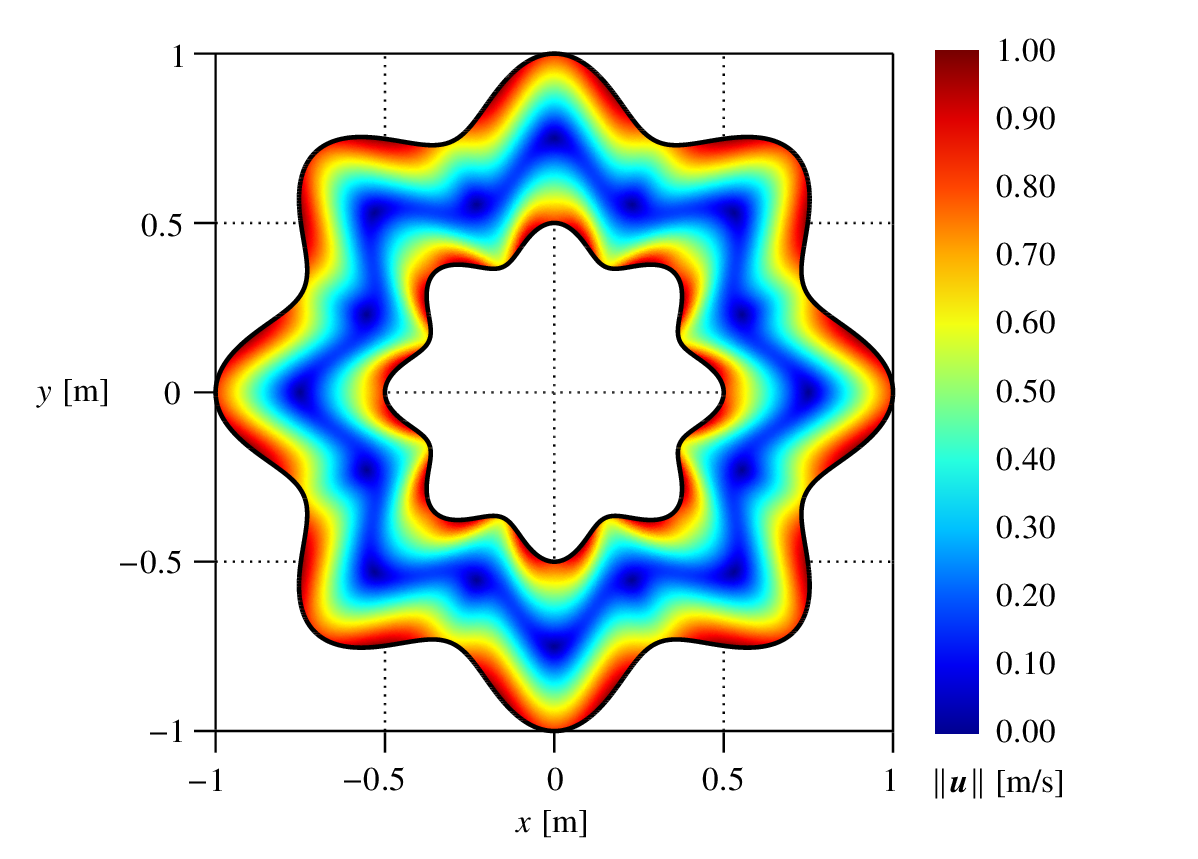}\\[0.2cm]
\small (a) Coarse mesh. & \small (b) Velocity magnitude.\\
\includegraphics[width=0.49\textwidth,trim=0cm 0cm 0cm 0cm,clip=true]{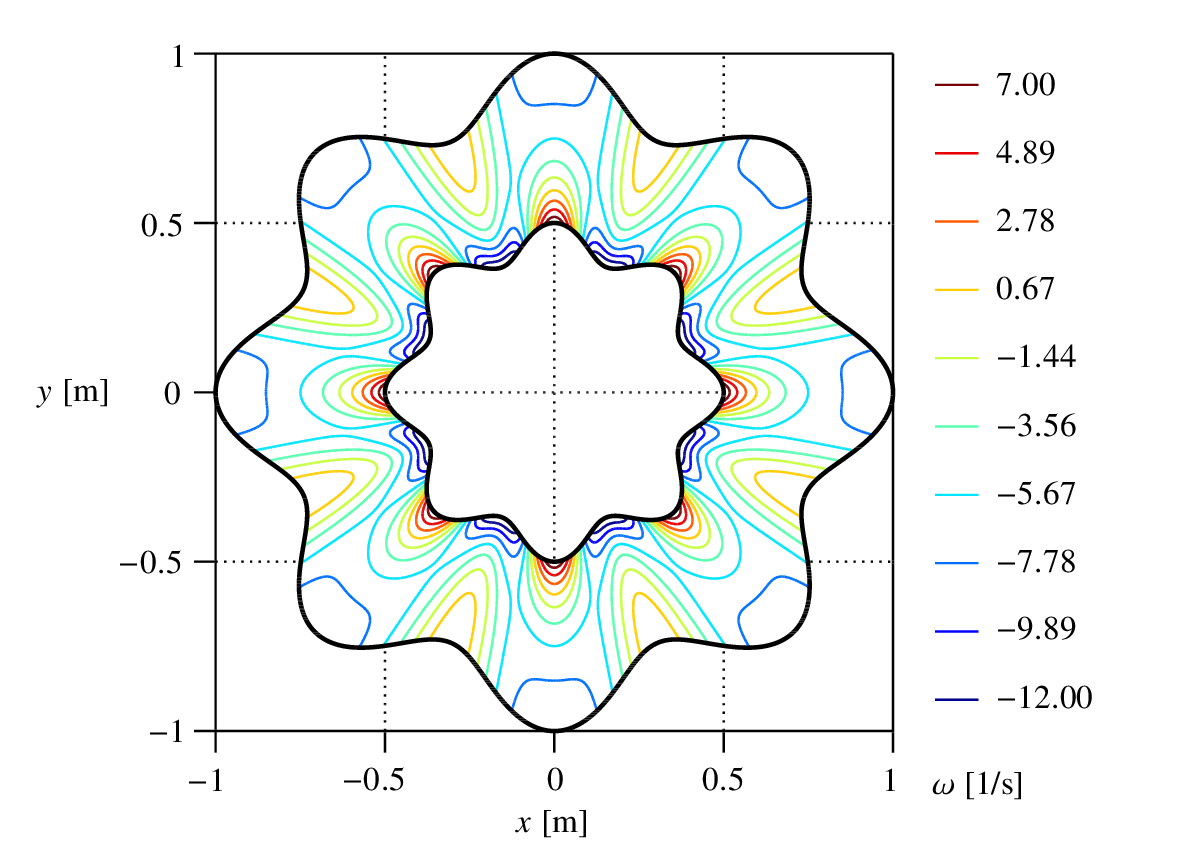} &
\includegraphics[width=0.49\textwidth,trim=0cm 0cm 0cm 0cm,clip=true]{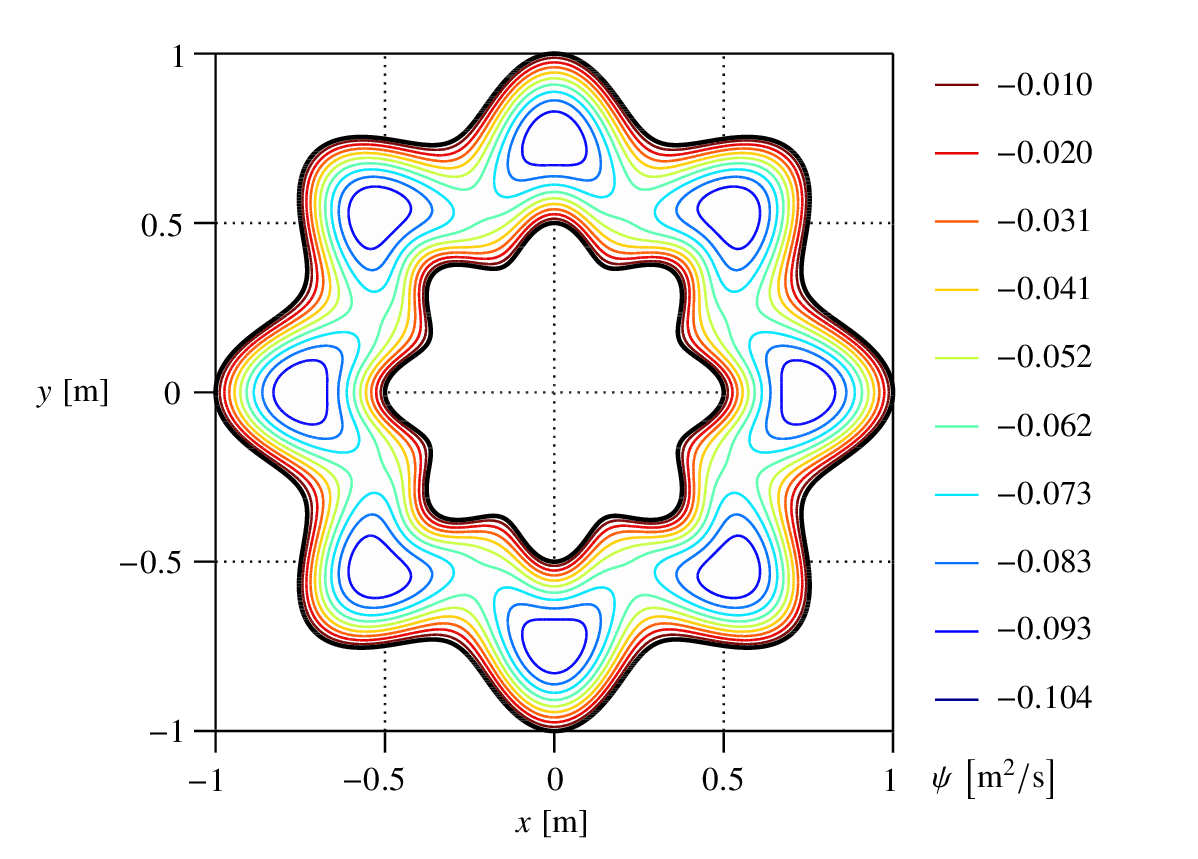}\\[0.2cm]
\small (c) Vorticity contours. & \small (d) Streamlines.
\end{tabular}
\caption{Coarse mesh and analytical solution for the flow in a rose-shaped domain test case (for proper interpretation of the colour scale, the reader is referred to the electronic version of this article).}
\label{fig:verification_benchmark_t3}
\end{figure}

For the domain, the interior physical boundary has parameters $r_{\textrm{I}}=1/2$~\si[per-mode=symbol]{\m}, $\alpha_{\textrm{I}}=8$, and $\beta_{\textrm{I}}=r_{\textrm{I}}/10$~\si[per-mode=symbol]{\m}, and centre at the origin.
The exterior physical boundary has parameters $r_{\textrm{E}}=1$~\si[per-mode=symbol]{\m}, $\alpha_{\textrm{E}}=8$, and $\beta_{\textrm{E}}=r_{\textrm{E}}/10$~\si[per-mode=symbol]{\m}, and centre at the origin.
The fluid has a reference linear velocity magnitude of $u=1$~\si[per-mode=symbol]{\m\per\second}, which corresponds to a counter-clockwise and clockwise rotation on the interior and exterior physical boundaries, respectively, a dynamic viscosity of $\mu=1$~\si[per-mode=symbol]{\kg\per\m\per\second}, and a density of $\rho=1$~\si[per-mode=symbol]{\kg\per\m^{3}}.
The normalization parameter is chosen as $a=0.6021$.
Successively finer uniform Delaunay triangular meshes are used to discretise the physical domain, and the simulations are carried out for the $\mathbb{P}_{d}$--$\mathbb{P}_{d+1}$ method with $d=1,3,5$.
Either one or two collocation points per boundary edge are considered to fulfil the prescribed boundary conditions on the vorticity and streamfunction polynomial reconstructions.
For more than one collocation point, the ROD method follows the same least-squares procedure considering more linear restrictions on the constraint functional.
Figure~\ref{fig:verification_benchmark_t3} illustrates the physical domain with a coarse mesh and the analytic solutions for the velocity, vorticity, and streamfunction fields.


The errors and convergence orders for the approximate vorticity, streamfunction, boundary vorticity, and velocity are reported in Tables~\ref{tab:verification_benchmark_t31_errorstable_psi},~\ref{tab:verification_benchmark_t31_errorstable_omega},~\ref{tab:verification_benchmark_t31_errorstable_omegabound}, and~\ref{tab:verification_benchmark_t31_errorstable_u}, respectively.
Regardless of the number of collocation points, the optimal second-, fourth-, and sixth-orders are achieved with $d=1,3,5$, respectively, in both error norms for the streamfunction and in the $L^{1}$-norm for the vorticity.
Moreover, more than the first-, third-, and fifth-orders are obtained with $d=1,3,5$, respectively, in both errors norm for the boundary vorticity, whereas the second-, fourth-, and sixth-order are obtained with $d=1,3,5$, respectively, in both errors norm for the velocity.
The results are in accordance with the previous test cases for the $\mathbb{P}_{d}$--$\mathbb{P}_{d+1}$ method, demonstrating the capability of the ROD method to preserve its effectiveness in recovering the optimal convergence orders in non-trivial curved physical domains.
Furthermore, comparable accuracy and convergence orders are obtained with one and two collocation points per boundary edge, providing strong evidence of the simplicity and efficiency of the proposed method.
Indeed, these results distinctly emphasise the versatility of the ROD method in handling arbitrary curved physical boundaries prescribed with general boundary conditions.

\begin{table}[!htb]
\centering
\footnotesize
\caption{Vorticity errors and convergence orders obtained in the flow in a rose-shaped domain test case.}
\label{tab:verification_benchmark_t31_errorstable_omega}
\resizeboxlarger{
\begin{tabular}{@{}r@{}r@{}rrrr@{}r@{}rrrr@{}r@{}rrrr@{}}
\toprule
& \phantom{aaa} & \multicolumn{4}{@{}l@{}}{$d=1$} & \phantom{aaa} & \multicolumn{4}{@{}l@{}}{$d=3$} & \phantom{aaa} & \multicolumn{4}{@{}l@{}}{$d=5$} \\
\cmidrule{3-6}\cmidrule{8-11}\cmidrule{13-16}
$N_{\textrm{C}}$ & & $E_{1}$ & $O_{1}$ & $E_{\infty}$ & $O_{\infty}$ & & $E_{1}$ & $O_{1}$ & $E_{\infty}$ & $O_{\infty}$ & & $E_{1}$ & $O_{1}$ & $E_{\infty}$ & $O_{\infty}$ \\
\midrule
\multicolumn{16}{@{}l}{1 collocation point} \\
\midrule
2,216 & & 8.84E$-$02 & --- & 8.31E$-$01 & --- & & 1.17E$-$02 & --- & 1.68E$-$01 & --- & & 9.38E$-$03 & --- & 1.14E$-$01 & --- \\
5,784 & & 3.12E$-$02 & 2.17 & 3.69E$-$01 & 1.69 & & 1.65E$-$03 & 4.08 & 4.07E$-$02 & 2.95 & & 4.39E$-$04 & 6.39 & 1.41E$-$02 & 4.37 \\
15,522 & & 1.19E$-$02 & 1.96 & 1.10E$-$01 & 2.45 & & 2.35E$-$04 & 3.95 & 6.61E$-$03 & 3.68 & & 3.28E$-$05 & 5.25 & 1.28E$-$03 & 4.85 \\
41,996 & & 4.32E$-$03 & 2.04 & 4.54E$-$02 & 1.78 & & 3.35E$-$05 & 3.91 & 1.27E$-$03 & 3.31 & & 1.14E$-$06 & 6.75 & 1.06E$-$04 & 5.02 \\
113,256 & & 1.61E$-$03 & 1.98 & 2.73E$-$02 & 1.03 & & 4.81E$-$06 & 3.91 & 2.27E$-$04 & 3.48 & & 6.07E$-$08 & 5.91 & 6.29E$-$06 & 5.69 \\
\midrule
\multicolumn{16}{@{}l}{2 collocation points} \\
\midrule
2,216 & & 8.32E$-$02 & --- & 5.70E$-$01 & --- & & 1.78E$-$02 & --- & 2.87E$-$01 & --- & & 1.36E$-$02 & --- & 2.19E$-$01 & --- \\
5,784 & & 3.10E$-$02 & 2.06 & 1.88E$-$01 & 2.31 & & 2.02E$-$03 & 4.54 & 4.70E$-$02 & 3.78 & & 5.68E$-$04 & 6.62 & 2.02E$-$02 & 4.98 \\
15,522 & & 1.19E$-$02 & 1.94 & 1.03E$-$01 & 1.23 & & 2.36E$-$04 & 4.34 & 6.79E$-$03 & 3.92 & & 3.60E$-$05 & 5.59 & 1.64E$-$03 & 5.08 \\
41,996 & & 4.32E$-$03 & 2.03 & 4.95E$-$02 & 1.46 & & 3.37E$-$05 & 3.91 & 1.32E$-$03 & 3.29 & & 1.14E$-$06 & 6.93 & 1.02E$-$04 & 5.59 \\
113,256 & & 1.61E$-$03 & 1.98 & 2.87E$-$02 & 1.10 & & 4.82E$-$06 & 3.92 & 2.64E$-$04 & 3.24 & & 6.06E$-$08 & 5.92 & 5.90E$-$06 & 5.74 \\
\bottomrule
\end{tabular}

}
\end{table}

\begin{table}[!htb]
\centering
\footnotesize
\caption{Streamfunction errors and convergence orders obtained in the flow in a rose-shaped domain test case.}
\label{tab:verification_benchmark_t31_errorstable_psi}
\resizeboxlarger{
\begin{tabular}{@{}r@{}r@{}rrrr@{}r@{}rrrr@{}r@{}rrrr@{}}
\toprule
& \phantom{aaa} & \multicolumn{4}{@{}l@{}}{$d=1$} & \phantom{aaa} & \multicolumn{4}{@{}l@{}}{$d=3$} & \phantom{aaa} & \multicolumn{4}{@{}l@{}}{$d=5$} \\
\cmidrule{3-6}\cmidrule{8-11}\cmidrule{13-16}
$N_{\textrm{C}}$ & & $E_{1}$ & $O_{1}$ & $E_{\infty}$ & $O_{\infty}$ & & $E_{1}$ & $O_{1}$ & $E_{\infty}$ & $O_{\infty}$ & & $E_{1}$ & $O_{1}$ & $E_{\infty}$ & $O_{\infty}$ \\
\midrule
\multicolumn{16}{@{}l}{1 collocation point} \\
\midrule
2,216 & & 9.98E$-$04 & --- & 2.07E$-$03 & --- & & 1.72E$-$04 & --- & 3.12E$-$04 & --- & & 8.24E$-$05 & --- & 2.06E$-$04 & --- \\
5,784 & & 4.87E$-$04 & 1.50 & 9.85E$-$04 & 1.55 & & 1.12E$-$05 & 5.71 & 3.03E$-$05 & 4.86 & & 4.35E$-$06 & 6.14 & 8.22E$-$06 & 6.72 \\
15,522 & & 2.07E$-$04 & 1.73 & 3.76E$-$04 & 1.95 & & 9.54E$-$07 & 4.98 & 3.31E$-$06 & 4.49 & & 1.06E$-$07 & 7.53 & 3.35E$-$07 & 6.48 \\
41,996 & & 5.98E$-$05 & 2.50 & 1.20E$-$04 & 2.30 & & 1.23E$-$07 & 4.12 & 4.55E$-$07 & 3.99 & & 5.01E$-$09 & 6.13 & 1.62E$-$08 & 6.09 \\
113,256 & & 2.32E$-$05 & 1.90 & 4.57E$-$05 & 1.94 & & 2.28E$-$08 & 3.40 & 5.43E$-$08 & 4.28 & & 1.36E$-$10 & 7.26 & 4.76E$-$10 & 7.11 \\
\midrule
\multicolumn{16}{@{}l}{2 collocation points} \\
\midrule
2,216 & & 1.18E$-$03 & --- & 2.26E$-$03 & --- & & 2.55E$-$04 & --- & 4.26E$-$04 & --- & & 8.61E$-$05 & --- & 2.06E$-$04 & --- \\
5,784 & & 4.56E$-$04 & 1.99 & 9.45E$-$04 & 1.81 & & 1.58E$-$05 & 5.79 & 3.83E$-$05 & 5.02 & & 4.97E$-$06 & 5.95 & 9.23E$-$06 & 6.48 \\
15,522 & & 2.03E$-$04 & 1.64 & 3.71E$-$04 & 1.89 & & 8.66E$-$07 & 5.88 & 3.12E$-$06 & 5.08 & & 1.20E$-$07 & 7.54 & 3.63E$-$07 & 6.56 \\
41,996 & & 5.88E$-$05 & 2.49 & 1.18E$-$04 & 2.30 & & 1.11E$-$07 & 4.13 & 4.24E$-$07 & 4.01 & & 4.74E$-$09 & 6.50 & 1.58E$-$08 & 6.29 \\
113,256 & & 2.30E$-$05 & 1.89 & 4.54E$-$05 & 1.93 & & 2.63E$-$08 & 2.90 & 6.14E$-$08 & 3.90 & & 1.27E$-$10 & 7.30 & 5.20E$-$10 & 6.89 \\
\bottomrule
\end{tabular}

}
\end{table}

\begin{table}[!htb]
\centering
\footnotesize
\caption{Boundary vorticity errors and convergence orders obtained in the flow in a rose-shaped domain test case.}
\label{tab:verification_benchmark_t31_errorstable_omegabound}
\resizeboxlarger{
\begin{tabular}{@{}r@{}r@{}rrrr@{}r@{}rrrr@{}r@{}rrrr@{}}
\toprule
& \phantom{aaa} & \multicolumn{4}{@{}l@{}}{$d=1$} & \phantom{aaa} & \multicolumn{4}{@{}l@{}}{$d=3$} & \phantom{aaa} & \multicolumn{4}{@{}l@{}}{$d=5$} \\
\cmidrule{3-6}\cmidrule{8-11}\cmidrule{13-16}
$N_{\textrm{B}}$ & & $E_{1}$ & $O_{1}$ & $E_{\infty}$ & $O_{\infty}$ & & $E_{1}$ & $O_{1}$ & $E_{\infty}$ & $O_{\infty}$ & & $E_{1}$ & $O_{1}$ & $E_{\infty}$ & $O_{\infty}$ \\
\midrule
\multicolumn{16}{@{}l}{1 collocation point} \\
\midrule
200 & & 2.04E$-$01 & --- & 9.22E$-$01 & --- & & 4.03E$-$02 & --- & 2.51E$-$01 & --- & & 5.25E$-$02 & --- & 1.98E$-$01 & --- \\
330 & & 8.00E$-$02 & 1.86 & 5.12E$-$01 & 1.17 & & 6.50E$-$03 & 3.64 & 5.61E$-$02 & 2.99 & & 3.57E$-$03 & 5.37 & 1.78E$-$02 & 4.81 \\
546 & & 3.09E$-$02 & 1.89 & 1.89E$-$01 & 1.98 & & 1.47E$-$03 & 2.95 & 8.07E$-$03 & 3.85 & & 3.19E$-$04 & 4.80 & 2.03E$-$03 & 4.32 \\
900 & & 1.37E$-$02 & 1.62 & 8.36E$-$02 & 1.63 & & 2.82E$-$04 & 3.31 & 1.50E$-$03 & 3.37 & & 1.78E$-$05 & 5.77 & 1.44E$-$04 & 5.30 \\
1,484 & & 6.58E$-$03 & 1.47 & 4.69E$-$02 & 1.15 & & 4.84E$-$05 & 3.52 & 3.13E$-$04 & 3.13 & & 1.27E$-$06 & 5.28 & 1.00E$-$05 & 5.32 \\
\midrule
\multicolumn{16}{@{}l}{2 collocation points} \\
\midrule
400 & & 1.97E$-$01 & --- & 1.03E$+$00 & --- & & 9.60E$-$02 & --- & 4.96E$-$01 & --- & & 1.13E$-$01 & --- & 5.42E$-$01 & --- \\
660 & & 7.57E$-$02 & 1.91 & 4.26E$-$01 & 1.75 & & 1.24E$-$02 & 4.08 & 7.57E$-$02 & 3.76 & & 6.77E$-$03 & 5.62 & 3.75E$-$02 & 5.33 \\
1,092 & & 3.12E$-$02 & 1.76 & 2.14E$-$01 & 1.37 & & 1.56E$-$03 & 4.13 & 9.01E$-$03 & 4.23 & & 5.95E$-$04 & 4.83 & 2.68E$-$03 & 5.24 \\
1,800 & & 1.37E$-$02 & 1.65 & 1.06E$-$01 & 1.40 & & 2.91E$-$04 & 3.35 & 1.77E$-$03 & 3.26 & & 2.57E$-$05 & 6.29 & 2.07E$-$04 & 5.13 \\
2,968 & & 6.57E$-$03 & 1.47 & 5.43E$-$02 & 1.34 & & 5.02E$-$05 & 3.52 & 3.75E$-$04 & 3.10 & & 1.52E$-$06 & 5.65 & 1.12E$-$05 & 5.83 \\
\bottomrule
\end{tabular}

}
\end{table}

\begin{table}[!htb]
\centering
\footnotesize
\caption{Velocity errors and convergence orders obtained in the flow in a rose-shaped domain test case.}
\label{tab:verification_benchmark_t31_errorstable_u}
\resizeboxlarger{
\begin{tabular}{@{}r@{}r@{}rrrr@{}r@{}rrrr@{}r@{}rrrr@{}}
\toprule
& \phantom{aaa} & \multicolumn{4}{@{}l@{}}{$d=1$} & \phantom{aaa} & \multicolumn{4}{@{}l@{}}{$d=3$} & \phantom{aaa} & \multicolumn{4}{@{}l@{}}{$d=5$} \\
\cmidrule{3-6}\cmidrule{8-11}\cmidrule{13-16}
$N_{\textrm{E}}$ & & $E_{1}$ & $O_{1}$ & $E_{\infty}$ & $O_{\infty}$ & & $E_{1}$ & $O_{1}$ & $E_{\infty}$ & $O_{\infty}$ & & $E_{1}$ & $O_{1}$ & $E_{\infty}$ & $O_{\infty}$ \\
\midrule
\multicolumn{16}{@{}l}{1 collocation point} \\
\midrule
3,424 & & 1.13E$-$02 & --- & 1.13E$-$01 & --- & & 2.16E$-$03 & --- & 4.61E$-$02 & --- & & 2.00E$-$03 & --- & 6.55E$-$02 & --- \\
8,841 & & 3.28E$-$03 & 2.61 & 3.49E$-$02 & 2.48 & & 2.72E$-$04 & 4.37 & 7.57E$-$03 & 3.81 & & 9.62E$-$05 & 6.39 & 4.21E$-$03 & 5.79 \\
23,556 & & 1.24E$-$03 & 1.99 & 1.76E$-$02 & 1.40 & & 3.57E$-$05 & 4.15 & 1.84E$-$03 & 2.89 & & 6.60E$-$06 & 5.47 & 5.33E$-$04 & 4.22 \\
63,444 & & 4.25E$-$04 & 2.16 & 5.12E$-$03 & 2.49 & & 4.67E$-$06 & 4.11 & 2.26E$-$04 & 4.23 & & 1.89E$-$07 & 7.18 & 2.03E$-$05 & 6.60 \\
170,626 & & 1.56E$-$04 & 2.03 & 2.07E$-$03 & 1.83 & & 1.01E$-$06 & 3.09 & 3.84E$-$05 & 3.59 & & 8.27E$-$09 & 6.32 & 1.43E$-$06 & 5.36 \\
\midrule
\multicolumn{16}{@{}l}{2 collocation points} \\
\midrule
3,424 & & 9.16E$-$03 & --- & 7.89E$-$02 & --- & & 3.42E$-$03 & --- & 6.18E$-$02 & --- & & 2.07E$-$03 & --- & 6.39E$-$02 & --- \\
8,841 & & 3.26E$-$03 & 2.18 & 3.50E$-$02 & 1.71 & & 3.56E$-$04 & 4.77 & 1.05E$-$02 & 3.73 & & 9.40E$-$05 & 6.52 & 4.26E$-$03 & 5.71 \\
23,556 & & 1.24E$-$03 & 1.98 & 1.76E$-$02 & 1.40 & & 3.57E$-$05 & 4.69 & 1.84E$-$03 & 3.57 & & 6.57E$-$06 & 5.43 & 5.35E$-$04 & 4.24 \\
63,444 & & 4.25E$-$04 & 2.16 & 5.13E$-$03 & 2.49 & & 4.66E$-$06 & 4.11 & 2.26E$-$04 & 4.23 & & 1.88E$-$07 & 7.17 & 2.03E$-$05 & 6.61 \\
170,626 & & 1.56E$-$04 & 2.03 & 2.07E$-$03 & 1.84 & & 1.01E$-$06 & 3.09 & 3.83E$-$05 & 3.59 & & 8.23E$-$09 & 6.32 & 1.43E$-$06 & 5.36 \\
\bottomrule
\end{tabular}

}
\end{table}


\subsection{Semielliptical lid-driven cavity}
\label{subsec:verification_benchmark_t4}

To finalise the numerical verification of the proposed method, the classical lid-driven cavity benchmark test case is addressed by replacing the rectangular cavity with a semiellipse centred at the origin, given implicitly as
\begin{equation}
\dfrac{x^{2}}{a^{2}}+\dfrac{y^{2}}{b^{2}}=1,
\end{equation}
where $a,b>0$ are the semi-axis lengths, or parametrically in polar coordinates as $\left(r,\theta\right)=\left(R\left(\theta\right),\theta\right)$ with $R\left(\theta\right)$ given as
\begin{equation}
R\left(\theta\right)=\dfrac{ab}{\sqrt{\left(a\sin\left(\theta\right)\right)^{2}+\left(b\cos\left(\theta\right)\right)^{2}}},
\end{equation}
which is more practical for projecting the edge midpoints onto the physical boundary after finding the corresponding angular coordinate.
The outward unit normal and tangential vectors on the semielliptical boundary are given as
\begin{equation}
\bm{n}=\dfrac{\left(\dfrac{2x}{a^{2}},\dfrac{2y}{b^{2}}\right)^{\textrm{T}}}{\sqrt{\left(\dfrac{2x}{a^{2}}\right)^{2}+\left(\dfrac{2y}{b^{2}}\right)^{2}}}
\qquad
\text{and}
\qquad
\bm{t}=\dfrac{\left(-\dfrac{2y}{b^{2}},\dfrac{2x}{a^{2}}\right)^{\textrm{T}}}{\sqrt{\left(\dfrac{2x}{a^{2}}\right)^{2}+\left(\dfrac{2y}{b^{2}}\right)^{2}}},
\end{equation}
respectively.
The straight top and semielliptical bottom physical boundaries are denoted as $\Gamma^{\textrm{T}}$ and $\Gamma^{\textrm{B}}$, respectively.

As in the classical lid-driven cavity test case in a rectangular domain, the top physical boundary is prescribed with a horizontal velocity, whereas the no-slip condition is imposed on the semielliptical physical boundary.
Thus, both the boundary streamfunction and its normal derivative are homogeneous on the whole physical boundary, apart from the top physical boundary where its normal derivative corresponds to the imposed horizontal velocity.
For the vorticity boundary condition, it is noted that the normal curvature of the semiellipse does not need to be determined since the no-slip boundary condition is imposed on the semielliptical physical boundary.

As is well-known, when prescribing a constant horizontal velocity on the lid, pressure and velocity singularities arise at each corner due to the incompatible boundary velocity prescribed on both sides, negatively affecting the convergence and the accuracy of any approximate solution~\cite{2019_kuhlmann}.
This phenomenon naturally arises in non-primitive variables as both the vorticity and streamfunction are derived from the velocity.
Thus, a regularised semielliptical lid-driven test case is firstly addressed, replacing the constant horizontal velocity with a polynomial function, given as
\begin{equation}
\bm{u}\left(x\right)=u\left(\left(x-a\right)^{4}\left(x+a\right)^{4}/a^{8},0\right)^{\textrm{T}}\qquad\text{on }\Gamma^{\textrm{T}}.
\end{equation}

In both the regularised and conventional lid-driven cavity test cases, a primary vortex arises, and its centre coordinates, along with the corresponding vorticity and streamfunction values, are used for assessing the accuracy and efficiency of the proposed method.
A reference solution is computed on a fine mesh to evaluate the errors $E_{x}$, $E_{y}$, $E_{\psi}$, and $E_{\omega}$ for the $x$-coordinate, $y$-coordinate, vorticity, and streamfunction, respectively, at the primary vortex centre.
A gradient descent method, supplied with polynomial reconstructions for the streamfunction, is employed to determine the primary vortex centre up to an appropriate tolerance.
Vorticity and streamfunction values at the vortex centre are also determined based on polynomial reconstructions.

Besides the numerical verification, a computational assessment is performed by reporting the number of fixed point and GMRES iterations to reach the corresponding stopping criteria, denoted as $N_{\text{FP}}$ and $N_{\text{GMRES}}$, respectively.
Moreover, the execution time (total time of the simulation) and the memory usage are also reported, denoted as $T_{\text{EXE}}$ and $M_{\text{USE}}$, respectively.
The GMRES method is supplemented with a preconditioning matrix and restarts every 200 iterations.

To assess the premises drawn in the introduction section that motivate the formulation of the incompressible Navier-Stokes equations based on vorticity and streamfunction, a comparison with the primitive formulation is carried out.
A very high-order accurate finite volume scheme on staggered meshes~\cite{2022_costa2} discretises the momentum and mass balance equations, and the ROD method is employed to preserve accuracy on the semielliptical physical boundary.
Then, the same simulations are run to compute velocity and pressure and a gradient descent method is employed to determine the primary vortex centre, where polynomial reconstructions for the velocity provide directly the search direction (perpendicular to the velocity vectors).
Only the errors for the primary vortex $x$- and $y$-coordinates are evaluated in this comparison.

\begin{figure}[!htb]
\centering
\includegraphics[width=0.49\textwidth,trim=0cm 0cm 4.0cm 0cm,clip=true]{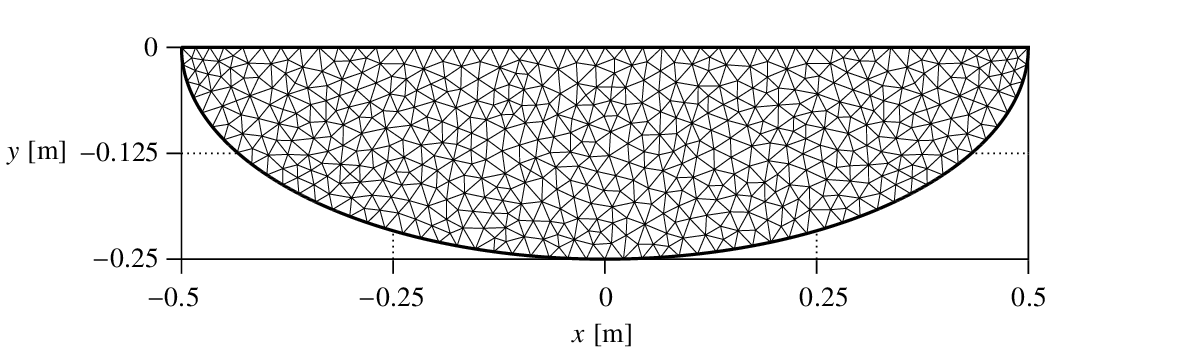}
\caption{Coarse mesh for the semielliptical lid-driven cavity test case.}
\label{fig:verification_benchmark_t4}
\end{figure}

For the domain, the semi-axes have lengths of $a=1/2$~\si[per-mode=symbol]{\m} and $b=1/4$~\si[per-mode=symbol]{\m}.
The fluid has a reference linear velocity magnitude of $u=1$~\si[per-mode=symbol]{\m\per\second}, a dynamic viscosity of either $\mu=1$~\si[per-mode=symbol]{\kg\per\m\per\second} or $\mu=0.01$~\si[per-mode=symbol]{\kg\per\m\per\second}, and a density of $\rho=1$~\si[per-mode=symbol]{\kg\per\m^{3}}, corresponding to Reynolds numbers of $Re=1$ and $Re=100$, respectively.
Successively finer uniform Delaunay triangular meshes are used to discretise the physical domain, and the simulations are carried out for the $\mathbb{P}_{d}$--$\mathbb{P}_{d+1}$ and $\mathbb{P}_{d}$--$\mathbb{P}_{d+2}$ methods with $d=1,3,5$.
A zero initial guess solution is set in all simulations.
Figure~\ref{fig:verification_benchmark_t4} illustrates the physical domain with a coarse mesh.


\subsubsection{Regularised semielliptical lid-driven cavity with $Re=1$}
\label{subsec:verification_benchmark_t41_re1}

\begin{figure}[!htb]
\centering
\begin{tabular}{@{}c@{}c@{}}
\includegraphics[width=0.49\textwidth,trim=0cm 0cm 0cm 0cm,clip=true]{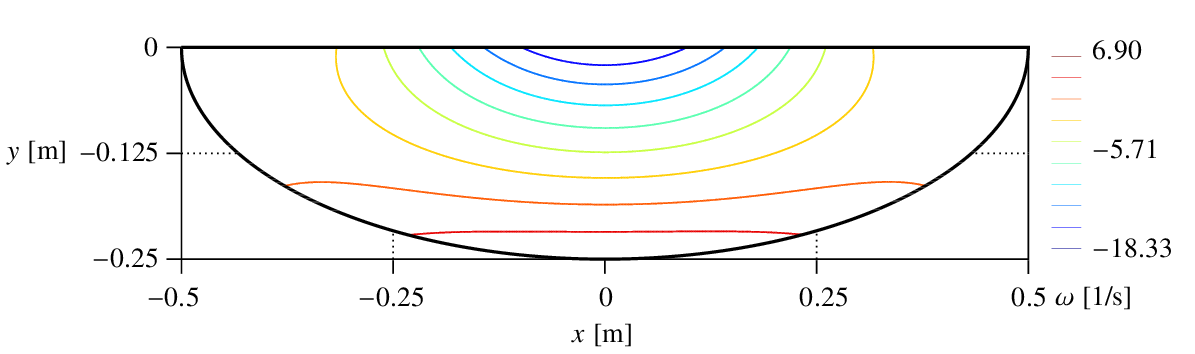} &
\includegraphics[width=0.49\textwidth,trim=0cm 0cm 0cm 0cm,clip=true]{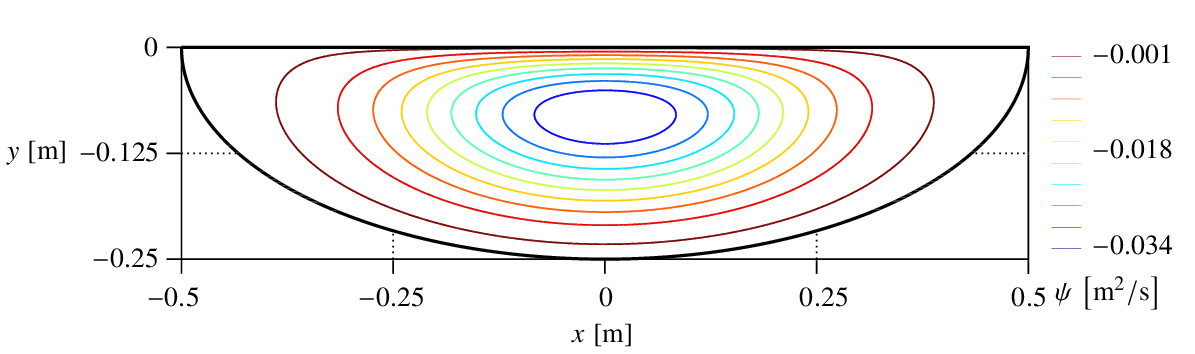}\\[0.2cm]
\small (a) Vorticity contours. & \small (b) Streamlines.
\end{tabular}
\caption{Reference solutions for the regularised semielliptical lid-driven cavity test case with $Re=1$ (for proper interpretation of the colour scale, the reader is referred to the electronic version of this article).}
\label{fig:verification_benchmark_t41_re1}
\end{figure}

Figure~\ref{fig:verification_benchmark_t41_re1} illustrates the associated reference solutions for the vorticity and streamfunction computed on the finest mesh with $d=5$.
The errors for the primary vortex centre coordinates, vorticity, and streamfunction are provided in Table~\ref{tab:verification_benchmark_t41_errorstable_re1_naive} when the naive method is employed for imposing the prescribed boundary conditions.
Although there is a substantial accuracy improvement from $d=1$ to $d=3$, the error convergence rates under mesh refinement do not improve further from $d=3$ to $d=5$.
Moreover, the coarser mesh with $d=3$ achieves comparable accuracy than the finest mesh with $d=1$.

Additionally, both the $\mathbb{P}_{d}$--$\mathbb{P}_{d+1}$ and $\mathbb{P}_{d}$--$\mathbb{P}_{d+2}$ methods provide similar accuracy regardless of the polynomial degree.
In turn, Table~\ref{tab:verification_benchmark_t41_errorstable_re1_rod} reports the errors for the same parameters when the ROD method is employed instead.
As observed, the error convergence rates under mesh refinement significantly improve with $d=3,5$ compared to the Naive method, whereas comparable accuracy and convergence are obtained with $d=1$.
In this case, the coarser mesh with $d=3$ achieves even smaller errors than the finest mesh with $d=1$, whereas $d=5$ further improves accuracy due to higher convergence rates than $d=3$.
In general, the $\mathbb{P}_{d}$--$\mathbb{P}_{d+2}$ method provides better accuracy than the $\mathbb{P}_{d}$--$\mathbb{P}_{d+1}$ method, although improvement ratios are inconsistent and vary significantly under mesh refinement and for each polynomial degree.
For the primitive formulation, comparable accuracy to the non-primitive formulation is achieved for the primary vortex centre coordinates, with improvement ratios below one order of magnitude relative to both methods.

\begin{table}[!htb]
\centering
\footnotesize
\caption{Errors obtained with the naive method in the regularised semielliptical lid-driven cavity test case with $Re=1$.}
\label{tab:verification_benchmark_t41_errorstable_re1_naive}
\resizeboxlarger{
\begin{tabular}{@{}rr@{}r@{}rrrr@{}@{}}
\toprule
$d$ & $N_{\textrm{C}}$ & \phantom{aaa} & $E_{x}$ [\%] & $E_{y}$ [\%] & $E_{\omega}$ [\%] & $E_{\psi}$ [\%] \\
\midrule
\multicolumn{7}{@{}l}{$\left(\psi,\omega\right)$/$\mathbb{P}_{d}$--$\mathbb{P}_{d+1}$} \\
\midrule
\multirow{4}{*}{$1$}
& 1,044 & & 1.20E$+$02 & 5.56E$-$01 & 1.08E$+$00 & 1.65E$-$01 \\
& 4,070 & & 8.39E$+$01 & 1.65E$-$01 & 3.25E$-$01 & 6.39E$-$02 \\
& 16,398 & & 1.31E$+$01 & 6.32E$-$02 & 1.00E$-$01 & 2.47E$-$02 \\
& 65,786 & & 3.01E$+$00 & 1.32E$-$02 & 2.41E$-$02 & 7.13E$-$03 \\
\midrule
\multirow{4}{*}{$3$}
& 1,044 & & 1.95E$+$00 & 2.18E$-$02 & 2.18E$-$02 & 2.18E$-$02 \\
& 4,070 & & 5.85E$-$02 & 5.70E$-$03 & 5.67E$-$03 & 5.60E$-$03 \\
& 16,398 & & 4.04E$-$03 & 1.51E$-$03 & 1.49E$-$03 & 1.41E$-$03 \\
& 65,786 & & 8.40E$-$04 & 3.78E$-$04 & 3.73E$-$04 & 3.52E$-$04 \\
\midrule
\multirow{4}{*}{$5$}
& 1,044 & & 2.47E$-$02 & 2.32E$-$02 & 2.30E$-$02 & 2.14E$-$02 \\
& 4,070 & & 1.17E$-$02 & 6.05E$-$03 & 5.98E$-$03 & 5.63E$-$03 \\
& 16,398 & & 2.47E$-$03 & 1.51E$-$03 & 1.49E$-$03 & 1.41E$-$03 \\
& 65,786 & & 6.24E$-$04 & 3.78E$-$04 & 3.74E$-$04 & 3.52E$-$04 \\
\midrule
\multicolumn{7}{@{}l}{$\left(\psi,\omega\right)$/$\mathbb{P}_{d}$--$\mathbb{P}_{d+2}$} \\
\midrule
\multirow{4}{*}{$1$}
& 1,044 & & 2.33E$+$02 & 2.88E$-$01 & 3.34E$-$01 & 3.30E$-$01 \\
& 4,070 & & 1.27E$+$02 & 6.59E$-$02 & 6.40E$-$02 & 1.02E$-$01 \\
& 16,398 & & 4.46E$+$01 & 4.69E$-$02 & 3.35E$-$02 & 2.44E$-$02 \\
& 65,786 & & 3.17E$+$01 & 1.78E$-$02 & 1.57E$-$02 & 6.13E$-$03 \\
\midrule
\multirow{4}{*}{$3$}
& 1,044 & & 2.37E$+$00 & 2.31E$-$02 & 2.29E$-$02 & 2.14E$-$02 \\
& 4,070 & & 1.04E$-$02 & 5.58E$-$03 & 5.52E$-$03 & 5.60E$-$03 \\
& 16,398 & & 1.81E$-$03 & 1.51E$-$03 & 1.49E$-$03 & 1.41E$-$03 \\
& 65,786 & & 5.25E$-$04 & 3.78E$-$04 & 3.73E$-$04 & 3.52E$-$04 \\
\midrule
\multirow{4}{*}{$5$}
& 1,044 & & 3.27E$-$02 & 2.31E$-$02 & 2.29E$-$02 & 2.15E$-$02 \\
& 4,070 & & 1.20E$-$02 & 6.05E$-$03 & 5.98E$-$03 & 5.63E$-$03 \\
& 16,398 & & 2.48E$-$03 & 1.51E$-$03 & 1.49E$-$03 & 1.41E$-$03 \\
& 65,786 & & 6.24E$-$04 & 3.78E$-$04 & 3.74E$-$04 & 3.52E$-$04 \\
\bottomrule
\end{tabular}

}
\end{table}

\begin{table}[!htb]
\centering
\footnotesize
\caption{Errors obtained with the ROD method in the regularised semielliptical lid-driven cavity test case with $Re=1$.}
\label{tab:verification_benchmark_t41_errorstable_re1_rod}
\resizeboxlarger{
\begin{tabular}{@{}rr@{}r@{}rrrr@{}@{}r@{}rrrr@{}}
\toprule
$d$ & $N_{\textrm{C}}$ & \phantom{aaa} & $E_{x}$ [\%] & $E_{y}$ [\%] & $E_{\omega}$ [\%] & $E_{\psi}$ [\%] & \phantom{aaa} & $N_{\textrm{FP}}$ & $N_{\textrm{GMRES}}$ & $T_{\textrm{EXE}}$ [s] & $M_{\textrm{USE}}$ [Gb] \\
\midrule
\multicolumn{11}{@{}l}{$\left(\psi,\omega\right)$/$\mathbb{P}_{d}$--$\mathbb{P}_{d+1}$} \\
\midrule
\multirow{4}{*}{$1$}
& 1,044 & & 1.20E$+$02 & 5.77E$-$01 & 1.10E$+$00 & 1.43E$-$01 & & 7 & 240 & 0.07 & 0.02 \\
& 4,070 & & 8.40E$+$01 & 1.71E$-$01 & 3.31E$-$01 & 5.83E$-$02 & & 7 & 490 & 0.31 & 0.06 \\
& 16,398 & & 1.31E$+$01 & 6.47E$-$02 & 1.02E$-$01 & 2.33E$-$02 & & 8 & 1,110 & 3.29 & 0.22 \\
& 65,786 & & 3.01E$+$00 & 1.35E$-$02 & 2.45E$-$02 & 6.78E$-$03 & & 9 & 2,850 & 38.40 & 0.85 \\
\midrule
\multirow{4}{*}{$3$}
& 1,044 & & 2.00E$+$00 & 1.34E$-$03 & 1.03E$-$03 & 3.09E$-$04 & & 7 & 380 & 0.19 & 0.03 \\
& 4,070 & & 6.86E$-$02 & 3.49E$-$04 & 3.11E$-$04 & 3.28E$-$05 & & 8 & 860 & 0.94 & 0.10 \\
& 16,398 & & 1.25E$-$02 & 1.00E$-$05 & 9.04E$-$06 & 2.03E$-$06 & & 9 & 1,630 & 8.04 & 0.42 \\
& 65,786 & & 2.15E$-$04 & 4.98E$-$07 & 2.52E$-$07 & 1.46E$-$07 & & 9 & 6,400 & 115.35 & 1.35 \\
\midrule
\multirow{4}{*}{$5$}
& 1,044 & & 6.28E$-$02 & 2.98E$-$05 & 1.22E$-$04 & 7.26E$-$05 & & 8 & 590 & 0.54 & 0.05 \\
& 4,070 & & 1.92E$-$03 & 7.79E$-$07 & 1.87E$-$06 & 1.59E$-$06 & & 8 & 1,050 & 2.97 & 0.20 \\
& 16,398 & & 2.50E$-$05 & 1.40E$-$08 & 2.91E$-$08 & 1.46E$-$08 & & 9 & 3,040 & 19.28 & 0.60 \\
& 65,786 & & --- & --- & --- & --- & & 9 & 9,860 & 240.24 & 2.27 \\
\midrule
\multicolumn{11}{@{}l}{$\left(\psi,\omega\right)$/$\mathbb{P}_{d}$--$\mathbb{P}_{d+2}$} \\
\midrule
\multirow{4}{*}{$1$}
& 1,044 & & 3.46E$+$01 & 4.26E$-$01 & 4.01E$-$01 & 4.95E$-$02 & & 7 & 330 & 0.10 & 0.02 \\
& 4,070 & & 7.44E$+$00 & 2.10E$-$01 & 1.95E$-$01 & 4.65E$-$03 & & 7 & 630 & 0.51 & 0.07 \\
& 16,398 & & 5.38E$+$00 & 6.63E$-$02 & 6.13E$-$02 & 8.73E$-$04 & & 8 & 1,610 & 4.87 & 0.24 \\
& 65,786 & & 4.80E$+$00 & 1.30E$-$02 & 1.30E$-$02 & 1.63E$-$04 & & 9 & 4,410 & 64.86 & 0.94 \\
\midrule
\multirow{4}{*}{$3$}
& 1,044 & & 1.46E$+$00 & 6.13E$-$03 & 6.89E$-$03 & 3.69E$-$04 & & 7 & 490 & 0.28 & 0.04 \\
& 4,070 & & 1.50E$-$02 & 4.74E$-$04 & 4.87E$-$04 & 3.08E$-$05 & & 8 & 1,110 & 1.36 & 0.11 \\
& 16,398 & & 6.84E$-$04 & 2.18E$-$06 & 1.43E$-$06 & 1.12E$-$06 & & 9 & 2,520 & 11.46 & 0.40 \\
& 65,786 & & 9.91E$-$05 & 4.18E$-$07 & 3.18E$-$07 & 3.75E$-$08 & & 9 & 7,960 & 155.00 & 1.51 \\
\midrule
\multirow{4}{*}{$5$}
& 1,044 & & 7.07E$-$02 & 1.61E$-$05 & 3.65E$-$05 & 5.44E$-$06 & & 8 & 1,120 & 0.87 & 0.05 \\
& 4,070 & & 1.97E$-$03 & 3.36E$-$07 & 1.13E$-$07 & 2.96E$-$07 & & 8 & 1,550 & 3.31 & 0.18 \\
& 16,398 & & 1.72E$-$05 & 1.62E$-$08 & 1.78E$-$08 & 1.34E$-$09 & & 9 & 3,700 & 25.63 & 0.68 \\
& 65,786 & & --- & --- & --- & --- & & 9 & 27,790 & 677.04 & 2.59 \\
\midrule
\multicolumn{11}{@{}l}{$\left(\bm{u},p\right)$} \\
\midrule
\multirow{4}{*}{$1$}
& 1,044 & & 1.05E$+$02 & 1.22E$+$00 & --- & --- & & 7 & 940 & 0.20 & 0.04 \\
& 4,070 & & 1.92E$+$01 & 3.70E$-$01 & --- & --- & & 7 & 2,080 & 1.94 & 0.13 \\
& 16,398 & & 8.49E$+$00 & 7.36E$-$02 & --- & --- & & 8 & 5,780 & 60.11 & 0.65 \\
& 65,786 & & 6.83E$-$01 & 2.20E$-$02 & --- & --- & & 8 & 21,440 & 2,121.02 & 3.59 \\
\midrule
\multirow{4}{*}{$3$}
& 1,044 & & 6.40E$-$01 & 2.08E$-$03 & --- & --- & & 7 & 810 & 0.32 & 0.05 \\
& 4,070 & & 2.58E$-$02 & 1.58E$-$04 & --- & --- & & 7 & 2,360 & 3.21 & 0.18 \\
& 16,398 & & 1.83E$-$03 & 9.74E$-$06 & --- & --- & & 7 & 6,870 & 86.42 & 0.85 \\
& 65,786 & & 1.20E$-$04 & 6.50E$-$07 & --- & --- & & 8 & 23,030 & 2,434.13 & 4.28 \\
\midrule
\multirow{4}{*}{$5$}
& 1,044 & & 8.27E$-$03 & 7.96E$-$05 & --- & --- & & 7 & 920 & 0.70 & 0.07 \\
& 4,070 & & 4.01E$-$04 & 1.25E$-$06 & --- & --- & & 7 & 2,550 & 5.96 & 0.27 \\
& 16,398 & & 1.51E$-$05 & 4.11E$-$08 & --- & --- & & 8 & 8,020 & 119.03 & 1.08 \\
& 65,786 & & --- & --- & --- & --- & & 8 & 24,710 & 2,877.39 & 5.53 \\
\bottomrule
\end{tabular}

}
\end{table}

In terms of computational efficiency, both the $\mathbb{P}_{d}$--$\mathbb{P}_{d+1}$ and $\mathbb{P}_{d}$--$\mathbb{P}_{d+2}$ methods in the non-primitive and primitive formulations require roughly the same number of fixed point iterations, which remains constant under mesh refinement and for higher polynomial degrees.
However, the number of GMRES iterations increases with mesh refinement and higher polynomial degrees due to larger systems of linear equations and higher coefficients matrix condition numbers, respectively.
For the same mesh, the $\mathbb{P}_{d}$--$\mathbb{P}_{d+2}$ method requires more GMRES iterations than the $\mathbb{P}_{d}$--$\mathbb{P}_{d+1}$ method since higher polynomial degrees are employed for streamfunction polynomial reconstructions in the former.
Consequently, the $\mathbb{P}_{d}$--$\mathbb{P}_{d+2}$ method also requires more execution time and leaves a slightly higher memory footprint, hence it is less efficient than the $\mathbb{P}_{d}$--$\mathbb{P}_{d+1}$ method, since both provide comparable accuracy.
For the primitive formulation, considerably more GMRES iterations are necessary than for the non-primitive formulation with the same mesh, and the difference between polynomial degrees is less significant.
Consequently, the primitive formulation requires considerably more execution time, hence it is less efficient than the non-primitive formulation since both formulations provide comparable accuracy.
It is also worth noting that the primitive formulation uses up to around four, three, and two times more memory than the non-primitive formulation with $d=1,3,5$, respectively, with the same mesh.

\begin{figure}[!htb]
\centering
\begin{tabular}{@{}c@{}c@{}}
\includegraphics[width=0.49\textwidth,trim=0cm 0cm 0cm 0cm,clip=true]{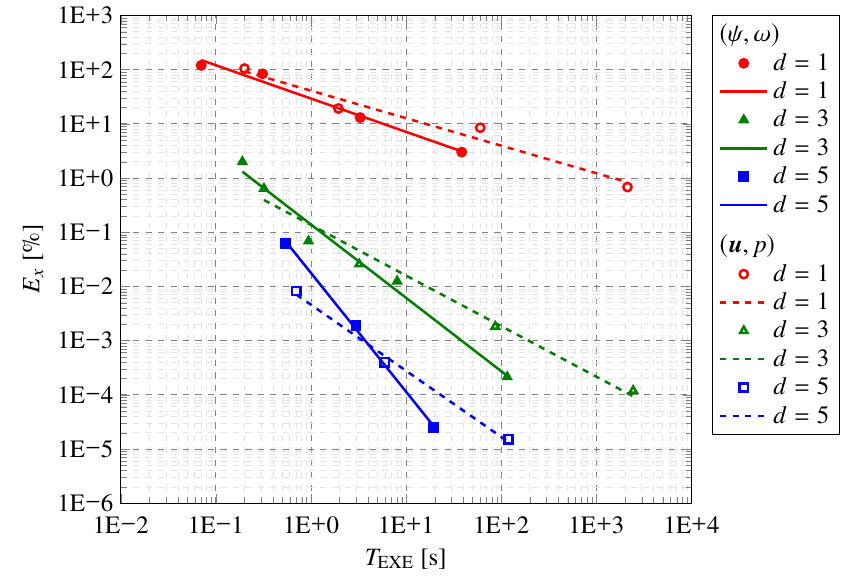} &
\includegraphics[width=0.49\textwidth,trim=0cm 0cm 0cm 0cm,clip=true]{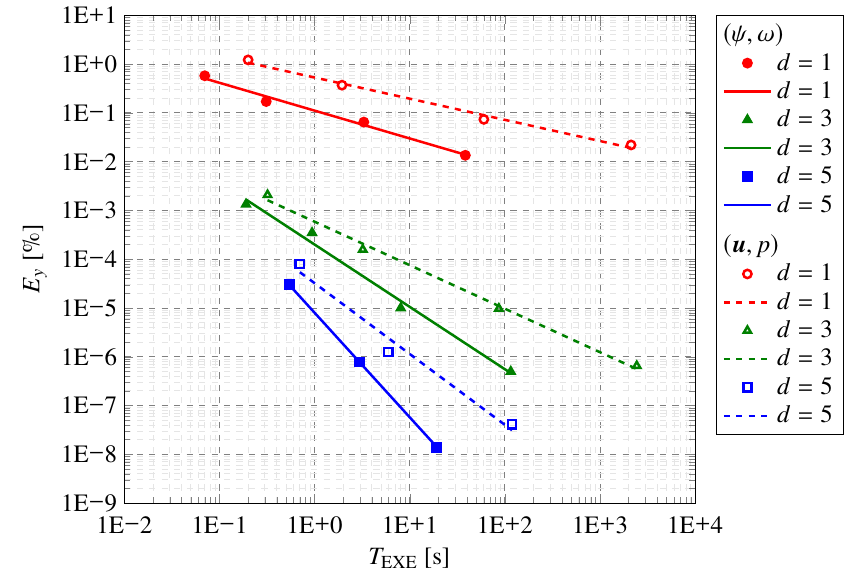}\\[0.2cm]
\small (a) $x$-coordinate. & \small (b) $y$-coordinate.\\
\includegraphics[width=0.49\textwidth,trim=0cm 0cm 0cm 0cm,clip=true]{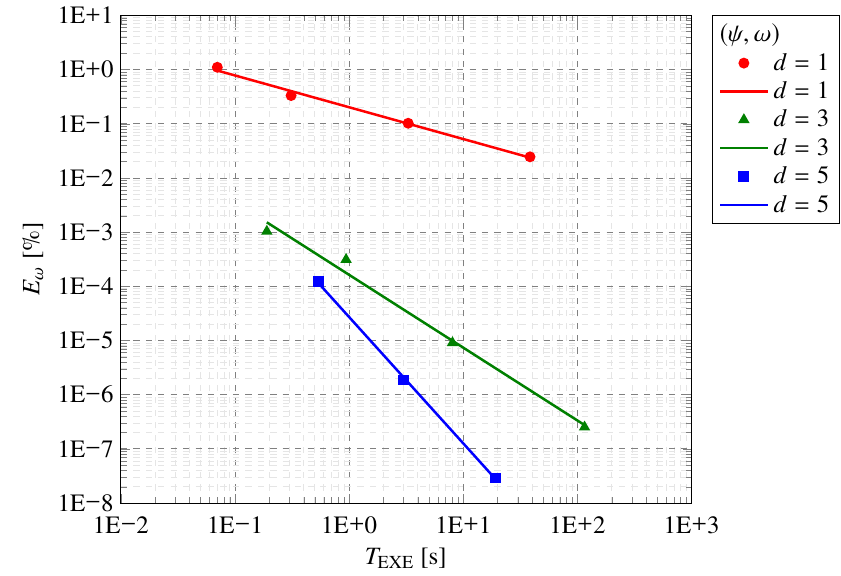} &
\includegraphics[width=0.49\textwidth,trim=0cm 0cm 0cm 0cm,clip=true]{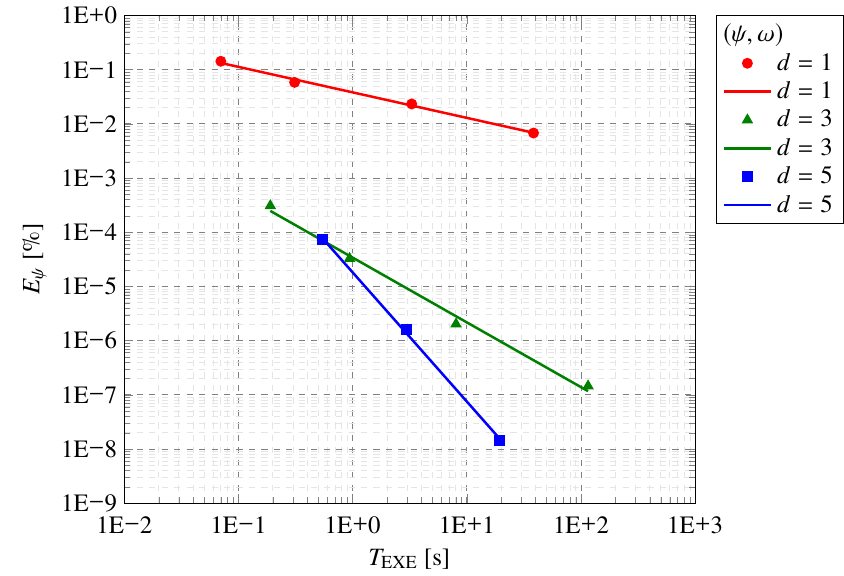}\\[0.2cm]
\small (c) Vorticity value. & \small (d) Streamfunction value.
\end{tabular}
\caption{Errors evolution as a function of execution time in the regularised semielliptical lid-driven cavity test case with $Re=1$ (for proper interpretation of the colour scale, the reader is referred to the electronic version of this article).}
\label{fig:verification_benchmark_t41_timeplot_re1}
\end{figure}

To provide better insight into the efficiency comparison, the error evolution as a function of execution time for the non-primitive formulation with the $\mathbb{P}_{d}$--$\mathbb{P}_{d+1}$ method and the primitive formulation is illustrated in Figure~\ref{fig:verification_benchmark_t41_timeplot_re1}.
Consider a target accuracy of $E_{x}=1$ \% and $E_{y}=0.01$ \% for the $x$- and $y$-coordinates of the primary vortex centre.
In that case, for the $x$-coordinate, the primitive formulation with $d=1$ requires $T_{\textrm{EXE}}\approx 1000$~s in comparison to $T_{\textrm{EXE}}\approx 100$~s for the non-primitive formulation with the same polynomial degree.
Similarly, for the $y$-coordinate, the primitive formulation with $d=1$ requires $T_{\textrm{EXE}}\approx 1500$~s, whereas the non-primitive formulation with the same polynomial degree requires $T_{\textrm{EXE}}\approx 50$~s.
Thus, the non-primitive formulation offers improvement ratios of one order of magnitude in the execution time.
Moreover, higher polynomial degrees further enhance this efficiency, as the same target accuracy is achieved in less than one second with $d=3$ in both formulations (comparable accuracy is obtained on coarser meshes), representing improvement ratios of three to four orders of magnitude in execution time compared to the primitive formulation with $d=1$.

\subsubsection{Regularised semielliptical lid-driven cavity with $Re=100$}
\label{subsec:verification_benchmark_t41_re100}

\begin{figure}[!htb]
\centering
\begin{tabular}{@{}c@{}c@{}}
\includegraphics[width=0.49\textwidth,trim=0cm 0cm 0cm 0cm,clip=true]{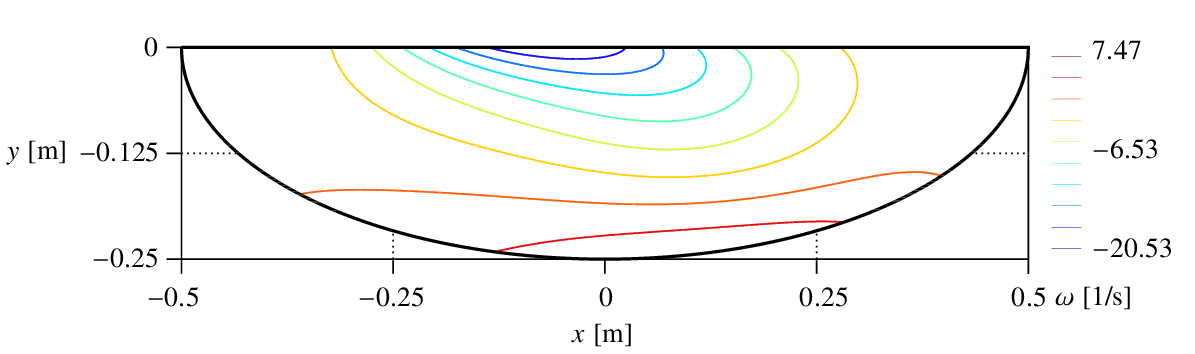} &
\includegraphics[width=0.49\textwidth,trim=0cm 0cm 0cm 0cm,clip=true]{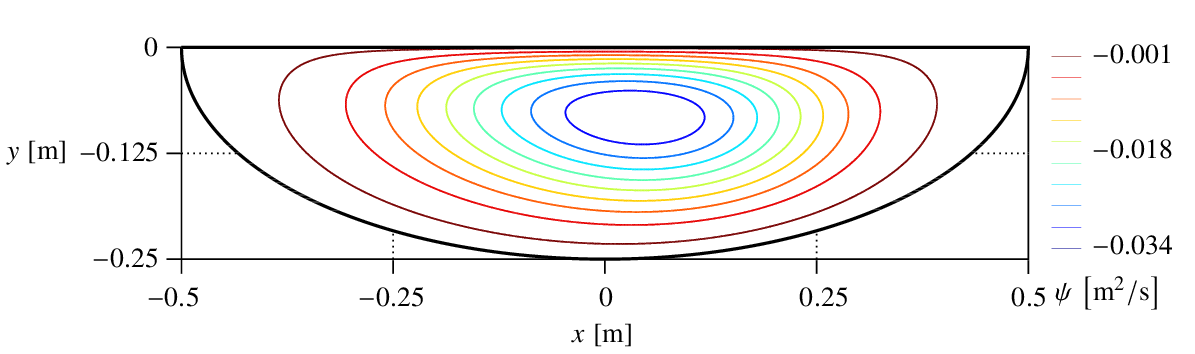}\\[0.2cm]
\small (a) Vorticity contours. & \small (b) Streamlines.
\end{tabular}
\caption{Reference solutions for the regularised semielliptical lid-driven cavity test case with $Re=100$ (for proper interpretation of the colour scale, the reader is referred to the electronic version of this article).}
\label{fig:verification_benchmark_t41_re100}
\end{figure}

Figure~\ref{fig:verification_benchmark_t41_re100} illustrates the reference solutions for the vorticity and streamfunction computed on the finest mesh with $d=5$.

The errors for the primary vortex centre coordinates, vorticity, and streamfunction are provided in Table~\ref{tab:verification_benchmark_t41_errorstable_re100_rod} when the ROD method is used to impose the prescribed boundary conditions.
In terms of accuracy and convergence, the $\mathbb{P}_{d}$--$\mathbb{P}_{d+1}$ and $\mathbb{P}_{d}$--$\mathbb{P}_{d+2}$ methods in the non-primitive formulation follow the same behaviour as for the previous test case with $Re=1$, with the latter generally providing better accuracy.
Conversely, while the primitive formulation in the previous test case provided accuracy improvement ratios always below an order of magnitude compared to the non-primitive formulation, it appears that increasing the Reynolds numbers favours the former formulation.
Indeed, improvement ratios up to around $20$ and $10$ are observed for the $\mathbb{P}_{d}$--$\mathbb{P}_{d+1}$ and $\mathbb{P}_{d}$--$\mathbb{P}_{d+2}$ methods, respectively, especially for high polynomial degrees.
In terms of computational efficiency, the $\mathbb{P}_{d}$--$\mathbb{P}_{d+2}$ method requires more GMRES iterations than the $\mathbb{P}_{d}$--$\mathbb{P}_{d+1}$ method with the same mesh, but the same number of fixed point iterations, following the same behaviour as previously with $Re=1$.
The main difference arises in the comparison with the primitive formulation, which requires fewer fixed point iterations than the non-primitive formulation, favouring its efficiency.


\begin{table}[!htb]
\centering
\footnotesize
\caption{Errors obtained with the ROD method in the regularised semielliptical lid-driven cavity test case with $Re=100$.}
\label{tab:verification_benchmark_t41_errorstable_re100_rod}
\resizeboxlarger{
\begin{tabular}{@{}rr@{}r@{}rrrr@{}@{}r@{}rrrr@{}}
\toprule
$d$ & $N_{\textrm{C}}$ & \phantom{aaa} & $E_{x}$ [\%] & $E_{y}$ [\%] & $E_{\omega}$ [\%] & $E_{\psi}$ [\%] & \phantom{aaa} & $N_{\textrm{FP}}$ & $N_{\textrm{GMRES}}$ & $T_{\textrm{EXE}}$ [s] & $M_{\textrm{USE}}$ [Gb] \\
\midrule
\multicolumn{11}{@{}l}{$\left(\psi,\omega\right)$/$\mathbb{P}_{d}$--$\mathbb{P}_{d+1}$} \\
\midrule
\multirow{4}{*}{$1$}
& 1,044 & & 4.02E$+$00 & 9.88E$-$01 & 1.73E$+$00 & 2.34E$-$01 & & 21 & 960 & 0.15 & 0.02 \\
& 4,070 & & 1.17E$+$00 & 2.63E$-$01 & 5.07E$-$01 & 1.19E$-$01 & & 22 & 1,880 & 0.83 & 0.06 \\
& 16,398 & & 4.44E$-$01 & 6.11E$-$02 & 1.25E$-$01 & 3.73E$-$02 & & 23 & 3,700 & 9.69 & 0.22 \\
& 65,786 & & 6.25E$-$02 & 1.54E$-$02 & 3.18E$-$02 & 9.99E$-$03 & & 23 & 8,360 & 108.67 & 0.85 \\
\midrule
\multirow{4}{*}{$3$}
& 1,044 & & 3.56E$-$01 & 7.65E$-$03 & 1.01E$-$01 & 3.37E$-$02 & & 22 & 1,450 & 0.42 & 0.04 \\
& 4,070 & & 1.78E$-$02 & 1.19E$-$03 & 2.66E$-$03 & 1.04E$-$03 & & 22 & 2,890 & 2.05 & 0.10 \\
& 16,398 & & 1.40E$-$03 & 1.02E$-$04 & 1.38E$-$04 & 6.34E$-$05 & & 23 & 6,330 & 22.19 & 0.36 \\
& 65,786 & & 7.07E$-$05 & 9.00E$-$06 & 7.12E$-$06 & 3.02E$-$06 & & 25 & 15,500 & 266.50 & 1.35 \\
\midrule
\multirow{4}{*}{$5$}
& 1,044 & & 1.72E$-$02 & 2.44E$-$03 & 8.74E$-$03 & 2.80E$-$03 & & 22 & 2,040 & 0.85 & 0.05 \\
& 4,070 & & 6.80E$-$04 & 3.89E$-$05 & 1.45E$-$04 & 7.81E$-$05 & & 22 & 4,400 & 4.67 & 0.16 \\
& 16,398 & & 9.80E$-$06 & 9.69E$-$07 & 1.20E$-$06 & 1.13E$-$06 & & 24 & 10,420 & 51.22 & 0.60 \\
& 65,786 & & --- & --- & --- & --- & & 24 & 19,940 & 456.98 & 2.27 \\
\midrule
\multicolumn{11}{@{}l}{$\left(\psi,\omega\right)$/$\mathbb{P}_{d}$--$\mathbb{P}_{d+2}$} \\
\midrule
\multirow{4}{*}{$1$}
& 1,044 & & 2.64E$+$00 & 8.31E$-$01 & 1.14E$+$00 & 3.98E$-$02 & & 23 & 1,190 & 0.24 & 0.02 \\
& 4,070 & & 7.93E$-$02 & 2.20E$-$01 & 2.66E$-$01 & 1.29E$-$02 & & 22 & 2,620 & 1.31 & 0.06 \\
& 16,398 & & 1.91E$-$02 & 2.89E$-$02 & 4.43E$-$02 & 5.06E$-$03 & & 23 & 5,210 & 13.84 & 0.24 \\
& 65,786 & & 1.39E$-$02 & 8.82E$-$03 & 1.17E$-$02 & 9.14E$-$04 & & 24 & 12,440 & 175.91 & 0.94 \\
\midrule
\multirow{4}{*}{$3$}
& 1,044 & & 2.31E$-$01 & 2.77E$-$03 & 2.31E$-$02 & 1.54E$-$02 & & 22 & 1,680 & 0.48 & 0.04 \\
& 4,070 & & 8.75E$-$03 & 4.69E$-$04 & 1.88E$-$03 & 7.62E$-$04 & & 23 & 3,520 & 2.86 & 0.11 \\
& 16,398 & & 8.15E$-$04 & 2.89E$-$05 & 8.46E$-$05 & 1.90E$-$05 & & 23 & 8,260 & 31.14 & 0.40 \\
& 65,786 & & 2.15E$-$05 & 3.24E$-$06 & 1.43E$-$06 & 1.77E$-$06 & & 24 & 19,141 & 356.02 & 1.51 \\
\midrule
\multirow{4}{*}{$5$}
& 1,044 & & 2.60E$-$02 & 1.38E$-$03 & 8.33E$-$03 & 3.83E$-$03 & & 22 & 2,350 & 1.36 & 0.07 \\
& 4,070 & & 4.87E$-$04 & 4.10E$-$05 & 1.19E$-$04 & 5.04E$-$05 & & 23 & 5,210 & 6.01 & 0.18 \\
& 16,398 & & 3.22E$-$06 & 8.00E$-$07 & 1.07E$-$06 & 4.17E$-$07 & & 24 & 11,540 & 61.76 & 0.68 \\
& 65,786 & & --- & --- & --- & --- & & 24 & 24,100 & 591.23 & 2.59 \\
\midrule
\multicolumn{11}{@{}l}{$\left(\bm{u},p\right)$} \\
\midrule
\multirow{4}{*}{$1$}
& 1,044 & & 5.68E$+$00 & 9.85E$-$01 & --- & --- & & 12 & 1,560 & 0.30 & 0.04 \\
& 4,070 & & 9.57E$-$01 & 2.48E$-$01 & --- & --- & & 12 & 3,180 & 2.95 & 0.13 \\
& 16,398 & & 1.84E$-$01 & 6.65E$-$02 & --- & --- & & 13 & 10,020 & 108.43 & 0.66 \\
& 65,786 & & 1.60E$-$02 & 1.77E$-$02 & --- & --- & & 13 & 28,990 & 2,840.35 & 3.59 \\
\midrule
\multirow{4}{*}{$3$}
& 1,044 & & 7.08E$-$02 & 2.39E$-$03 & --- & --- & & 11 & 1,430 & 0.31 & 0.05 \\
& 4,070 & & 7.07E$-$03 & 5.44E$-$05 & --- & --- & & 12 & 3,340 & 4.45 & 0.18 \\
& 16,398 & & 3.40E$-$04 & 6.19E$-$06 & --- & --- & & 12 & 10,420 & 130.03 & 0.85 \\
& 65,786 & & 2.74E$-$05 & 1.18E$-$06 & --- & --- & & 13 & 34,340 & 3,557.23 & 4.28 \\
\midrule
\multirow{4}{*}{$5$}
& 1,044 & & 7.66E$-$03 & 3.70E$-$04 & --- & --- & & 11 & 1,520 & 0.88 & 0.07 \\
& 4,070 & & 2.98E$-$04 & 5.64E$-$06 & --- & --- & & 12 & 3,991 & 8.37 & 0.27 \\
& 16,398 & & 4.44E$-$06 & 1.61E$-$07 & --- & --- & & 13 & 11,710 & 176.37 & 1.15 \\
& 65,786 & & --- & --- & --- & --- & & 13 & 35,060& 4,006.00 & 5.53 \\
\bottomrule
\end{tabular}

}
\end{table}

\begin{figure}[!htb]
\centering
\begin{tabular}{@{}c@{}c@{}}
\includegraphics[width=0.49\textwidth,trim=0cm 0cm 0cm 0cm,clip=true]{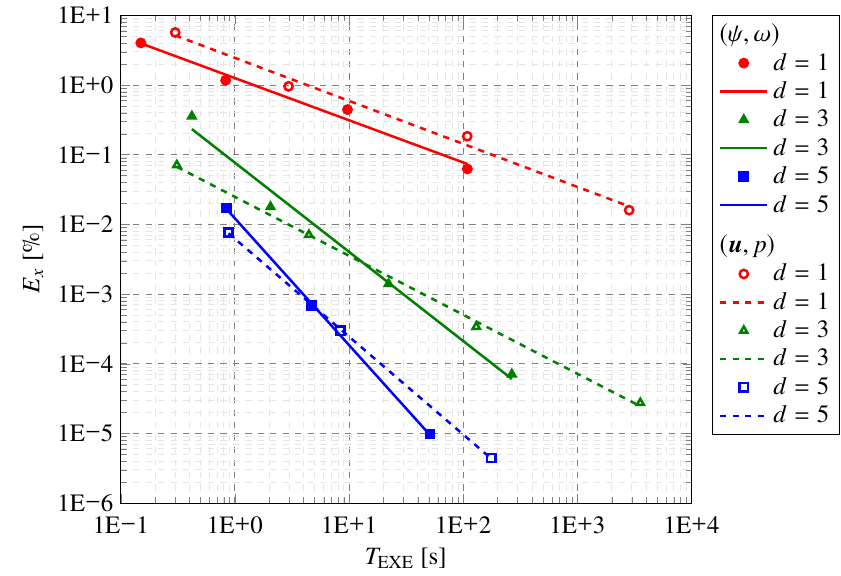} &
\includegraphics[width=0.49\textwidth,trim=0cm 0cm 0cm 0cm,clip=true]{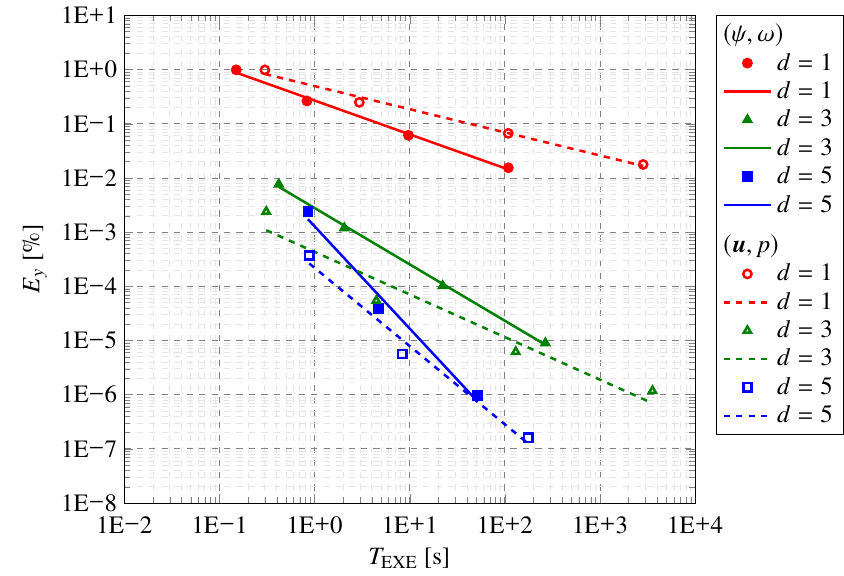}\\[0.2cm]
\small (a) $x$-coordinate. & \small (b) $y$-coordinate.\\
\includegraphics[width=0.49\textwidth,trim=0cm 0cm 0cm 0cm,clip=true]{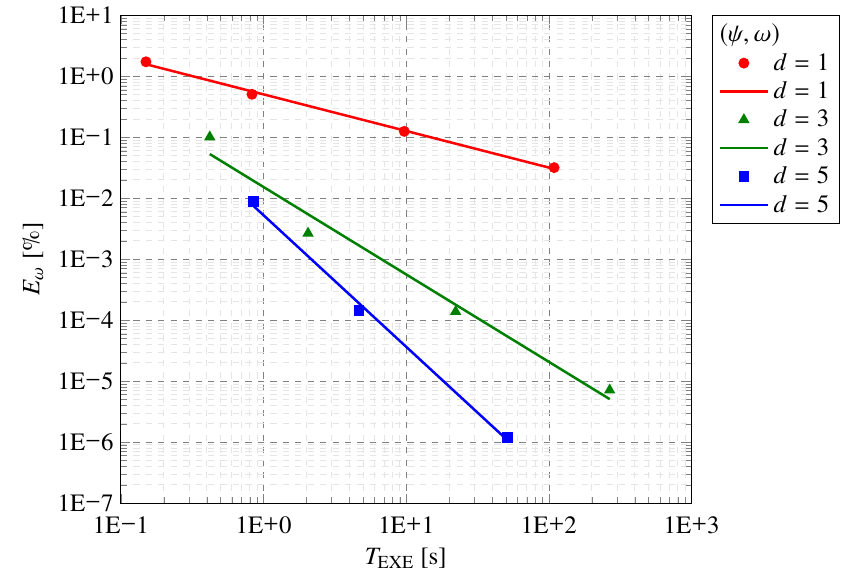} &
\includegraphics[width=0.49\textwidth,trim=0cm 0cm 0cm 0cm,clip=true]{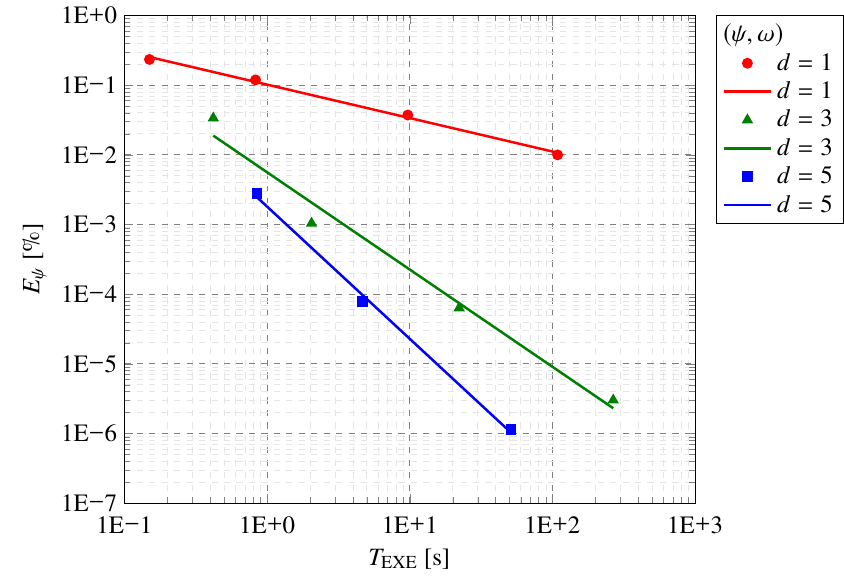}\\[0.2cm]
\small (c) Vorticity value. & \small (d) Streamfunction value.
\end{tabular}
\caption{Errors evolution as a function of execution time in the regularised semielliptical lid-driven cavity test case with $Re=100$ (for proper interpretation of the colour scale, the reader is referred to the electronic version of this article).}
\label{fig:verification_benchmark_t41_timeplot_re100}
\end{figure}

To provide better insight into the efficiency comparison, the evolution of the error as a function of execution time for the non-primitive formulation with the $\mathbb{P}_{d}$--$\mathbb{P}_{d+1}$ method and the primitive formulation is illustrated in Figure~\ref{fig:verification_benchmark_t41_timeplot_re100}.
As observed, although the errors for the primary vortex centre coordinates decrease faster as a function of execution time in non-primitive formulation, the crossover point only occurs at significantly lower error levels with $d=3,5$, especially for the $y$-coordinate.
Indeed, both the improved accuracy and the faster fixed point convergence of the primitive formulation compared to the non-primitive formulation, contribute to better computational performance with high polynomial degrees.
In conclusion, the non-primitive formulation can still enhance computational efficiency over the primitive formulation, but only when a very high accuracy for the primary vortex centre is required or with $d=1$.

\subsubsection{Conventional semielliptical lid-driven cavity with $Re=1$}
\label{subsec:verification_benchmark_t42_re1}

\begin{figure}[!htb]
\centering
\begin{tabular}{@{}c@{}c@{}}
\includegraphics[width=0.49\textwidth,trim=0cm 0cm 0cm 0cm,clip=true]{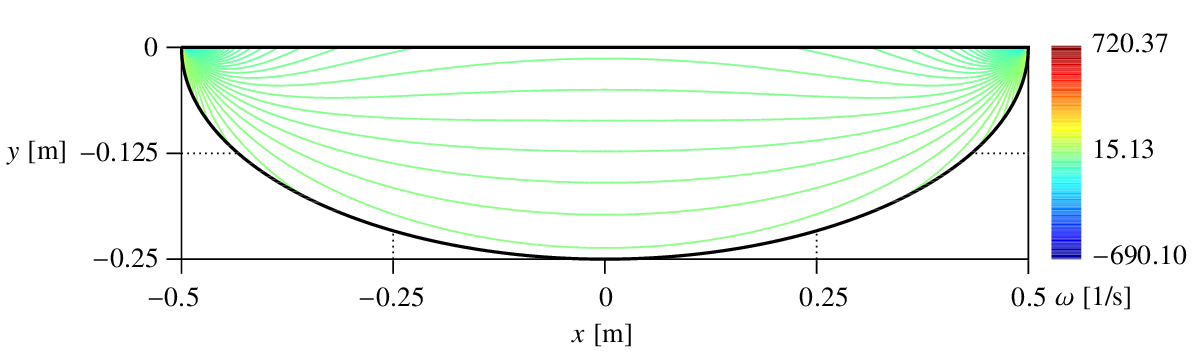} &
\includegraphics[width=0.49\textwidth,trim=0cm 0cm 0cm 0cm,clip=true]{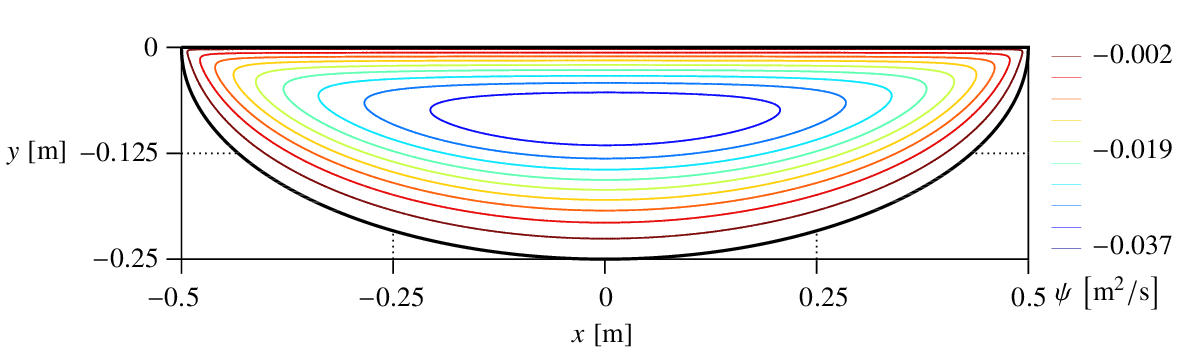}\\[0.2cm]
\small (a) Vorticity contours. & \small (b) Streamlines.
\end{tabular}
\caption{Reference solutions for the conventional semielliptical lid-driven cavity test case with $Re=1$ (for proper interpretation of the colour scale, the reader is referred to the electronic version of this article).}
\label{fig:verification_benchmark_t42_re1}
\end{figure}

Figure~\ref{fig:verification_benchmark_t42_re1} illustrates the reference solutions for the vorticity and streamfunction computed on the finest mesh with $d=5$.

The errors for the primary vortex centre coordinates, vorticity, and streamfunction are provided in Table~\ref{tab:verification_benchmark_t42_errorstable_re1_rod} when the ROD method is employed to impose the prescribed boundary conditions.
In terms of accuracy and convergence, the $\mathbb{P}_{d}$--$\mathbb{P}_{d+1}$ method in the non-primitive formulation follows essentially the same behaviour as for the regularised test case, except $d=5$ that converges at almost the same rate as $d=3$ and, therefore, offers little to no accuracy improvement.
This behaviour is linked to the singularity points at the cavity corners, producing non-physical oscillations in the approximate solution, which exacerbate as the polynomial degree increases.
Nevertheless, polynomial reconstructions with $d=3$ do not appear to be affected, hence the same convergence behaviour as for the regularised test case is observed.
For the $\mathbb{P}_{d}$--$\mathbb{P}_{d+2}$ method, omitted for compactness, the same consequence is observed with $d=5$, whereas with $d=1,3$ the same observations as for the regularised test case apply.
Conversely, the singularity issue has a more pronounced impact in the primitive formulation, with $d=3,5$ exhibiting the same convergence rates as $d=1$ for the error in the primary vortex centre $x$-coordinate, and even lower convergence rates and higher errors for the primary vortex centre $y$-coordinate.
As such, the numerical accuracy benefit of using higher polynomial degrees is diminished, making the primitive formulation more computationally efficient with $d=1$ than with $d=3,5$.
Indeed, it appears that the non-primitive formulation is more robust and proficient in withstanding the non-physical oscillations produced at the cavity corners than the primitive formulation.


\begin{table}[!htb]
\centering
\footnotesize
\caption{Errors obtained with the ROD method in the conventional semielliptical lid-driven cavity test case with $Re=1$.}
\label{tab:verification_benchmark_t42_errorstable_re1_rod}
\resizeboxlarger{
\begin{tabular}{@{}rr@{}r@{}rrrr@{}@{}r@{}rrrr@{}}
\toprule
$d$ & $N_{\textrm{C}}$ & \phantom{aaa} & $E_{x}$ [\%] & $E_{y}$ [\%] & $E_{\omega}$ [\%] & $E_{\psi}$ [\%] & \phantom{aaa} & $N_{\textrm{FP}}$ & $N_{\textrm{GMRES}}$ & $T_{\textrm{EXE}}$ [s] & $M_{\textrm{USE}}$ [Gb] \\
\midrule
\multicolumn{11}{@{}l}{$\left(\psi,\omega\right)$/$\mathbb{P}_{d}$--$\mathbb{P}_{d+1}$} \\
\midrule
\multirow{4}{*}{$1$}
& 1,044 & & 8.72E$+$02 & 8.63E$-$01 & 1.26E$+$00 & 1.55E$-$01 & & 7 & 230 & 0.07 & 0.02 \\
& 4,070 & & 3.16E$+$02 & 2.05E$-$01 & 3.12E$-$01 & 5.80E$-$02 & & 7 & 480 & 0.31 & 0.06 \\
& 16,398 & & 2.57E$+$01 & 6.03E$-$02 & 8.78E$-$02 & 1.50E$-$02 & & 8 & 1,070 & 3.17 & 0.22 \\
& 65,786 & & 7.87E$+$00 & 1.94E$-$02 & 2.64E$-$02 & 3.73E$-$03 & & 9 & 2,580 & 35.21 & 0.85 \\
\midrule
\multirow{4}{*}{$3$}
& 1,044 & & 3.91E$+$00 & 6.78E$-$04 & 6.42E$-$04 & 7.13E$-$05 & & 7 & 400 & 0.22 & 0.03 \\
& 4,070 & & 6.37E$-$02 & 8.41E$-$05 & 8.79E$-$05 & 9.83E$-$06 & & 8 & 810 & 0.90 & 0.10 \\
& 16,398 & & 1.93E$-$03 & 3.22E$-$06 & 2.39E$-$06 & 4.43E$-$07 & & 8 & 1,820 & 7.57 & 0.36 \\
& 65,786 & & 1.53E$-$04 & 9.29E$-$08 & 1.44E$-$07 & 4.51E$-$09 & & 9 & 5,870 & 106.31 & 1.35 \\
\midrule
\multirow{4}{*}{$5$}
& 1,044 & & 1.13E$-$01 & 8.88E$-$05 & 2.31E$-$04 & 5.46E$-$05 & & 7 & 440 & 0.69 & 0.06 \\
& 4,070 & & 2.85E$-$02 & 9.75E$-$07 & 3.52E$-$06 & 1.28E$-$06 & & 8 & 990 & 2.94 & 0.20 \\
& 16,398 & & 1.69E$-$03 & 1.50E$-$07 & 6.64E$-$07 & 1.54E$-$07 & & 9 & 2,280 & 19.84 & 0.72 \\
& 65,786 & & --- & --- & --- & --- & & 9 & 7,320 & 183.97 & 2.27 \\
\midrule
\multicolumn{11}{@{}l}{$\left(\bm{u},p\right)$} \\
\midrule
& 1,044 & & 2.02E$+$02 & 1.16E$+$00 & --- & --- & & 7 & 840 & 0.19 & 0.04 \\
& 4,070 & & 9.18E$+$01 & 3.77E$-$01 & --- & --- & & 7 & 2,090 & 2.09 & 0.13 \\
& 16,398 & & 3.41E$+$01 & 7.74E$-$02 & --- & --- & & 7 & 5,980 & 65.33 & 0.66 \\
& 65,786 & & 6.85E$+$00 & 1.44E$-$02 & --- & --- & & 8 & 18,160 & 1,763.87 & 3.59 \\
\midrule
\multirow{4}{*}{$3$}
& 1,044 & & 2.90E$+$00 & 2.23E$+$00 & --- & --- & & 6 & 730 & 0.30 & 0.05 \\
& 4,070 & & 3.87E$-$01 & 1.12E$+$00 & --- & --- & & 7 & 2,310 & 3.20 & 0.18 \\
& 16,398 & & 1.75E$-$01 & 5.75E$-$01 & --- & --- & & 7 & 6,910 & 88.17 & 0.85 \\
& 65,786 & & 3.45E$-$02 & 2.68E$-$01 & --- & --- & & 8 & 20,420 & 2,181.51 & 4.41 \\
\midrule
\multirow{4}{*}{$5$}
& 1,044 & & 2.36E$+$01 & 1.96E$+$00 & --- & --- & & 7 & 2,030 & 0.82 & 0.06 \\
& 4,070 & & 6.09E$+$00 & 1.22E$+$00 & --- & --- & & 7 & 2,430 & 6.38 & 0.30 \\
& 16,398 & & 2.42E$+$00 & 6.38E$-$01 & --- & --- & & 7 & 7,970 & 126.64 & 1.15 \\
& 65,786 & & 4.73E$-$01 & 3.13E$-$01 & --- & --- & & 8 & 25,130 & 2,987.98 & 5.90 \\
\bottomrule
\end{tabular}

}
\end{table}

\begin{figure}[!htb]
\centering
\begin{tabular}{@{}c@{}c@{}}
\includegraphics[width=0.49\textwidth,trim=0cm 0cm 0cm 0cm,clip=true]{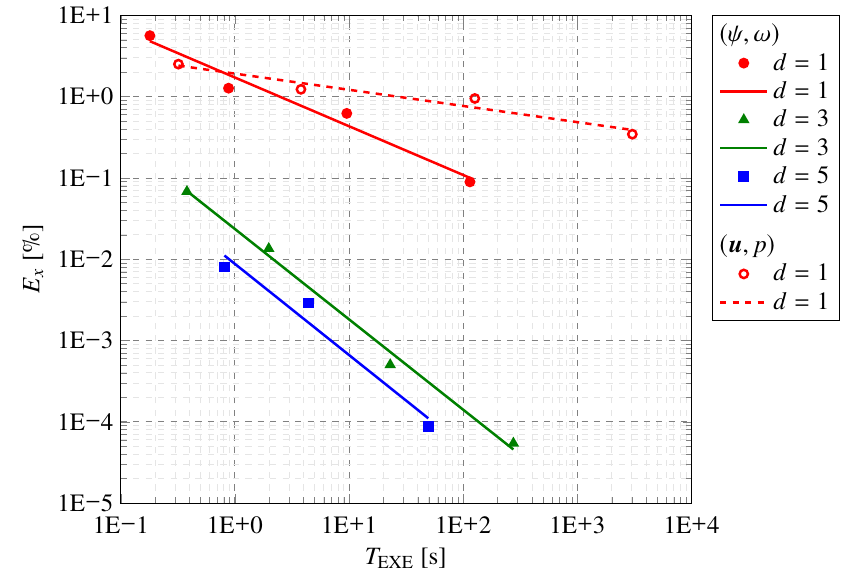} &
\includegraphics[width=0.49\textwidth,trim=0cm 0cm 0cm 0cm,clip=true]{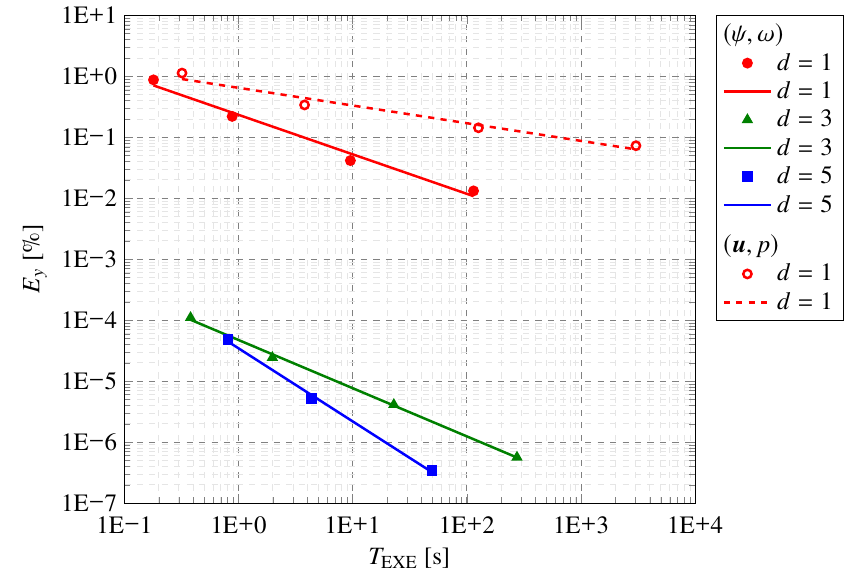}\\[0.2cm]
\small (a) $x$-coordinate. & \small (b) $y$-coordinate.\\
\includegraphics[width=0.49\textwidth,trim=0cm 0cm 0cm 0cm,clip=true]{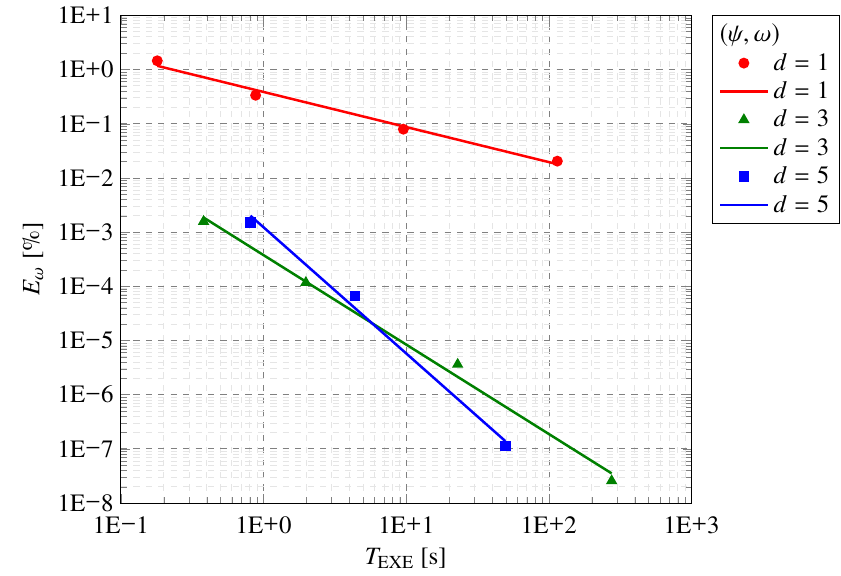} &
\includegraphics[width=0.49\textwidth,trim=0cm 0cm 0cm 0cm,clip=true]{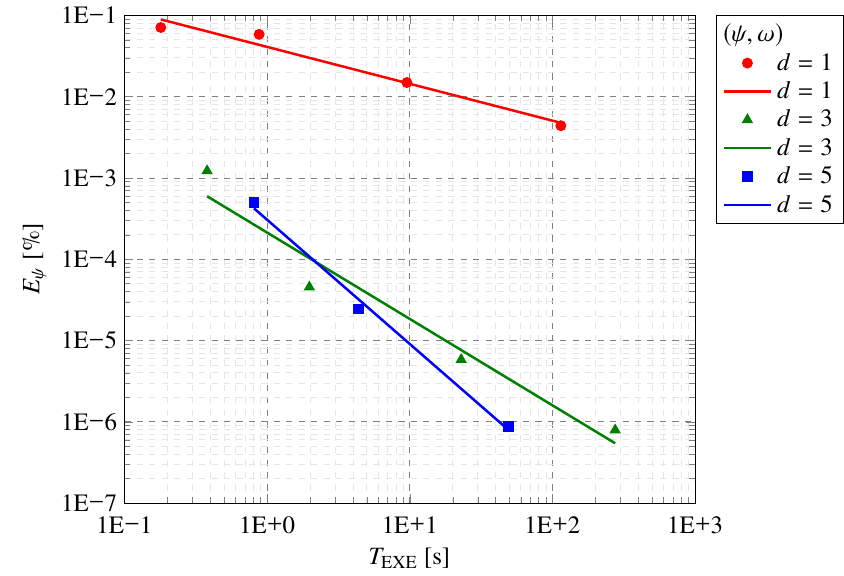}\\[0.2cm]
\small (c) Vorticity value. & \small (d) Streamfunction value.
\end{tabular}
\caption{Errors evolution as a function of execution time in the conventional semielliptical lid-driven cavity test case with $Re=100$ (for proper interpretation of the colour scale, the reader is referred to the electronic version of this article).}
\label{fig:verification_benchmark_t42_timeplot_re100}
\end{figure}

To provide better insight into the efficiency comparison, the evolution of the error as a function of execution time for the non-primitive formulation with the $\mathbb{P}_{d}$--$\mathbb{P}_{d+1}$ method and the primitive formulation is shown in Figure~\ref{fig:verification_benchmark_t42_timeplot_re100}.
For the primitive formulation, only $d=1$ is reported.
As observed, both $d=3,5$ offer the same computational efficiency for the non-primitive formulation, and are significantly superior to the primitive formulation with $d=1$.

\begin{figure}[!htb]
\centering
\begin{tabular}{@{}c@{}c@{}}
\includegraphics[width=0.49\textwidth,trim=0cm 0cm 0cm 0cm,clip=true]{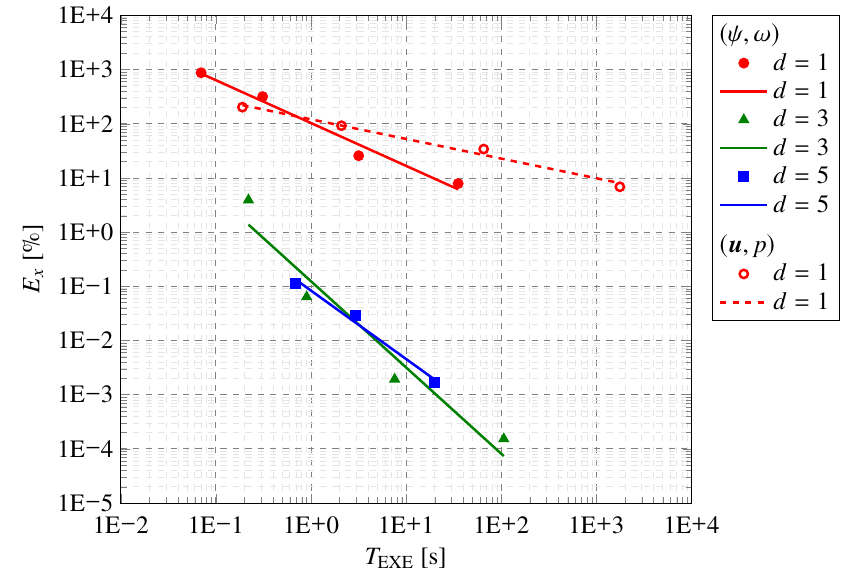} &
\includegraphics[width=0.49\textwidth,trim=0cm 0cm 0cm 0cm,clip=true]{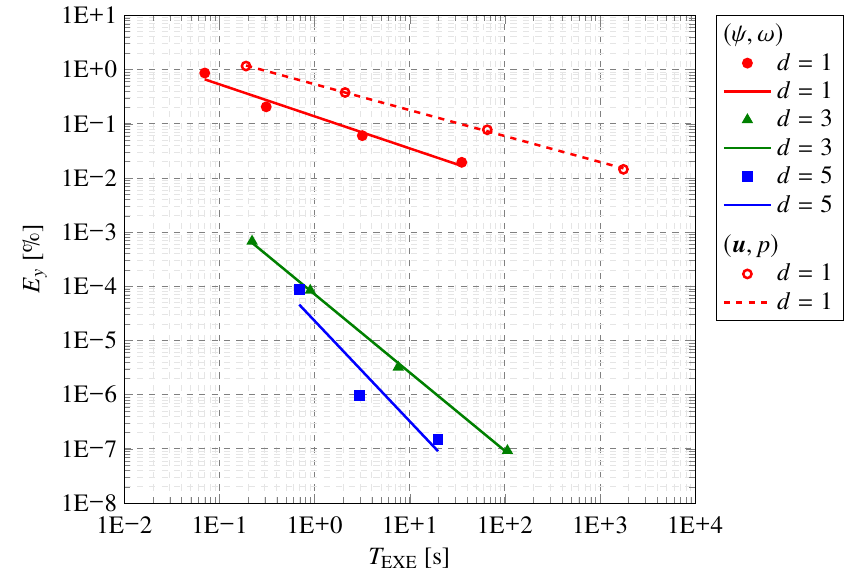}\\[0.2cm]
\small (a) $x$-coordinate. & \small (b) $y$-coordinate.\\
\includegraphics[width=0.49\textwidth,trim=0cm 0cm 0cm 0cm,clip=true]{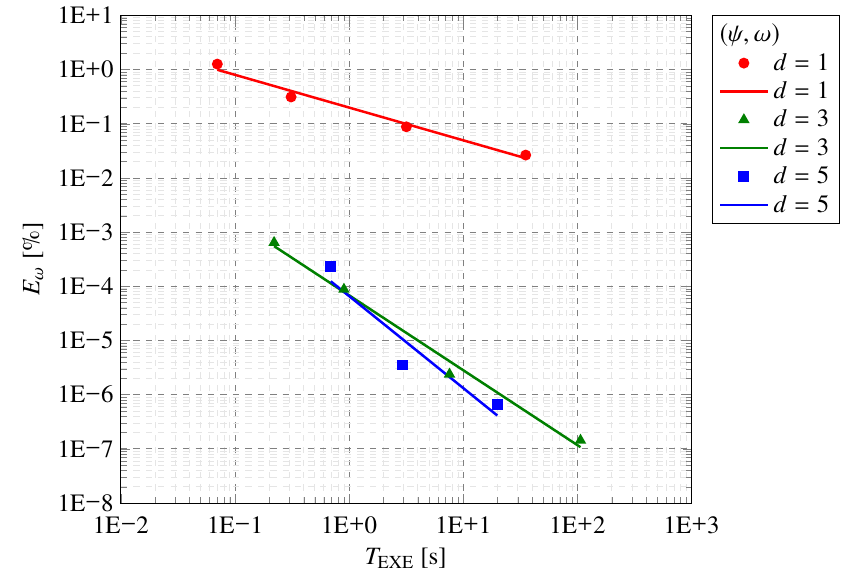} &
\includegraphics[width=0.49\textwidth,trim=0cm 0cm 0cm 0cm,clip=true]{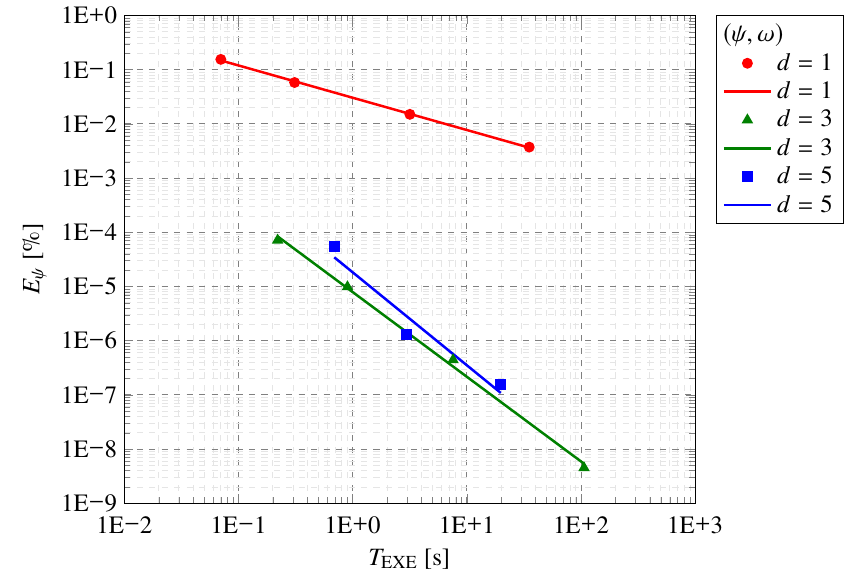}\\[0.2cm]
\small (c) Vorticity value. & \small (d) Streamfunction value.
\end{tabular}
\caption{Errors evolution as a function of execution time in the conventional semielliptical lid-driven cavity test case with $Re=1$ (for proper interpretation of the colour scale, the reader is referred to the electronic version of this article).}
\label{fig:verification_benchmark_t42_timeplot_re1}
\end{figure}


\subsubsection{Conventional semielliptical lid-driven cavity with $Re=100$}
\label{subsec:verification_benchmark_t42_re100}

\begin{figure}[!htb]
\centering
\begin{tabular}{@{}c@{}c@{}}
\includegraphics[width=0.49\textwidth,trim=0cm 0cm 0cm 0cm,clip=true]{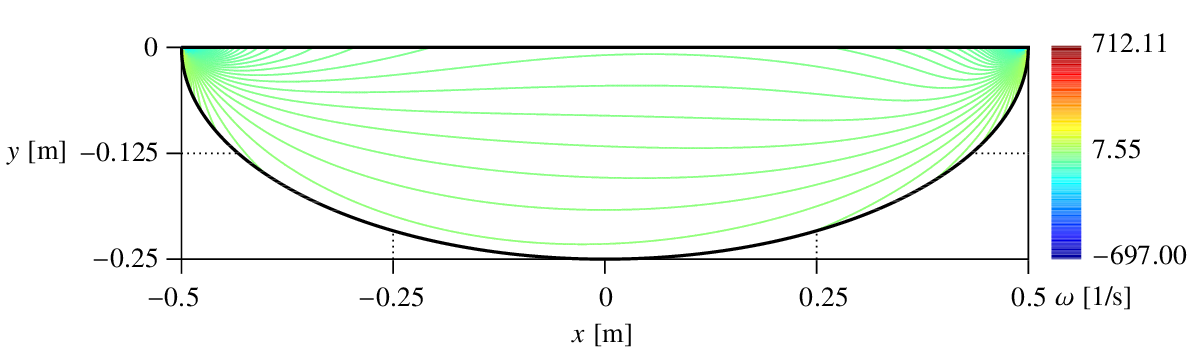} &
\includegraphics[width=0.49\textwidth,trim=0cm 0cm 0cm 0cm,clip=true]{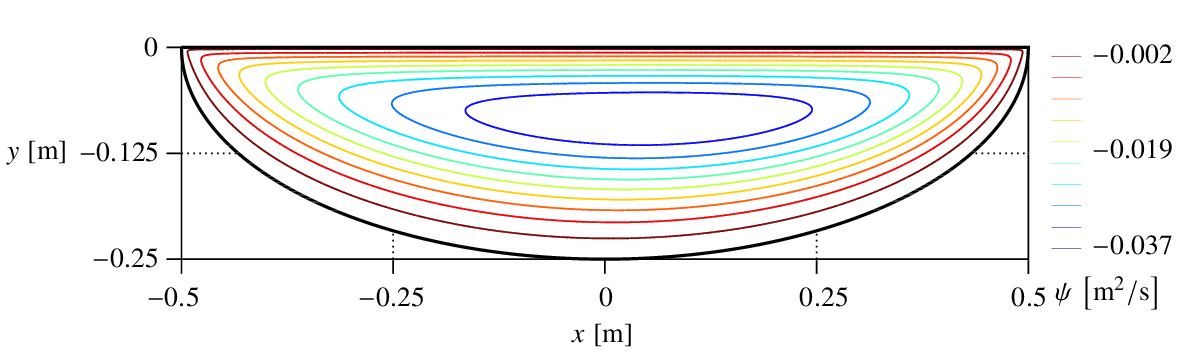}\\[0.2cm]
\small (a) Vorticity contours. & \small (b) Streamlines.
\end{tabular}
\caption{Reference solutions for the conventional semielliptical lid-driven cavity test case with $Re=100$ (for proper interpretation of the colour scale, the reader is referred to the electronic version of this article).}
\label{fig:verification_benchmark_t42_re100}
\end{figure}

Figure~\ref{fig:verification_benchmark_t42_re100} illustrates the associated reference solutions for the vorticity and streamfunction computed on the finest mesh with $d=5$.

The errors for the primary vortex centre coordinates, vorticity, and streamfunction are provided in Table~\ref{tab:verification_benchmark_t42_errorstable_re100_rod} when the ROD method is used to impose the prescribed boundary conditions.
Figure~\ref{fig:verification_benchmark_t41_timeplot_re100} provides a better insight into the efficiency comparison, showing errors evolution as a function of execution time for the non-primitive formulation with the $\mathbb{P}_{d}$--$\mathbb{P}_{d+1}$ method and the primitive formulation with $d=1$.
This final test case supports the conclusions drawn from the previous test case with $Re=1$, indicating that the primitive formulation is only effective with $d=1$, whereas the non-primitive formulation continues to provide the expected convergence rates and accuracy improvement with $d=3$.
Hence, it also becomes more computationally efficient, either with $d=1$ and $d=3$, when compared to the primitive formulation.


\begin{table}[!htb]
\centering
\footnotesize
\caption{Errors obtained with the ROD method in the conventional semielliptical lid-driven cavity test case with $Re=100$.}
\label{tab:verification_benchmark_t42_errorstable_re100_rod}
\resizeboxlarger{
\begin{tabular}{@{}rr@{}r@{}rrrr@{}@{}r@{}rrrr@{}}
\toprule
$d$ & $N_{\textrm{C}}$ & \phantom{aaa} & $E_{x}$ [\%] & $E_{y}$ [\%] & $E_{\omega}$ [\%] & $E_{\psi}$ [\%] & \phantom{aaa} & $N_{\textrm{FP}}$ & $N_{\textrm{GMRES}}$ & $T_{\textrm{EXE}}$ [s] & $M_{\textrm{USE}}$ [Gb] \\
\midrule
\multicolumn{11}{@{}l}{$\left(\psi,\omega\right)$/$\mathbb{P}_{d}$--$\mathbb{P}_{d+1}$} \\
\midrule
\multirow{4}{*}{$1$}
& 1,044 & & 5.62E$+$00 & 8.74E$-$01 & 1.45E$+$00 & 7.09E$-$02 & & 24 & 1,130 & 0.18 & 0.02 \\
& 4,070 & & 1.27E$+$00 & 2.20E$-$01 & 3.35E$-$01 & 5.83E$-$02 & & 24 & 2,030 & 0.88 & 0.06 \\
& 16,398 & & 6.22E$-$01 & 4.15E$-$02 & 7.93E$-$02 & 1.49E$-$02 & & 24 & 3,560 & 9.55 & 0.22 \\
& 65,786 & & 8.94E$-$02 & 1.32E$-$02 & 2.04E$-$02 & 4.39E$-$03 & & 26 & 8,770 & 114.51 & 0.85 \\
\midrule
\multirow{4}{*}{$3$}
& 1,044 & & 6.79E$-$02 & 1.10E$-$04 & 1.57E$-$03 & 1.22E$-$03 & & 24 & 1,570 & 0.38 & 0.03 \\
& 4,070 & & 1.35E$-$02 & 2.44E$-$05 & 1.17E$-$04 & 4.54E$-$05 & & 24 & 2,720 & 1.98 & 0.11 \\
& 16,398 & & 5.03E$-$04 & 4.12E$-$06 & 3.62E$-$06 & 5.81E$-$06 & & 25 & 6,730 & 22.93 & 0.36 \\
& 65,786 & & 5.45E$-$05 & 5.64E$-$07 & 2.57E$-$08 & 7.92E$-$07 & & 26 & 15,980 & 275.02 & 1.35 \\
\midrule
\multirow{4}{*}{$5$}
& 1,044 & & 8.02E$-$03 & 4.76E$-$05 & 1.48E$-$03 & 4.96E$-$04 & & 23 & 1,971 & 0.81 & 0.05 \\
& 4,070 & & 2.90E$-$03 & 5.25E$-$06 & 6.60E$-$05 & 2.44E$-$05 & & 26 & 3,030 & 4.38 & 0.19 \\
& 16,398 & & 8.81E$-$05 & 3.38E$-$07 & 1.13E$-$07 & 8.82E$-$07 & & 26 & 10,101 & 49.43 & 0.60 \\
& 65,786 & & --- & --- & --- & --- & & 26 & 22,020 & 503.91 & 2.26 \\
\midrule
\multicolumn{11}{@{}l}{$\left(\bm{u},p\right)$} \\
\midrule
& 1,044 & & 2.51E$+$00 & 1.13E$+$00 & --- & --- & & 12 & 1,650 & 0.32 & 0.04 \\
& 4,070 & & 1.23E$+$00 & 3.39E$-$01 & --- & --- & & 12 & 3,980 & 3.79 & 0.14 \\
& 16,398 & & 9.49E$-$01 & 1.43E$-$01 & --- & --- & & 14 & 11,940 & 126.39 & 0.68 \\
& 65,786 & & 3.45E$-$01 & 7.27E$-$02 & --- & --- & & 14 & 31,701 & 3,029.64 & 3.58 \\
\midrule
\multirow{4}{*}{$3$}
& 1,044 & & 1.82E$+$00 & 2.20E$+$00 & --- & --- & & 12 & 1,670 & 0.51 & 0.05 \\
& 4,070 & & 9.10E$-$01 & 1.11E$+$00 & --- & --- & & 13 & 4,190 & 5.41 & 0.18 \\
& 16,398 & & 4.69E$-$01 & 5.69E$-$01 & --- & --- & & 14 & 12,700 & 156.03 & 0.85 \\
& 65,786 & & 2.18E$-$01 & 2.70E$-$01 & --- & --- & & 14 & 36,320 & 3,779.81 & 4.28 \\
\midrule
\multirow{4}{*}{$5$}
& 1,044 & & 1.74E$+$00 & 2.31E$+$00 & --- & --- & & 13 & 1,620 & 0.91 & 0.07 \\
& 4,070 & & 9.37E$-$01 & 1.25E$+$00 & --- & --- & & 13 & 4,570 & 9.29 & 0.27 \\
& 16,398 & & 5.31E$-$01 & 6.32E$-$01 & --- & --- & & 14 & 13,700 & 207.78 & 1.15 \\
& 65,786 & & 2.62E$-$01 & 3.17E$-$01 & --- & --- & & 14 & 40,610 & 4,633.66 & 5.53 \\
\bottomrule
\end{tabular}

}
\end{table}

\section{Conclusions}
\label{sec:conclusions}

In the present work, a finite volume method is proposed to solve the steady-state incompressible Navier-Stokes equations in non-primitive variables in domains with arbitrary smooth curved boundaries.
Firstly, the appropriate boundary conditions for the vorticity and streamfunction are derived from the prescribed boundary velocity.
Multiply connected domains are also carefully discussed since prescribing the appropriate boundary conditions requires a more involved derivation.
For the numerical scheme, the physical fluxes are approximated from polynomial reconstructions computed in the least-squares sense based on the approximate piecewise cell mean-values.
Polygonal meshes are generated with conventional meshing algorithms to discretise the curved domain, leading to a geometrical mismatch between the curved physical boundaries and the associated piecewise linear representation.
In this regard, the ROD method is employed to overcome the accuracy deterioration and achieve very high-orders of convergence for the approximate solution.
The method requires only a set of points on the physical boundary and avoids the difficulties of generating curved meshes and discretising on curved elements.
The resulting system of linear equations is solved using a coupled approach, and a fixed point algorithm iteratively computes an approximate solution for the non-linear Navier-Stokes problem.

The accuracy, convergence order, robustness, and stability of the proposed method were assessed with a comprehensive numerical benchmark of test cases, with increasing boundary shape complexity.
The proposed method achieves second-, fourth-, and sixth-orders of convergence for the approximate vorticity and streamfunction with $d=1,3,5$, respectively, except for the vorticity in the $L ^{\infty}$-norm, which generally achieves first-, third-, and fifth-orders of convergence for the same degrees.
Moreover, employing a higher polynomial degree for the streamfunction polynomial reconstructions that impose the prescribed boundary conditions can further improve the accuracy and the convergence order of the method.

The numerical and computational performance of the method in a more classical and practical scenario is also assessed with the semielliptical lid-driven cavity benchmark and compared to the same simulations performed with a primitive formulation.
In the regularised test case, the non-primitive formulation provides comparable accuracy to the primitive formulation for low Reynolds numbers ($Re=1$), but the latter is generally an order of magnitude more accurate with higher polynomial degrees for higher Reynolds numbers ($Re=100$).
In the conventional test case, the primitive formulation fails to recover the optimal convergence rates for higher polynomial degrees, whereas the non-primitive formulation mitigates the non-physical oscillations produced at the cavity corners for $d=1,3$.
Finally, the performance of the fixed point algorithm depends solely on the Reynolds number and, therefore, obtaining a very high-order of convergence requires roughly the same number of fixed point iterations as the low-order of convergence methods.
The non-primitive formulation typically requires more fixed point iterations than the primitive formulation for high Reynolds numbers, although it still requires fewer GMRES iterations in total.
More importantly, the proposed method recovers the optimal convergence orders in domains with arbitrary curved boundaries, while preserving the computational efficiency superiority that is expected with higher-orders of convergence.

In conclusion, the non-primitive formulation demonstrates a significant potential to solve the incompressible Navier-Stokes equations more efficiently than the primitive formulation for fluid flow problems at low Reynolds numbers and, particularly, problems with singular points in the solution at any Reynolds number.
This potential is further enhanced by the proposed method to achieve very high-orders of convergence, even in domains with arbitrary curved boundaries, providing the same results as the low-order of convergence with improvement ratios of several orders of magnitude in execution time.
Further developments of the proposed method include extensions to inlet/permeable boundary conditions and three-dimensional fluid flow problems, both requiring a careful formulation of the governing equations and appropriate boundary conditions.

\section*{Use of AI tools declaration}

The authors declare they have not used Artificial Intelligence (AI) tools in the creation of this article.

\section*{Acknowledgements}

R. Costa and J.M. N\'obrega acknowledge the financial support by FEDER -- Fundo Europeu de Desenvolvimento Regional, through COMPETE 2020 -- Programa Operational Fatores de Competitividade, and the National Funds through FCT -- Funda\c c\~ao para a Ci\^encia e a Tecnologia, project no. UIDB/05256/2020 and UIDP/05256/2020.

R. Costa also acknowledges the financial support of Funda\c c\~ao para a Ci\^encia e a Tecnologia (FCT) through the Contract-Program signed between the Foundation for Science and Technology (FCT) and the University of Minho, within the scope of the Individual Scientific Employment Stimulus Call -- 6th Edition (CEEC IND6ed), project no. 2023.09481.CEECIND/CP2841/CT0008.

S. Clain acknowledges the financial support by FEDER -- Fundo Europeu de Desenvolvimento Regional, through COMPETE 2020 -- Programa Operational Fatores de Competitividade, and the National Funds through FCT -- Funda\c c\~ao para a Ci\^encia e a Tecnologia, project no. UIDB/00324/2020.

G.J. Machado acknowledges the financial support of the Portuguese Foundation for Science and Technology (FCT) in the framework of the Strategic Funding, project no. UID/00013/2020 and UIDB/04650/2020.

The authors also acknowledge the infrastructure support of Search-ON2 -- Revitalization of HPC infrastructure of UMinho, project no. NORTE-07-0162-FEDER-000086, co-funded by the North Portugal Regional Operational Programme (ON.2 -- O Novo Norte), under the National Strategic Reference Framework (NSRF), through the European Regional Development Fund (ERDF).



\spacing{0.5}
\setlength{\bibsep}{-0.05cm plus 0.0cm}
\small


\spacing{1.0}
\setlength{\bibsep}{0.0cm plus 0.0cm}
\normalfont
\appendix
\section{Derivation of the vorticity transport equation}
\label{appendix:derivation_of_the_vorticity_transport_equation}

Using the Cartesian coordinate system $\bm{x}\coloneqq\left(x,y,z\right)$, the steady-state three-dimensional Navier-Stokes equations for a Newtonian fluid with constant dynamic viscosity $\mu>0$ and constant density $\rho>0$ consists of the momentum and mass balance equations, given as
\begin{alignat}{4}
\left(\bm{u}\cdot\nabla\right)\bm{u}-\nu\nabla^{2}\bm{u}+\frac{1}{\rho}\nabla p\;=\;&\bm{f},\label{eq:appendix_momentum_equation_1}\\
\nabla\cdot\bm{u}\;=\;&0,\label{eq:appendix_continuity_equation_1}
\end{alignat}
respectively, where $\nu=\mu/\rho$ is the constant kinematic viscosity, $\bm{u}\coloneqq\bm{u}\left(\bm{x}\right)\coloneqq\left(u_{x},u_{y},u_{z}\right)$ with $u_{x}\coloneqq u_{x}\left(\bm{x}\right)$, $u_{y}\coloneqq u_{y}\left(\bm{x}\right)$, and $u_{z}\coloneqq u_{z}\left(\bm{x}\right)$ is the unknown velocity function, $p\coloneqq p\left(\bm{x}\right)$ is the unknown pressure function, while $\bm{f}\coloneqq\bm{f}\left(\bm{x}\right)\coloneqq\left(f_{x},f_{y},f_{z}\right)$ with $f_{x}\coloneqq f_{x}\left(\bm{x}\right)$,$f_{y}\coloneqq f_{y}\left(\bm{x}\right)$, and $f_{z}\coloneqq f_{z}\left(\bm{x}\right)$ is a given body force function per unit volume acting on the continuum, for instance, the gravitational force.

Taking the curl of the momentum balance equation~\cref{eq:appendix_momentum_equation_1}, and using the definition of the vorticity vector in three-dimensions, $\bs{\omega}\left(\bm{x}\right)=\nabla\times\bm{u}\left(\bm{x}\right)$, and the identities
\begin{equation}
\nabla\times\left(\left(\bm{u}\cdot\nabla\right)\bm{u}\right)=\left(\bs{\omega}\cdot\nabla\right)\bm{u}-\left(\bm{u}\cdot\nabla\right)\bs{\omega},
\qquad
\nabla\times\left(\nu\nabla^{2}\bm{u}\right)=\nu\nabla^{2}\left(\nabla\times\bm{u}\right)=\nu\nabla^{2} \bs{\omega},
\qquad
\text{and}
\qquad
\nabla\times\nabla p=\bm{0},
\end{equation}
yields the vorticity vector transport equation, given as
\begin{equation}
\left(\bm{u}\cdot\nabla\right)\bs{\omega}-\left(\bs{\omega}\cdot\nabla\right)\bm{u}-\nu\nabla^{2}\bs{\omega}=\nabla\times\bm{f},
\end{equation}
where $\left(\bs{\omega}\cdot\nabla\right)\bm{u}$ represents vortex stretching, an important phenomenon in three-dimensional fluid flows.
For planar fluid flows in the $xy$-plane, the vorticity vector is perpendicular to the plane of motion such that the vortex stretching term vanishes, as well as the components of the vorticity vector transport equation parallel to the plane of motion.
Therefore, in two-dimensional fluid flows, the vorticity vector transport equation reduces to a scalar equation (the perpendicular component) for the scalar vorticity, defined as $\omega\left(\bm{x}\right)=\partial u_{y}\left(\bm{x}\right)/\partial x-\partial u_{x}\left(\bm{x}\right)/\partial y$, given as
\begin{equation}
\left(\bm{u}\cdot\nabla\right)\bs{\omega}-\nu\nabla^{2}\omega\;=\;f,
\end{equation}
where $f\coloneqq f\left(\bm{x}\right)=\partial f_{y}\left(\bm{x}\right)/\partial x-\partial f_{x}\left(\bm{x}\right)/\partial y$, which can be rewritten in conservative form using the identity $\left(\bm{u}\cdot\nabla\right)\omega=\nabla\cdot\left(\bm{u}\omega\right)-\omega\left(\nabla\cdot\bm{u}\right)$ and the incompressibility constraint $\nabla\cdot\bm{u}=0$ as
\begin{equation}
\nabla\cdot\left(\bm{u}\omega\right)-\nu\nabla^{2}\omega\;=\;f.
\end{equation}

\section{Derivation of the compatibility condition}
\label{appendix:derivation_of_the_compatibility_condition}

Following the notations and definition for local coordinate system introduced in Section~\ref{subsec:boundary_conditions_local_coordinate_system}, the equation for the tangential balance momentum along boundary subset $\Gamma^{k}$, $k=1,\ldots,K$, in terms of normal and tangential velocities is given as
\begin{equation}
\Biggl(u_{\eta}\frac{\partial}{\partial\eta}+u_{\xi}\frac{\partial}{\partial\xi}\Biggr)u_{\xi}-\nu\Biggl(\frac{\partial^{2}}{\partial\eta^{2}}+\frac{\partial^{2}}{\partial\xi^{2}}\Biggr)u_{\xi}+\frac{1}{\rho}\frac{\partial p}{\partial\xi}=f_{\xi},\qquad\text{on }\Gamma^{k},\quad k=1,\ldots,K,
\end{equation}
where $f_{\xi}\coloneqq f_{\xi}\left(\bs{\eta}\right)=\bm{f}\left(\bs{x}\right)\cdot\bm{t}\left(\bs{x}\right)$ is the tangential body force function per unit area acting on the continuum.

The convective term in the tangential balance momentum can be rewritten as
\begin{equation}
\Biggl(u_{\eta}\frac{\partial}{\partial\eta}+u_{\xi}\frac{\partial}{\partial\xi}\Biggr)u_{\xi}
=u_{\eta}\omega+u_{\eta}\frac{\partial u_{\eta}}{\partial\xi}+u_{\xi}\frac{\partial u_{\xi}}{\partial\xi}
=u_{\eta}\omega+\frac{\partial}{\partial\xi}\Biggl(\frac{1}{2}u_{\eta}^{2}\Biggr)+\frac{\partial}{\partial\xi}\Biggl(\frac{1}{2}u_{\xi}^{2}\Biggr),\qquad\text{on }\Gamma^{k},\quad k=1,\ldots,K.
\end{equation}
where the vorticity definition in terms of the local coordinate system~\eqref{eq:boundary_conditions_vorticity_local_definition} was used to derive the identity $\partial u_{\xi}/\partial\eta=\omega+\partial u_{\eta}/\partial\xi$.
Similarly, the viscous term in the tangential balance momentum can be rewritten as
\begin{equation}
-\nu\Biggl(\frac{\partial^{2}}{\partial\eta^{2}}+\frac{\partial^{2}}{\partial\xi^{2}}\Biggr)u_{\xi}
=-\nu\Biggl(\frac{\partial^{2}u_{\xi}}{\partial\eta^{2}}+\frac{\partial}{\partial\xi}\Biggl(-\frac{\partial u_{\eta}}{\partial\eta}\Biggr)\Biggr)
=-\nu\frac{\partial}{\partial\eta}\Biggl(\frac{\partial u_{\xi}}{\partial\eta}-\frac{\partial u_{\eta}}{\partial\xi}\Biggr)
=-\nu\frac{\partial\omega}{\partial\eta},\qquad\text{on }\Gamma^{k},\quad k=1,\ldots,K.
\end{equation}
where the mass conservation equation~\eqref{eq:mathematical_formulation_continuity_equation_1} was used to derive the identity $\partial u_{\xi}/\partial\xi=-\partial u_{\eta}/\partial\eta$.

Gathering the rewritten convective and viscous terms, the equation for the tangential balance momentum along boundary subset $\Gamma^{k}$, $k=1,\ldots,K$, can be rewritten as
\begin{equation}
u_{\eta}\omega+\frac{\partial}{\partial\xi}\Biggl(\frac{1}{2}u_{\eta}^{2}\Biggr)+\frac{\partial}{\partial\xi}\Biggl(\frac{1}{2}u_{\xi}^{2}\Biggr)-\nu\frac{\partial\omega}{\partial\eta}+\frac{1}{\rho}\frac{\partial p}{\partial\xi}=f_{\xi},\qquad\text{on }\Gamma^{k},\quad k=1,\ldots,K.
\end{equation}
Then, integrating the previous equation along boundary subset $\Gamma^{k}$, yields
\begin{equation}
\oint_{\Gamma^{k}}\Biggl(u_{\eta}\omega+\frac{\partial}{\partial\xi}\Biggl(\frac{1}{2}u_{\eta}^{2}\Biggr)+\frac{\partial}{\partial\xi}\Biggl(\frac{1}{2}u_{\xi}^{2}\Biggr)-\nu\frac{\partial\omega}{\partial\eta}+\frac{1}{\rho}\frac{\partial p}{\partial\xi}\Biggr)\diff{s}=\oint_{\Gamma^{k}}f_{\xi}\diff{s},\quad k=1,\ldots,K.
\end{equation}
Finally, using the gradient theorem for the integral of the tangential derivative along a closed path, the previous equation simplifies to
\begin{equation}
\oint_{\Gamma^{k}}\Biggl(u_{\eta}\omega-\nu\frac{\partial\omega}{\partial\eta}\Biggr)\diff{s}=\oint_{\Gamma^{k}}f_{\xi}\diff{s},\quad k=1,\ldots,K.
\end{equation}
or, equivalently, written in the global coordinate system as
\begin{equation}
\oint_{\Gamma^{k}}\left(\bm{u}\omega-\nu\nabla\omega\right)\cdot\bm{n}\diff{s}=\oint_{\Gamma^{k}}\bm{f}\cdot\bm{t}\diff{s},\quad k=1,\ldots,K.
\end{equation}

\section{Calculation of line curvature}
\label{appendix:calculation_of_line_curvature}

Consider a twice-differentiable plane curve represented parametrically in Cartesian coordinates as $\gamma\coloneqq\gamma\left(t\right)=\left(X,Y\right)$ with $X\coloneqq X\left(t\right)$ and $Y\coloneqq Y\left(t\right)$.
Assuming that the derivative $d\gamma\left(t\right)/dt$ is defined, differentiable, and nowhere equal to the zero vector on the domain, then the normal curvature as a function of the curve parameter, $\kappa\coloneqq\kappa\left(t\right)$, can be determined as
\begin{equation}
\kappa=\frac{\left|X'Y''-X''Y'\right|}{\left(\left(X'\right)^2+\left(Y'\right)^2\right)^{3/2}},
\end{equation}
where $X'\coloneqq X'\left(t\right)$ and $Y'\coloneqq Y'\left(t\right)$ are the first derivatives and $X''\coloneqq X''\left(t\right)$ and $Y''\coloneqq Y''\left(t\right)$ are the second derivatives of $\gamma\left(t\right)$ with respect to the parameter $t$, that is
\begin{equation}
X'=\frac{dX}{dt},
\qquad
Y'=\frac{dY}{dt},
\qquad
X''=\frac{d^{2}X}{dt^{2}},
\qquad
\text{and}
\qquad
Y''=\frac{d^{2}Y}{dt^{2}}.
\end{equation}

Consider a twice-differentiable plane curve represented parametrically in polar coordinates as $\gamma\coloneqq\gamma\left(\theta\right)=\left(R,\theta\right)$ with $R\coloneqq R\left(\theta\right)$, that is, the polar angle is the curve parameter and the radius of the curve is a function of the polar angle.
This particular curve can be easily parametrised in Cartesian coordinates as $\gamma\left(\theta\right)=\left(X,Y\right)$ with $X\coloneqq X\left(\theta\right)=R\left(\theta\right)\cos\left(\theta\right)$ and $Y\coloneqq Y\left(\theta\right)=R\left(\theta\right)\sin\left(\theta\right)$.
Thus, substituting $X$ and $Y$ in the normal curvature formula for a general parametrisation and making the necessary simplifications, the normal curvature as a function of the polar angle, $\kappa\coloneqq\kappa\left(\theta\right)$, is given as
\begin{equation}
\kappa=\frac{\left|R^{2}+2\left(R'\right)^{2}-RR''\right|}{\left(R^{2}+\left(R'\right)^{2}\right)^{3/2}},
\end{equation}
where $R'\coloneqq R'\left(\theta\right)$ and $R''\coloneqq R''\left(\theta\right)$ are the first and second derivatives of $R\left(\theta\right)$ with respect to the polar angle, that is
\begin{equation}
R'=\frac{dR}{d\theta}
\qquad
\text{and}
\qquad
R''=\frac{d^{2}R}{d\theta^{2}}.
\end{equation}



\begin{thebibliography}{}





\bibitem{1933_thom}{A. Thom,
The flow past circular cylinders at low speeds,
Proceedings of the Royal Society of London. Series A, Containing Papers of a Mathematical and Physical Character 141(845) (1933) 651--669.}

\bibitem{1954_woods}{L.C. Woods,
A note on the numerical solution of fourth order differential equations,
Aeronaut. Q. 5(4) (1954) 176--184.}


\bibitem{2017_costa1}{R. Costa, S. Clain, G.J. Machado,
A sixth-order finite volume scheme for the steady-state incompressible Stokes equations on staggered unstructured meshes,
J. Comput. Phys. 349 (2017) 501–-527.}

\bibitem{2017_costa2}{R. Costa, S. Clain, G.J. Machado, R. Loub\`ere,
A very high-order accurate staggered finite volume scheme for the stationary incompressible Navier–Stokes and Euler equations on unstructured meshes,
J. Sci. Comput. 71 (2017) 1375–-1411.}




\bibitem{1978_tuann}{S.Y. Tuann, M.D. Olson,
Review of computing methods for recirculating flows,
J. Comput. Phys. 29(1) (1978) 1--19.}


\bibitem{1983_schreiber}{R. Schreiber, H.B. Keller,
Driven cavity flows by efficient numerical techniques,
J. Comput. Phys. 49(2) (1983) 310--333.}

\bibitem{1989_goodrich}{J.W. Goodrich, W.Y. Soh,
Time-dependent viscous incompressible Navier-Stokes equations: the finite difference Galerkin formulation and streamfunction algorithms,
J. Comput. Phys. 84(1) (1989) 207–241.}

\bibitem{1990_goodrich_1}{J.W. Goodrich,
An unsteady time-asymptotic flow in the square driven cavity,
Technical report tech. mem. 103141, NASA (1990).}

\bibitem{1990_goodrich_2}{J.W. Goodrich, K. Gustafson, K. Halasi,
Hopf bifurcation in the driven cavity,
J. Comput. Phys. 90 (1990) 219–261.}







\bibitem{1963_fromm_1}{J.E. Fromm,
A method for computing nonsteady, incompressible, viscous fluid flows,
No. LA-2910. Los Alamos Scientific Lab., N. Mex., 1963.}

\bibitem{1963_fromm_2}{J.E. Fromm, F.H. Harlow,
Numerical solution of the problem of vortex street development,
Phys. Fluids 6(7) (1963) 975--982.}

\bibitem{1965_pearson}{C.E. Pearson,
A computational method for viscous flow problems,
J. Fluid Mech. 21(4) (1965) 611--622.}

\bibitem{1974_orszag}{S.A. Orszag, M. Israeli,
Numerical simulation of viscous incompressible flows,
Annu. Rev. Fluid Mech. 6(1) (1974) 281--318.}

\bibitem{1979_gupta}{M.M. Gupta, R.P. Manohar,
Boundary approximations and accuracy in viscous flow computations,
J. Comput. Phys. 31(2) (1979) 265--288.}


\bibitem{1972_cheng}{R.T.S. Cheng,
Numerical solution of the Navier‐Stokes equations by the finite element method,
Phys. Fluids 15(12) (1972) 2098--2105.}

\bibitem{1973_baker}{A.J. Baker,
Finite element solution algorithm for viscous incompressible fluid dynamics,
Int. J. Numer. Meth. Eng. 6(1) (1973) 89--101.}

\bibitem{1973_taylor}{C. Taylor, P. Hood,
A numerical solution of the Navier-Stokes equations using the finite element technique,
Comput. Fluids 1(1) (1973) 73--100.}

\bibitem{1988_tezduyar}{T.E. Tezduyar, R. Glowinski, J. Liou,
Petrov‐Galerkin methods on multiply connected domains for the vorticity‐stream function formulation of the incompressible Navier‐Stokes equations,
Int. J. Numer. Meth. Fluids 8(10) (1988) 1269--1290.}

\bibitem{1994_comini}{G. Comini, M. Manzan, C. Nonino,
Finite element solution of the streamfunction‐vorticity equations for incompressible two‐dimensional flows,
Int. J. Numer. Meth. Fluids 19(6) (1994) 513--525.}


\bibitem{1989_kettleborough}{C.F. Kettleborough, S.R. Husain, C. Prakash,
Solution of fluid flow problems with the vorticity-streamfunction formulation and the control-volume-based finite-element method,
Numer. Heat Transf. B 16(1) (1989) 31--58.}

\bibitem{1991_elkaim}{D. Elkaim, M. Reggio, R. Camarero,
Numerical solution of reactive laminar flow by a control-volume based finite-element method and the vorticity-streamfunction formulation,
Numer. Heat Transf. B 20(2) (1991) 223--240.}


\bibitem{1972_davis}{R.T. Davis,
Numerical solution of the Navier-Stokes equations for symmetric laminar incompressible flow past a parabola,
J. Fluid Mech. 51(3) (1972) 417--433.}

\bibitem{1978_campion-renson}{A. Campion‐Renson, M.J. Crochet,
On the stream function‐vorticity finite element solutions of Navier‐Stokes equations,
Int. J. Numer. Meth. Eng. 12(12) (1978) 1809--1818.}

\bibitem{1981_dhatt}{G. Dhatt, B.K. Fomo, C. Bourque,
A $\psi$–$\omega$ finite element formulations for the Navier‐Stokes equations,
Int. J. Numer. Meth. Eng. 17(2) (1981) 199--212.}

\bibitem{1984_gunzburger}{M.D. Gunzburger, J.S. Peterson,
On the finite element approximation of the stream function-vorticity equations,
NASA STI/Recon Technical Report N 84 (1984) 32758.}

\bibitem{1987_peeters}{M.F. Peeters, W.G. Habashi, E.G. Dueck,
Finite element stream function‐vorticity solutions of the incompressible Navier‐Stokes equations,
Int. J. Numer. Meth. Fluids 7(1) (1987) 17--27.}


\bibitem{1981_quartapelle_1}{L. Quartapelle, F. Valz‐Gris,
Projection conditions on the vorticity in viscous incompressible flows,
Int. J. Numer. Meth. Fluids 1(2) (1981) 129--144.}

\bibitem{1981_quartapelle_2}{L. Quartapelle,
Vorticity conditioning in the computation of two-dimensional viscous flows,
J. Comput. Phys. 40(2) (1943) 453--477.}

\bibitem{1984_quatapelle}{L. Quartapelle, M. Napolitano,
A method for solving the factorized vorticity‐stream function equations by finite elements,
Int. J. Numer. Meth. Fluids 4(2) (1984) 109--125.}

\bibitem{1989_anderson}{C.R. Anderson,
Vorticity boundary conditions and boundary vorticity generation for two-dimensional viscous incompressible flows,
J. Comput. Phys. 80(1) (1989) 72--97.}

\bibitem{1993_quartapelle}{L. Quartapelle,
Numerical Solution of the Incompressible Navier-Stokes Equations,
Springer, 1993.}


\bibitem{1996_e_1}{E. Weinan, J.G. Liu,
Vorticity boundary condition and related issues for finite difference schemes,
J. Comput. Phys. 124(2) (1996) 368--382.}

\bibitem{1996_e_2}{E. Weinan, J.G. Liu,
Finite difference schemes for incompressible flows in vorticity formulations,
In Esaim: Proceedings 1 (1996) 181--195. EDP Sciences.}

\bibitem{1991_gresho}{P.M. Gresho,
Some interesting issues in incompressible fluid dynamics, both in the continuum and in numerical simulation,
Adv. Appl. Mech. 28 (1991) 45--140.}

\bibitem{1999_napolitano}{M. Napolitano, G. Pascazio, L. Quartapelle,
A review of vorticity conditions in the numerical solution of the $\zeta$-$\psi$ equations,
Comput. Fluids 28.2 (1999) 139--185.}





\bibitem{2001_calhoun}{D. Calhoun,
A Cartesian grid method for solving the two-dimensional streamfunction-vorticity equations in irregular regions,
J. Comput. Phys. 176(2) (2002) 231--275.}

\bibitem{2003_li}{Z. Li, C. Wang,
A fast finite difference method for solving Navier-Stokes equations on irregular domains,
Commun. Math. Sci. 1 (2003) 180–196.}


\bibitem{1989_dennis}{S.C.R. Dennis, J.D. Hudson,
Compact h4 finite-difference approximations to operators of Navier-Stokes type,
J. Comput. Phys. 85(2) (1989) 390--416.}

\bibitem{1991_gupta}{M.M. Gupta,
High accuracy solutions of incompressible Navier-Stokes equations,
J. Comput. Phys. 93(2) (1991) 343--359.}

\bibitem{1995_li}{M. Li, T. Tang, B. Fornberg,
A compact fourth‐order finite difference scheme for the steady incompressible Navier‐Stokes equations,
Int. J. Numer. Meth. Fluids 20(10) (1995) 1137--1151.}

\bibitem{1995_spotz}{W.F. Spotz, G.F. Carey,
High‐order compact scheme for the steady stream‐function vorticity equations,
Int. J. Numer. Meth. Eng. 38(20) (1995) 3497--3512.}

\bibitem{1996_e_3}{E. Weinan, J.G. Liu,
Essentially compact schemes for unsteady viscous incompressible flows,
J. Comput. Phys. 126(1) (1996) 122--138.}

\bibitem{2001_li}{M. Li, T. Tang,
A compact fourth-order finite difference scheme for unsteady viscous incompressible flows,
J. Sci. Comput. 16 (2001) 29--45.}



\bibitem{2001_kupferman}{R. Kupferman,
A central-difference scheme for a pure stream function formulation of incompressible viscous flow,
SIAM J. Sci. Comput. 23(1) (2001) 1--18.}

\bibitem{2005_ben-artzi}{M. Ben-Artzi, J.P. Croisille, D. Fishelov, S. Trachtenberg,
A pure-compact scheme for the streamfunction formulation of Navier–Stokes equations,
J. Comput. Phys. 205(2) (2005) 640--664.}

\bibitem{2005_gupta}{M.M. Gupta, J.C. Kalita,
A new paradigm for solving Navier–Stokes equations: streamfunction–velocity formulation,
J. Comput. Phys. 207(1) (2005) 52--68.}

\bibitem{2006_ben-artzi}{M. Ben-Artzi, J.P. Croisille, D. Fishelov,
Convergence of a compact scheme for the pure streamfunction formulation of the unsteady Navier–Stokes system,
SIAM J. Numer. Anal. 44(5) (2006) 1997--2024.}

\bibitem{2010_ben-artzi}{M. Ben-Artzi, J.P. Croisille, D. Fishelov,
A high order compact scheme for the pure-streamfunction formulation of the Navier-Stokes equations,
J. Sci. Comput. 42(2) (2010) 216--250.}

\bibitem{2017_fishelov}{D. Fishelov,
A new fourth-order compact scheme for the Navier–Stokes equations in irregular domains,
Comput. Math. Appl. 74(1) (2017) 6--25.}


\bibitem{2018_costa}{R. Costa, S. Clain, R. Loub\`ere, G.J. Machado,
Very high-order accurate finite volume scheme on curved boundaries for the two-dimensional steady-state convection-diffusion equation with Dirichlet condition,
Appl. Math. Model. 54 (2018) 752--767.}

\bibitem{2019_costa1}{R. Costa, J.M. N\'obrega, S. Clain, G.J. Machado, R. Loub\`ere,
Very high-order accurate finite volume scheme for the convection-diffusion equation with general boundary conditions on arbitrary curved boundaries,
Int. J. Numer. Methods Eng. 117(2) (2019) 188--220.}

\bibitem{2019_costa2}{R. Costa, J.M. N\'obrega, S. Clain, G.J. Machado,
Very high-order accurate polygonal mesh finite volume scheme for conjugate heat transfer problems with curved interfaces and imperfect contacts,
Comput. Methods Appl. Mech. Eng. 357 (2019) 112560.}

\bibitem{2021_costa}{R. Costa, J.M. N\'obrega, S. Clain, G.J. Machado,
Efficient very high-order accurate polyhedral mesh finite volume scheme for 3D conjugate heat transfer problems in curved domains,
J. Comput. Phys. 445 (2021) 110604.}

\bibitem{2022_costa1}{R. Costa, J.M. N\'obrega, S. Clain, G.J. Machado,
High-order accurate conjugate heat transfer solutions with a finite volume method in anisotropic meshes with application in polymer processing,
Int. J. Numer. Eng. 123(4) (2022) 1146--1185.}

\bibitem{2022_costa2}{R. Costa, S. Clain, G.J. Machado, J.M. N\'obrega,
Very high-order accurate finite volume scheme for the steady-state incompressible Navier–Stokes equations with polygonal meshes on arbitrary curved boundaries,
Comput. Methods Appl. Mech. Eng. 396 (2022) 115064.}

\bibitem{2023_costa}{R. Costa, S. Clain, G.J. Machado, J.M. N\'obrega, H. Beir\~ao da Veiga, F. Crispo,
\textit{Imposing general slip conditions on curved boundaries for 3D incompressible flows with a very high-order accurate finite volume scheme on polygonal meshes},
Comput. Meth. Appl. Mech. Eng. 415 (2023) 116274.}

\bibitem{2024_santos}{M. Santos, A. Ara\'ujo, S. Barbeiro, S. Clain, R. Costa, G.J. Machado,
\textit{Very high-order accurate discontinuous Galerkin method for curved boundaries with polygonal meshes},
J. Sci. Comput. 100 (2024) 66.}



\bibitem{1971_wells}{D.E. Wells, E.J. Krakiwsky,
The method of least-squares,
Lecture notes 18, Department of Surveying Engineering, University of New Brunswick, Fredericton (1971).}

\bibitem{1982_bertsekas}{D.P. Bertsekas,
Constrained optimization and Lagrange multiplier methods,
1st Edition, Academic Press (1982).}



\bibitem{1979_girault}{V. Girault, P.A. Raviart,
Finite element approximation of the Navier-Stokes equations (Vol. 749),
Berlin: Springer (1979).}



\bibitem{1950_wannier}{G.H. Wannier,
A contribution to the hydrodynamics of lubrication,
Q. Appl. Math. 8(1) (1950) 1--32.}

\bibitem{2019_kuhlmann}{H.C. Kuhlmann, F. Roman\`o,
The lid-driven cavity, in Computational Modelling of Bifurcations and Instabilities in Fluid Dynamics, A. Gelfgat (Ed.) (2019) 233--309.}

%
%
%
%
%
%
%
%
%
%
%
%
%
%
%
%
%
%
%
%
%
%
%
%
%
%
%
%
%
%
%

\end{thebibliography}
\end{document}